\documentclass[11pt]{article}
\pdfoutput=1
\usepackage{amsmath,amssymb,amsfonts,a4wide,graphicx,bm,times,bbm}
\usepackage{mcite}
\usepackage{tikz}
\usepackage{hyperref}
\usepackage{amsthm} 
\usepackage{braket}
\usepackage{mathtools}
\usepackage{cite}
\usetikzlibrary{plotmarks}

\makeatletter \let\old@startsection=\@startsection
\renewcommand{\@startsection}[6]
{\old@startsection{#1}{#2}{#3}{#4}{#5}{#6\mathversion{bold}}}
\makeatother

\makeatletter

\@addtoreset{equation}{section}
\makeatother
\let\refOld\ref
\renewcommand{\ref}[1]{(\refOld{#1})}

\usetikzlibrary{decorations.pathreplacing,decorations.markings}
\tikzset{
  on each segment/.style={
    decorate,
    decoration={
      show path construction,
      moveto code={},
      lineto code={
        \path [#1]
        (\tikzinputsegmentfirst) -- (\tikzinputsegmentlast);
      },
      curveto code={
        \path [#1] (\tikzinputsegmentfirst)
        .. controls
        (\tikzinputsegmentsupporta) and (\tikzinputsegmentsupportb)
        ..
        (\tikzinputsegmentlast);
      },
      closepath code={
        \path [#1]
        (\tikzinputsegmentfirst) -- (\tikzinputsegmentlast);
      },
    },
  },
  mid arrow/.style={postaction={decorate,decoration={
        markings,
        mark=at position .5 with {\arrow[#1]{stealth}}
      }}},
}

 \newcommand{\braA}[1]{{_{A\!\!}}\bra{#1}}
 \newcommand{\braB}[1]{{_{B\!\!}}\bra{#1}}
 \newcommand{\braS}[1]{{ _{\otimes\!\!}\bra{#1}}}
 \newcommand{\ketA}[1]{{\ket{#1}_A}}
 \newcommand{\ketB}[1]{{\ket{#1}_B}}
 \newcommand{\ketS}[1]{{\ket{#1}_{\otimes}}}
 \newcommand{\dket}[1]{\ket{#1}\rangle}
 \newcommand{\dbra}[1]{\langle\bra{#1}}
 \newcommand{\superp}[2]{\genfrac{}{}{0pt}{}{#1}{#2}}
 
 \newcommand{\mat}[4]{\begin{bsmallmatrix}
 #1 & #2\\
 #3 & #4
 \end{bsmallmatrix}}
 
 \def\d{\delta}

 \def\p{\partial}
 
 \def\a{\alpha}
 \def\b{\beta}
 \def\g{\gamma}
 \def\d{\delta}
 \def\e{\epsilon}
 
 \def\th{\theta}
 
 \def\k{\kappa}
 \def\l{\lambda}

 \def\s{\sigma}
 \def\t{\tau}
 \def\th{\theta}

 \def\D{\Delta}

 \def\O{\Omega}
 \def\o{\omega }
 \def\u{\upsilon}
 \def\U{\Upsilon}
\def\CA{{\mathcal{A}}}

\def\CF{{\mathcal{F}}}
\def\CG{{\mathcal{G}}}

\def\CL{{\mathcal{L}}}
\def\CM{{\mathcal{M}}}
\def\CN{{\mathcal{N}}}
\def\CO{{\mathcal{O}}}
\def\CP{{\mathcal{P}}}

\def\CR{{\mathcal{R}}}
\def\CS{{\mathcal{S}}}
\def\CT{{\mathcal{T}}}
\def\CU{{\mathcal{U}}}

\def\CY{{\mathcal{Y}}}
\def\CZ{{\mathcal{Z}}}
\def\la{\left\langle}
\def\ra{\right\rangle}

\def\bc{{\bar{c}}}

\def\implies{\quad\Rightarrow\quad}

\def\vphi{\varphi}

\def\Zv{\mathcal{Z}_\text{vect.}}
\def\Zbf{\mathcal{Z}_\text{bfd.}}

\def\Zf{\mathcal{Z}_\text{fund.}}
\def\Zaf{\mathcal{Z}_\text{a.f.}}

\def\ZCS{\mathcal{Z}_\text{CS}}

\def\res{\mathop{\text{Res}}}

\def\baY{\langle\langle\vec{v},\vec{\lambda}|}

\def\qf{\mathfrak{q}}

\def\vac{\emptyset}

\def\hg{\hat\gamma}

\def\bCF{\bar{\mathcal{F}}}

\def\bd{{\bar d}}
\def\bc{{\bar c}}

\def\bt{{\bar \tau}}

\def\CYY{\mathcal{Y}_{\vec\lambda}}
\def\PsiY{\Psi_{\vec\lambda}}
\def\tup{\tilde{\upsilon}}
\def\tu{{\tilde{u}}}
\def\tv{{\tilde{v}}}
\def\bell{\bar\ell}
\def\Zol{\mathcal{Z}_\text{1-loop}}

\def\rf{\mathfrak{r}}
\def\Sf{\mathfrak{S}}
\def\bt{{\bar\tau}}
\def\bo{{\bar\omega}}
\def\DIM{\mathcal{A}_{\text{DIM}}}

\begin{document}
\begin{titlepage}
\vspace*{-2cm}
	\begin{flushright}
		KIAS-Q18023
	\end{flushright}
	
	\vskip 22mm
	
	\begin{center}
		{\huge\bf Fiber-base duality from the algebraic perspective}
	\end{center}
	\vskip 30mm
		\begin{center}
			{\Large J.-E. Bourgine$^\dagger$}
			\\[.4cm]
			{\em {}$^\dagger$Korea Institute for Advanced Study (KIAS)}\\
			{\em Quantum Universe Center (QUC)}\\
			{\em 85 Hoegiro, Dongdaemun-gu, Seoul, South Korea}
			\\[.4cm]
		\end{center}
	\vfill
	\begin{abstract}
		Quiver 5D $\CN=1$ gauge theories describe the low-energy dynamics on webs of $(p,q)$-branes in type IIB string theory. S-duality exchanges NS5 and D5 branes, mapping $(p,q)$-branes to branes of charge $(-q,p)$, and, in this way, induces several dualities between 5D gauge theories. On the other hand, these theories can also be obtained from the compactification of topological strings on a Calabi-Yau manifold, for which the S-duality is realized as a fiber-base duality. Recently, a third point of view has emerged in which 5D gauge theories are engineered using algebraic objects from the Ding-Iohara-Miki (DIM) algebra. Specifically, the instanton partition function is obtained as the vacuum expectation value of an operator $\CT$ constructed by gluing the algebra's intertwiners (the equivalent of topological vertices) following the rules of the toric diagram/brane web. Intertwiners and $\CT$-operators are deeply connected to the co-algebraic structure of the DIM algebra. We show here that S-duality can be realized as the twist of this structure by Miki's automorphism.
	\end{abstract}
	\vfill
\end{titlepage}

%
%

\section{Introduction}
The most remarkable achievement of string theory is arguably the derivation of non-perturbative dualities between quantum field theories. The network of dualities obtained in this way is particularly rich among four dimensional (class $\CS$) $\CN=2$ super Yang-Mills (SYM) theories, and their five dimensional $\CN=1$ uplifts. Indeed, these theories sustain correspondences with both classical and quantum integrable systems \cite{Gorsky1995,Donagi1995,Nekrasov2009}, but also with 2D Liouville/Toda conformal field theories (or q-deformed conformal blocks in the 5D scenario) through the celebrated AGT conjecture \cite{Alday2010,Wyllard2009,Awata2009}. Furthermore, these supersymmetric gauge theories can be realized as the low-energy limit of brane dynamics, and the application of (type IIB strings theory) S-duality on this construction generates dualities among them \cite{Witten1997,Aharony1997a,Bao2011}. From the 5D gauge theory perspective, S-duality implies an enhancement of global symmetries at the UV fixed points \cite{Mitev2014}, a fact that was first predicted by Seiberg in \cite{Seiberg1996} (and generalized further in \cite{Morrison1996,Intriligator1997}). This enhancement has been explicitly observed by computation of BPS quantities (conformal index in \cite{Kim2012}, Nekrasov partition functions in \cite{Mitev2014}).

This paper explores the realization of S-duality, also called \textit{fiber-base duality} (see below), among 5D $\CN=1$ quiver gauge theories using an algebraic formalism developed recently in the series of papers \cite{Mironov2016,Mironov2016a,Awata2016,Awata2016a,Awata2016b,Bourgine2017b,Bourgine2017c}, and based on the Ding-Iohara-Miki (DIM) algebra \cite{Ding1997,Miki2007}. This formalism, referred here as \textit{algebraic engineering}, highlights the integrable properties of the gauge theories'  BPS sector, and, at the same time, expresses in a very elegant manner the covariance properties under the (q-deformed) W-algebra at the root of AGT correspondence. In fact, we have two main motivations for our study. The first one is the perspective to export the S-duality tool to integrable systems and 2D conformal field theories (CFTs). In this way, a systematic derivation of spectral dualities between integrable systems (such as the one observed in \cite{Mukhin2005,Bulycheva2012,Chen2013,Mironov2013}), as well as non-trivial relations between 2D conformal blocks \cite{Mitev2014,Nedelin2017}, could be obtained. Our second motivation is the possibility to exploit the algebraic formalism to write the proofs of the non-trivial (and non-perturbative) relations that S-duality implies among gauge theories quantities (partition functions, BPS Wilson loops,...). Here, we propose to address these two issues using an automorphism of the DIM algebra, namely \textit{Miki's automorphism} \cite{Miki2007}, to twist the coalgebraic structure responsible for integrability.\footnote{As this paper was almost completed, we received the preprint \cite{Nieri2018} in which a similar program is proposed.}

In order to understand better the role of Miki's automorphism, it is necessary to review first the \textit{$(p,q)$-brane} construction of these gauge theories \cite{Aharony1997,Aharony1997a}. In type IIB string theory, a $(p,q)$-brane is a bound state of $p$ D5 and $q$ NS5-branes. It is a 5-brane that fills the dimensions 0-4 (where the gauge theory background lies) and forms a segment in the 56-plane oriented along the direction of angle $\th_{56}\sim \arctan(p/q)$. Hence, with this convention, pure NS5-branes are drawn horizontally (in the `5' direction) and pure D5-branes vertically. The brane-web represents the configuration of $(p,q)$-branes in the 56-plane, and characterizes the underlying $\CN=1$ theory. To be specific, we consider here 5D $\CN=1$ theories on the Omega-deformed background $\mathbb{R}^2_{\e_1}\times\mathbb{R}_{\e_2}^2\times S_R^1$. The corresponding partition functions are one-loop exact, but contain a tower of non-perturbative (instanton) corrections that have been computed by localization in \cite{Nekrasov2003}. Alternatively, the partition function can also be derived as a topological string amplitude \cite{Aganagic2005}. Away from the self-dual case $\e_1+\e_2=0$, it is necessary to employ the refined version of the topological vertex introduced in \cite{Iqbal2007,Awata2005,Taki2007}. The topological string is compactified on a certain Calabi-Yau threefold for which the toric skeleton (or, simply, toric diagram) coincides with the $(p,q)$-brane web \cite{Leung1998}. As a result, the brane-web provides the gluing rules for the topological vertices associated to each $(p,q)$-brane junction. 

In the $(p,q)$-brane web construction, the action of S-duality exchanges D5 and NS5-branes, and more generally acts on the branes' charges by sending $(p,q)$ to $(q,-p)$. This operation effectively rotates the $(p,q)$-branes web by 90$^\circ$, and thereby induces a duality between the gauge theories associated to the two brane webs, the original and the rotated ones. In the topological strings setup, the rotation of the toric diagram exchanges fiber and base for the Calabi-Yau, which coined the term \textit{fiber-base duality} \cite{Katz1997}.

The \textit{algebraic engineering} of the gauge theory is based on a realization by Awata, Feigin and Shiraishi \cite{Awata2011} of the (refined) topological vertex as an intertwiner of the DIM algebra.\footnote{The use of such intertwiners has been popularized by the Japanese school of Integrability, see for instance the book of Jimbo and Miwa \cite{Jimbo1994}.} This algebra depends on two parameters $q_1$ and $q_2$ that encode the dependence on the background parameters: $q_1=e^{R\e_1}$ and $q_2=e^{R\e_2}$. The algebraic engineering is, in fact, a reformulation of the $(p,q)$-brane construction in algebraic terms: to each brane of charge $(p,q)$ corresponds a representation characterized by the two levels $(\ell,\bell)=(q,p)$, and to each (trivalent) brane junction, or topological vertex, is associated an operator intertwining between three representations. The gluing rules of intertwiners also follow the brane-web picture, the gluing being simply realized as a product of operators in the shared representation. In this way, it is possible to associate to each brane-web an operator $\CT$ acting on a tensor product of representations. This operator is the essential object of the construction. Its vacuum expectation value (v.e.v.) reproduces the instanton partition function of the gauge theory \cite{Bourgine2017b,Bourgine2017c}. Besides, it can further be used to compute other BPS quantities (like the qq-characters \cite{Bourgine2017b}).

In \cite{Miki2007}, Miki describes an automorphism of the DIM algebra that acts on the two central charges $(c,\bc)$ by sending them to $(-\bc,c)$. As a result, a representation $\rho$ with levels $(\rho(c),\rho(\bc))=(\ell,\bell)$ composed with this automorphism becomes a representation $\rho'$ of different levels $(\rho'(c),\rho'(\bc))=(-\bell,\ell)$. Since the two levels encode the branes charges, the composition by Miki's automorphism renders the S-duality transformation in the algebraic formalism \cite{Awata2016}.\footnote{In fact, the whole $\text{SL}(2,\mathbb{Z})$ invariance of type IIB string theory can be realized in this way, but we will restrict ourselves to the element $\CS$.} Building on this fact, we investigate here the transformation of intertwiners, and their composite objects ($\CT$-operators, Lax matrices).\footnote{We call \textit{Lax matrix} the evaluation of the universal $\CR$-matrix in two specified representations. The $\CR$-matrix is the central object of quantum integrability, it is defined as an intertwiner between the coproduct and its permutation $\D'$, i.e. $\CR\D=\D'\CR$. On the other hand, (Baxter's) $\CT$-operators commute with the action of the algebra. These objects can be constructed using trivalent intertwiners, they will be defined more rigorously in the main body of the paper.} In this process, a twist of the coproduct by Miki's automorphism appears naturally, an the S-duality is reformulated as an equivalence between quantities derived from the two different coalgebraic structures. Note that this approach is essentially different from the one employed by Awata and Kanno in \cite{Awata2009a}, and in which S-duality relations follow from a change of the preferred direction for the topological vertex. In our case, the preferred direction is fixed, it corresponds to the vertical representation and it is associated to D5-branes once and for all. In practice, the main difference is that, in addition to changing the way topological vertices are coupled, we also need to implement the rotation of the vertices, and thus introduce new types of intertwiners associated to the rotated representations.

This paper is organized as follows. In the section two, we introduce the DIM algebra, describe the vertical (D5) and horizontal (NS5, NS5+D5) representations, and discuss the action of Miki's automorphism. The notion of intertwiner is reviewed in the third section, their transformation properties are investigated at a general level, and many new types of intertwiners are constructed. These results are used in the section four and five to analyze the S-transformation of Lax matrices and $\CT$-operators (respectively). Three important examples will be provided: the (resolved) conifold, and the pure $U(1)$ and $U(2)$ gauge theories. In each example, the S-duality relation appearing in gauge theory follows from the equality between the v.e.v. of algebraic quantities derived from the two coalgebraic structures. The transformation of the vacuum state is a key point in this derivation, and must be studied case by case. The technical parts of the discussions have been gathered in the appendices. The appendix A presents advanced properties of Miki's automorphism, it includes expressions for the S-dual generators, discusses their action in the vertical module, and investigate the possibility to refine the definition of Miki's automorphism using grading operators. The second appendix is a simple reminder on the building blocks entering in the construction of Nekrasov partition functions. Appendix C shows the construction of a new type of intertwiners, but also introduces new operators in the Cartan sector, and defines a set of coherent state for the Fock modules of horizontal representations. Finally, the appendix D contains the calculations relevant to the S-transformation of vacua.

\section{Algebra, representations and automorphisms}
\subsection{Presentation of the Ding-Iohara-Miki algebra}
The DIM algebra $\DIM$ is, in fact, the quantum toroidal algebra of $\mathfrak{gl}_1$. It can be formulated in terms of four currents 
\begin{equation}\label{DIM_currents}
x^\pm(z)=\sum_{k\in\mathbb{Z}}z^{-k}x^\pm_k,\quad \psi^\pm(z)=\sum_{k\geq0}z^{\mp k}\psi_{\pm k}^\pm,
\end{equation} 
together with a central element $\hg$, that satisfy
\begin{align}
\begin{split}\label{def_DIM}
&[\psi^\pm(z),\psi^\pm(w)]=0,\quad \psi^+(z)\psi^-(w)=\dfrac{g(\hg z/w)}{g(\hg^{-1}z/w)}\psi^-(w)\psi^+(z),\\
&\psi^+(z)x^\pm(w)=g(\hg^{\pm1/2}z/w)^{\pm1}x^\pm(w)\psi^+(z),\quad \psi^-(z)x^\pm(w)=g(\hg^{\mp1/2}z/w)^{\pm1}x^\pm(w)\psi^-(z),\\
&x^\pm(z)x^\pm(w)=g(z/w)^{\pm1}x^\pm(w)x^\pm(z),\\
&[x^+(z),x^-(w)]=\k\left(\d(\hg^{-1}z/w)\psi^+(\hg^{1/2}w)-\d(\hg z/w)\psi^-(\hg^{-1/2}w)\right).
\end{split}
\end{align}
In this formulation, $\k$ is a $\mathbb{C}$-number and $g(z)$ a function that can be expressed in terms of the three complex parameters $q_1,q_2,q_3$ constraint under the relation $q_1q_2q_3=1$,
\begin{equation}\label{def_g}
\k=\dfrac{(1-q_1)(1-q_2)}{(1-q_1q_2)},\quad g(z)=\prod_{\a=1,2,3}\dfrac{1-q_\a z}{1-q_\a^{-1}z}.
\end{equation}
The function $g(z)$ obeys the `unitarity' property $g(z)g(1/z)=1$ necessary to the consistency of the relations \ref{def_DIM}. The form of this algebra directly follows from the gauge theory background $\mathbb{R}^2_{\e_1}\times\mathbb{R}_{\e_2}^2\times S_R^1$, and the parameters are identified as $(q_1,q_2)=(e^{R\e_1},e^{R\e_2})$. Deformations of this algebra have been introduced to treat different backgrounds: an elliptic deformation for 6D gauge theories \cite{Foda2018}, a higher rank version (quantum toroidal $\mathfrak{gl}_n$) for the 5D background with orbifold \cite{Awata2017}, and a degenerate version for 4D $\CN=2$ gauge theories \cite{Bourgine2018a}.

The DIM algebra possesses two central charges denoted $c$ and $\bc$. The first one corresponds to the central element $\hg$ that can be written $\hg=\g^c$ with the shortcut notation $\g=q_3^{1/2}$. The second one is associated to the zero modes $\psi_0^\pm$ of the currents $\psi^\pm(z)$: these two modes are also central, and can be written $\psi_0^\pm=\g^{\mp\bar c}$.\footnote{The requirement $\psi_0^+=(\psi_0^-)^{-1}$ fixes partially the invariance of the algebra under a rescaling of the generators. The remaining invariance $x^\pm(z)\to\o^{\pm1}x^\pm(z)$ is associated to an automorphism generated by a grading element, it will be discussed below.} Hence, representations are labeled by two (integer) levels $(\ell,\bell)$ corresponding to the value of the central elements
\begin{equation}
\rho_v^{(\ell,\bell)}(c)=\ell,\quad \rho_v^{(\ell,\bell)}(\bar c)=\bell.
\end{equation}
Vector spaces equipped with a representation $\rho^{(\ell,\bell)}_v$ of levels $(\ell,\bell)$ and weight $v$ will be denoted $(\ell,\bell)_v$. Weights will later be associated to the (exponentiated) positions of the branes.

The subalgebra generated by the elements $\psi_{\pm k}^\pm$ and the central element $\hg$ can be seen as the analogue of the Cartan subalgebra of standard Lie algebras. It is sometimes useful to express the Cartan generators $\psi_{\pm k}^\pm$ in terms of the modes $a_k$ of an Heisenberg subalgebra \cite{Miki2007},
\begin{equation}\label{def_ak}
\psi^\pm(z)=\psi_0^\pm\exp\left(\pm\sum_{k>0}z^{\mp k}a_{\pm k}\right).
\end{equation}
Then, the relations $\psi\psi$, $\psi x$, and $[x,x]$ given in \ref{def_DIM} take a simpler form:
\begin{align}
\begin{split}\label{com_ak}
[a_k,a_l]=(\hg^k-\hg^{-k})c_k\d_{k+l},\quad [a_k,x_l^\pm]=\pm\hg^{\mp |k|/2}c_k x_{l+k}^\pm,\\
[x^+_k,x^-_l]=\left\{
\begin{array}{l}
\k\hg^{(k-l)/2}\psi^+_{k+l},\quad k+l>0\\
\k\hg^{(k-l)/2}\psi_0^+-\k\hg^{-(k-l)/2}\psi_0^-,\quad k+l=0\\
-\k\hg^{-(k-l)/2}\psi^-_{k+l},\quad k+l<0.\\
\end{array}
\right.
\end{split}
\end{align}
On the other hand, the q-commutation relations $x^\pm x^\pm$ remain fairly complicated. The coefficients $c_k$ in the r.h.s. of the commutators arise from the expansion of the function $g(z)$ in \ref{def_DIM}:
\begin{equation}\label{exp_g}
[g(z)]_{\pm}=\exp\left(\pm\sum_{k>0}z^{\mp k}c_{\pm k}\right),\quad c_k=-\dfrac1{k}\prod_{\a=1,2,3}(1-q_\a^k).
\end{equation} 
Here $[g(z)]_+$ refers to the expansion of the function $g(z)$ at infinity, while $[g(z)]_-$ denotes its expansion in a neighborhood of zero. It follows from the unitarity property of $g(z)$ that the coefficients $c_k$ obey $c_k=c_{-k}$.

\begin{figure}
\begin{center}
\includegraphics[width=4cm]{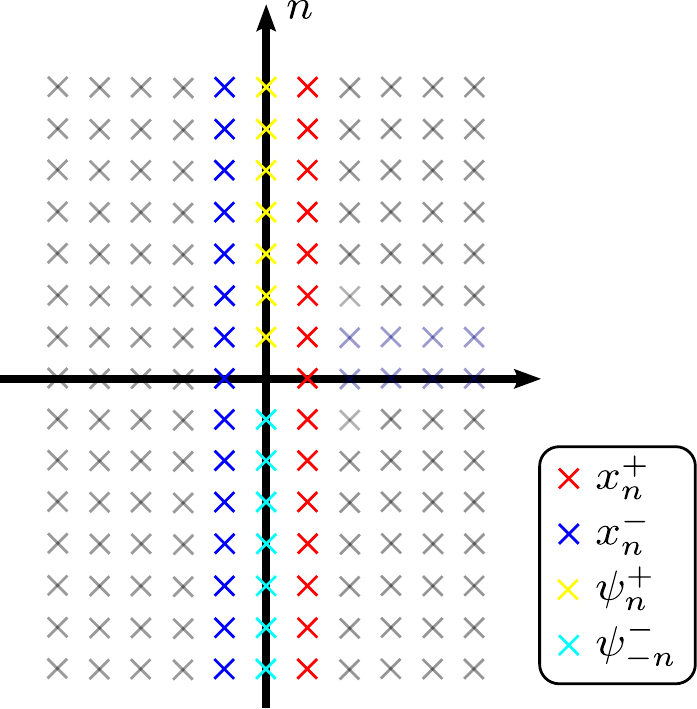}
\end{center}
\caption{DIM generators represented according to their degree $(x,y)=(\bd,-d)\in\mathbb{Z}\otimes\mathbb{Z}$.}
\label{fig_DIM}
\end{figure}

In addition, it is possible to associate two different degrees to the generators of the DIM algebra using the grading operators $d$ and $\bd$ obeying the commutation relations
\begin{equation}\label{act_gradings}
[d,x_k^\pm]=-kx_k^\pm,\quad [d,\psi_{\pm k}^\pm]=\mp k\psi_{\pm k}^\pm,\quad [\bd,x_k^\pm]=\pm x_k^\pm,\quad [\bd,\psi_{\pm k}^\pm]=0.
\end{equation}
These relations assign the degrees $(\bd,-d)\sim(\pm1,k)$ and $(\bd,-d)\sim(0,\pm k)$ respectively to the modes $x^\pm_k$ and $\psi_{\pm k}^\pm$, while central elements are of degrees $(0,0)$. Operators of higher degree under $\bd$ are obtained from the commutation of the modes $x_k^\pm$. Thus, DIM operators can be classified according to their degrees $(\bd,-d)\in \mathbb{Z}\times\mathbb{Z}$, and the generators have been represented in figure \ref{fig_DIM}. Furthermore, the grading operators can be used to define two different automorphisms,
\begin{equation}\label{def_gradings}
\t_\o(e)=\o^d e\o^{-d},\quad \bt_{\bo}(e)=\bo^{\bd}e\bo^{-\bd},
\end{equation}
for any element $e$ of the DIM algebra $\CA_{\text{DIM}}$. These automorphisms act on the modes of the Drinfeld currents as follows:
\begin{equation}\label{prop_gradings}
\t_\o(x^\pm(z))=x^\pm(\o z),\quad\t_\o(\psi^\pm(z))=\psi^\pm(\o z),\quad \text{and}\quad \bt_\bo(x^\pm(z))=\bo^{\pm1}x^\pm(z),\quad \bt_\bo(\psi^\pm(z))=\psi^\pm(z).
\end{equation}
In effect, $\t_\o$ describes the freedom of rescaling the variable $z$ of the currents, while $\bt_\bo$ generates a rescaling of the currents $x^\pm(z)$. These two automorphisms are discussed in greater details in the appendix (section \ref{AppA3}). In string theory, automorphisms of the DIM algebra correspond to geometric transformations of the (56)-plane (see for instance \cite{Bourgine2017b}, section 2.6). It is shown in appendix that the action of $\t_\o$ corresponds to an overall translation along the NS5-direction, while the action of $\bt_\bo$ encodes the translation along the perpendicular direction (D5-branes).

\subsection{Miki's automorphism}
The automorphism $\CS$ discovered by Miki in \cite{Miki2007} is fully determined by its action on the central charges $(c,\bc)\to (-\bc,c)$ and on the four generators
\begin{equation}\label{Miki_init}
a_1\to(\g-\g^{-1})x_0^+\to -a_{-1}\to -(\g-\g^{-1})x_0^-\to a_1.
\end{equation}
It is readily seen that this automorphism is of degree four (i.e. $\CS^4=1$). Formulas for the transformation of the generators $x_k^\pm$, $\psi_{\pm k}^\pm$ are provided in appendix \ref{AppA}. However, since they involve repeated commutations, they quickly become cumbersome to use. Apart from the modes $a_{\pm1}$ and $x_0^\pm$ appearing in \ref{Miki_init}, the four modes $x_1^\pm$ and $x_{-1}^\pm$ are the only ones that are mapped to modes of the Drinfeld currents \ref{DIM_currents}:
\begin{equation}
x_1^+\to\g^{-(c+\bc)/2}x_{-1}^+\to -x_{-1}^-\to-\g^{(c+\bc)/2}x_1^-\to x_1^+.
\end{equation}

Miki's automorphism also acts on the grading operators, sending $(d,\bd)$ to $(-\bd,d)$. This transformation follows from the requirement of keeping the commutation relations \ref{act_gradings} invariant. As a consequence, an element $e\in\DIM$ of degrees $(d_e,\bd_e)$ is mapped to another element with degrees $(\bd_e,-d_e)$. It results that the automorphism $\CS$ acts as a clockwise rotation of angle $90^\circ$ on the representation of the generators (c.f. figure \ref{fig_DIM}).

As we mentioned in the introduction, it is possible to define a new representation by the composition $\rho_v^{(\ell,\bell)}\circ\CS$. For short, we will call it the \textit{$\CS$-dual representation}. Due to the transformation property of central charges, this representation has levels $(-\bell,\ell)$. In string theory, the $(p,q)$-branes are uniquely characterized by their charges (and position), and so we expect that the representations associated to them will also be fully characterized by their levels (and weight). Thus, from the physics perspective, we expect that the $\CS$-dual representation $\rho_v^{(\ell,\bell)}\circ\CS$ identifies with a given representation $\rho_{\tilde{v}}^{(-\bell,\ell)}$, up to a possible change of basis:
\begin{equation}\label{prop_S}
\rho^{(\ell,\bell)}_v(\CS\cdot e)=\CM_{\CS}^{(\ell,\bell)-1}\rho_{\tilde{v}}^{(-\bell,\ell)}(e) \CM_{\CS}^{(\ell,\bell)},\quad\text{with}\quad \CM_\CS^{(\ell,\bell)}:(\ell,\bell)_v\to (-\bell,\ell)_{\tilde{v}}.
\end{equation}
In this relation, the image of the element $e\in\DIM$ under the automorphism $\CS$ has been denoted $\CS\cdot e$, and $\CM_{\CS}^{(\ell,\bell)}$ is an infinite (yet graded) matrix. Note also that while transformation of levels is known, on the contrary the transformation of the weights, formally denoted here $\tilde{v}=\CS\cdot v$, depends on the explicit form of the representation and has to be resolved case by case. Mathematically, the assumption \ref{prop_S} is rather challenging, and needs to be properly established. We will see below that the matrices $\CM_{\CS}^{(\ell,\bell)}$ can be constructed explicitly for the representations relevant to our construction. Furthermore, these matrices can be used to define the transformation matrices $\CM_{\CS^k}^{(\ell,\bell)}$ of arbitrary powers $\CS^k$ of Miki's automorphism using relations of the form
\begin{equation}
\CM_{\CS^2}^{(\ell,\bell)}=\CM_{\CS}^{(-\bell,\ell)}\CM_{\CS}^{(\ell,\bell)},\quad \CM_{\CS^2}^{(-\ell,-\bell)}=\CM_{\CS^2}^{(\ell,\bell)-1},\quad\text{and}\quad\CM_{\CS^{-1}}^{(\ell,\bell)}=\CM_{\CS}^{(\bell,-\ell)-1}.
\end{equation} 

\subsection{Twisted coproducts}
\subsubsection{Twist by an automorphism}
The DIM algebra has the structure of a Hopf algebra with the Drinfeld coproduct defined as
\begin{align}
\begin{split}\label{Drinfeld_coproduct}
&\D(x^+(z))=x^+(z)\otimes 1+\psi^-(\hg_{(1)}^{1/2}z)\otimes x^+(\hg_{(1)}z),\\
&\D(x^-(z))=x^-(\hg_{(2)} z)\otimes \psi^+(\hg_{(2)}^{1/2}z)+1\otimes x^-(z),\\
&\D(\psi^\pm(z))=\psi^\pm(\hg_{(2)}^{\pm1/2}z)\otimes\psi^\pm(\hg_{(1)}^{\mp1/2}z),\quad \D(a_k)=a_k\otimes \hg^{-|k|/2}+\hg^{|k|/2}\otimes a_k,\\
&\D(\hg)=\hg\otimes\hg,\quad \D(c)=c\otimes 1+1\otimes c,\quad \D(\bc)=\bc\otimes1+1\otimes \bc.
\end{split}
\end{align}
We denoted here $\hg_{(1)}=\hg\otimes1$ and $\hg_{(2)}=1\otimes\hg$. The coalgebraic structure also contains an antipode and a co-unit but since we do not need them in this paper, we will not give their expression. This Hopf algebra is quasi-triangular, which implies the existence of a universal R-matrix $\CR$ that intertwines $\D$ with the opposite coproduct $\D'$ obtained by permutation ($\D'=\CP\D\CP$ with $\CP(a\otimes b)=(b\otimes a)\CP$, $\CP^2=1$):
\begin{equation}\label{def_R}
\CR\D=\D'\CR,\quad (\D\otimes 1)\CR=\CR_{13}\CR_{23},\quad (1\otimes\D)\CR=\CR_{13}\CR_{12}.
\end{equation}
The expression of the universal R-matrix with generic $c$ and $\bc$ follows from the Corollary 1.2 in \cite{Negut2013} (see also \cite{Maulik2012,Awata2016,Awata2016b,Fukuda2017}).\footnote{I would like to thank A. Negut for this clarification.} 

Miki's automorphism preserves both the linear and multiplicative structure of the algebra, therefore it can be used to twist the co-algebraic structure \cite{Khoroshkin1994}. The twisted coproduct
\footnote{This coproduct $\D_\CS$ differs from $\D$ as it is possible to compute, for instance,
\begin{equation}
\D(x_0^+)=x_0^+\otimes1+\sum_{k>0}\g^{-k(c\otimes1)/2}\ \psi_{-k}^-\otimes x_k^+\quad\text{and}\quad \D_\CS(x_0^+)=x_0^+\otimes\g^{-\bc/2}+\g^{\bc/2}\otimes x_0^+.
\end{equation}}
\begin{equation}\label{def_DCS}
\D_\CS(e)=(\CS^{-1}\otimes\CS^{-1})\D(\CS\cdot e),\quad \forall e\in\CA_{\text{DIM}},
\end{equation} 
defines a new quasi-triangular Hopf algebra with the universal R-matrix $\CR_\CS=(\CS^{-1}\otimes\CS^{-1})\CR$ and a twisted antipode. The coproduct $\D_\CS$ appears naturally in the transformation of trivalent intertwiners (a.k.a. topological vertices). Indeed, the intertwining property involves a representation $\rho$ obtained from two other representations $\rho_1$ and $\rho_2$ by coproduct, that is $\rho=(\rho_1\otimes\rho_2)\D$. Then, the $\CS$-dual representation $\rho'=\rho\circ\CS$ can be written in terms of $\rho_i'=\rho_i\circ\CS$ ($i=1,2$) using the twisted coproduct $\rho'=(\rho_1'\otimes\rho_2')\D_\CS$.

We can define in a similar way a twisted coproduct $\D_{\CS^k}$ for any power $\CS^k$ of Miki's automorphism. The action of $\CS^2$ on the Drinfeld currents \ref{DIM_currents} takes a simple form:
\begin{equation}\label{action_CSCS}
\CS^2\cdot x^\pm(z)=-x^\mp(z^{-1}),\quad \CS^2\cdot\psi^\pm(z)=\psi^{\mp}(z^{-1}),\quad \CS^2\cdot(c,\bc)=(-c,-\bc),\quad \CS^2\cdot a_k=-a_{-k}.
\end{equation}
As a result, it is possible to write down explicitly the action of the twisted coproduct $\D_{\CS^2}$ on these generators. Remarkably, this twisted coproduct coincides with the opposite coproduct $\D'$ involved in the definition of the universal R-matrix. This leads to interpret the automorphism $\CS$ as a sort of square root of the R-matrix. This result should have tremendous consequences for integrable systems built upon DIM algebra \cite{Feigin2015,Feigin2016,Feigin2016a}. We hope to come back to this important issue in a future publication.

\subsubsection{Twist by a two-tensor}
The so-called \textit{Drinfeld twist} provides another way to deform a coalgebraic structure \cite{Reshetikhin1990,Khoroshkin1994}. In this case, the coproduct is twisted by an invertible two-tensor $\CF$ that should satisfy the following properties:
\begin{equation}
\CF_{12}(\D\otimes1)\CF=\CF_{23}(1\otimes\D)\CF,\quad (\varepsilon\otimes1)\CF=(1\otimes\varepsilon)\CF=1.
\end{equation}
Here we used the standard notation $\CF_{ij}$ for the action of $\CF$ in $i$th and $j$th tensor space and $\varepsilon$ denotes the co-unit. The twisted coproduct $\CF\D\CF^{-1}$ defines a new quasi-triangular Hopf algebra with universal R-matrix $\CF^\ast\CR\CF^{-1}$, $\CF^\ast=\CF_{21}$ denoting the permutation of the two-tensor, i.e. $\CP\CF\CP$. Although a general statement does not seem to exist, it has been observed in some cases that the twisting by an automorphism can also be realized as a two-tensor twist. For instance, this is the case for Lusztig's automorphism of $U_q(\mathfrak{g})$ where $\mathfrak{g}$ is a simple, finite-dimensional, Lie algebra \cite{Khoroshkin1994}. This is also the case for the twist by the automorphism $\CS^2$, since $\D_{\CS^2}$ identifies with the opposite coproduct, and the corresponding two-tensor $\CF_{\CS^2}$ is simply the universal R-matrix $\CR$. In fact, the existence of the two tensor $\CF_\CS$ has been prooved in \cite{Feigin2016} (lemma A.5) in the restricted case $c=0$.

In this paper, it will be convenient, although not essential, to assume that the $\CS-$twisted coproduct $\D_\CS$ defined in \ref{def_DCS} can also be obtained by a Drinfeld twist with the two-tensor $\CF_\CS$:
\begin{equation}\label{def_FS}
\D_{\CS}=\CF_\CS\D\CF_\CS^{-1},\quad \CR_\CS=\CF_{\CS}^\ast\CR\CF_{\CS}^{-1},
\end{equation}
where $\CF_\CS^\ast=\CF_{\CS21}$ is the permutation of $\CF_\CS$. This assumption will provide us a general intuition on the transformation of algebraic objects. However, strictly speaking, we do not need the existence of the universal object $\CF_\CS$, but only its realization in specific representations. The latter will be constructed explicitly below, up to a normalization factor, using a product of intertwiners.

Several interesting properties follow from our assumption \ref{def_FS}. First, the opposite coproduct $\D'$ can be twisted in the same manner,
\begin{equation}
\D_\CS'(e)=(\CS^{-1}\otimes\CS^{-1})\D'(\CS\cdot e)=\CF_\CS^\ast\D'(e)\CF_\CS^{\ast-1}.
\end{equation} 
Moreover, since $\CS^4=1$, we notice that $\D'_\CS=\D_{\CS^3}=\D_{\CS^{-1}}$, which implies the existence of the two tensor $\CF_{\CS^{-1}}=\CF_\CS^\ast\CR$ associated with the twist by the automorphism $\CS^{-1}$, and the corresponding twisted R-matrix is $\CR_{\CS^{-1}}=\CR_\CS^{-1}$. Similarly, the identification $\D'_{\CS^{-1}}=\D_{\CS}$ provides the permuted relation $\CF_{\CS^{-1}}^\ast=\CF_{\CS}\CR^{-1}$. Finally, inverting these relations, it is possible to express the universal R-matrices solely in terms of $\CF_\CS$ and $\CF_{\CS^{-1}}$ (and their permutation), and re-interpret them as factors in a certain decomposition,
\begin{equation}
\CR=\CF_{\CS^{-1}}^{\ast-1}\CF_\CS=\CF_\CS^{\ast-1}\CF_{\CS^{-1}},\quad \CR_\CS=\CF_\CS^\ast\CF_{\CS^{-1}}^{\ast-1}=\CF_{\CS^{-1}}\CF_\CS^{-1}.
\end{equation} 

\subsection{Representations}
In the algebraic engineering, vertical representations are associated to (multiple) D5-branes, and horizontal representations to NS5-branes (possibly dressed by extra D5-branes). These representations have already been presented in several papers, we reproduce them here for consistency and follow the conventions employed in \cite{Bourgine2017c}. Then, we investigate the $\CS$-dual representations and derive the expression of several matrices $\CM_\CS^{(\ell,\bell)}$.

\subsubsection{Vertical representations}
The modules of vertical representations, denoted here $(0,m)_{\vec v}$ were introduced in \cite{feigin2011quantum}. They are infinite dimensional vector spaces with a basis of states $\ket{\vec{v}, \vec\lambda}\rangle$ parameterized by an $m$-tuple Young diagram $\vec \l=(\l^{(1)},\cdots,\l^{(m)}$), and on which the Drinfeld currents act as follows:
\begin{align}\label{def_vert_rep}
\begin{split}
&\rho_{\vec v}^{(0,m)}(x^+(z))\ket{\vec v,\vec\lambda}\rangle =\sum_{x\in A(\vec\lambda)}\delta(z/\chi_x)\res_{z=\chi_x}\dfrac1{z\CY_{\vec\lambda}(z)}|\vec v,\vec\lambda+x\rangle\rangle,\\
&\rho_{\vec v}^{(0,m)}(x^-(z))\ket{\vec 
v,\vec\lambda}\rangle=\gamma^{-m}\sum_{x\in R(\vec\lambda)}\delta(z/\chi_x)\res_{z=\chi_x}z^{-1}\CY_{\vec\lambda}(q_3^{-1}z)\ket{\vec v,\vec \lambda-x}\rangle,\\
&\rho_{\vec v}^{(0,m)}(\psi^\pm(z))\ket{\vec v,\vec \lambda}\rangle=\gamma^{-m}\left[\Psi_{\vec\lambda}(z)\right]_\pm\ket{\vec v,\vec\lambda}\rangle.
\end{split}
\end{align}
These representations have levels $(\ell,\bell)=(0,m)$, they are lowest weight representations, with the vacuum $\dket{\vec v,\vac}$ corresponding to empty Young diagrams, and the lowest weights given by the vector $\vec v=(v_1,\cdots,v_m)$. They are associated to $m$ parallel D5-branes, with (exponentiated) positions $v_1,\cdots,v_m$ along the axis '5' of the 56-plane. In the gauge theory, the level $m$ coincides with the rank of the gauge group $U(m)$, and the weight with the exponentiated Coulomb branch v.e.v..

The modes of the currents $x^\pm(z)$ behave as instantons creation/annihilation operators, and their action involves a summation over the sets of boxes $A(\vec\l)$ (resp. $R(\vec\l)$) that can be added to $\vec\l$ (or removed from $\vec\l$). To each box $x=(l,i,j)$ in $\vec\l$ with coordinates $(i,j)$ in the $l$th Young diagram, has been associated the complex parameter $\chi_x=v_lq_1^{i-1}q_2^{j-1}\in\mathbb{C}$. The vertical action \ref{def_vert_rep} has been written using the functions
\begin{equation}\label{def_CYY}
\Psi_{\vec\lambda}(z)=\frac{\CY_{\vec\lambda}(q_3^{-1}z)}{\CY_{\vec\lambda}(z)},\quad \CY_{\vec\lambda}(z)=\frac{\prod_{x\in A(\vec\lambda)}1-\chi_x/z}{\prod_{x\in R(\vec\lambda)}1-\chi_x/(q_3z)}.
\end{equation} 
These two functions possess an alternative expression as a product over the box content of each Young diagram:
\begin{equation}\label{def_S}
\PsiY(z)=\prod_{l=1}^m\dfrac{1-q_3v_l/z}{1-v_l/z}\prod_{x\in\vec\l}g(z/\chi_x),\quad \CYY(z)=\prod_{l=1}^m\left(1-\dfrac{v_l}{z}\right)\prod_{x\in\vec\l}S(\chi_x/z),\quad S(z)=\dfrac{(1-q_1z)(1-q_2z)}{(1-z)(1-q_1q_2z)}.
\end{equation} 
The function $S(z)$ is related to the function $g(z)$ defined in \ref{def_g}, and involved in the definition of the algebra, through the relation $g(z)=S(z)/S(z^{-1})$. It obeys the functional identity $S(q_3z)=S(z^{-1})$.

By expanding the functions $\PsiY(z)$ both around $z=\infty$ and $z=0$, it is possible to deduce the representation of the modes $a_k$ describing the Cartan sector:
\begin{equation}
\rho^{(0,m)}_{\vec v}(a_k)|\vec v,\vec\l\rangle\rangle=c_k\left(\sum_{x\in\vec\l}\chi_x^k-\dfrac1{(1-q_1^k)(1-q_2^k)}\sum_{l=1}^mv_l^k\right)|\vec v,\vec\l\rangle\rangle,\quad k\in\mathbb{Z}\setminus\{0\}.
\end{equation}
This expression involves the coefficients $c_k$ defined in \ref{exp_g} from the two expansions of the function $g(z)$.

It is also necessary to introduce a dual basis $\dbra{\vec v,\vec\l}$ for the vertical modules, which is orthogonal (but not orthonormal) to the basis of states $\dket{\vec v,\vec\l}$:
\begin{equation}\label{scalar_vert}
\langle\langle\vec v,\vec\l|\vec v,\vec\l'\rangle\rangle=\d_{\vec\l,\vec\l'}\ a_{\vec\l}^{-1},\quad a_{\vec\l}=\Zv(\vec v,\vec\l)\prod_{l=1}^m(-\g v_l)^{-|\vec\l|}\prod_{x\in\vec\l}\chi_x^m.
\end{equation}
The scalar product involves the quantity $\Zv(\vec v,\vec\l)$ that coincides with the vector multiplet contribution to the Nekrasov instanton partition function. A brief reminder on the various building blocks for these partition functions can be found in appendix \ref{AppB}. Note that in the case of $m=1$, the (inverse) norms $a_\l$ are in fact independent of the weight $v$ of the representation. The non-trivial norm is introduced here in order to simplify the expression of the contragredient representation $\hat\rho$ that acts on the dual basis in such a way that
\begin{equation}\label{def_hatrho}
\left(\baY\hat\rho^{(0,m)}(e)\right)|\vec v,\vec\l'\rangle\rangle=\baY\left(\rho^{(0,m)}(e)|\vec v,\vec\l'\rangle\rangle\right),\quad\forall e\in\CA_{\text{DIM}}.
\end{equation}
With our choice of normalization, the action of $\hat\rho$ reads just like \ref{def_vert_rep}, but with $x^\pm(z)$ replaced by $-x^\mp(z)$. 

Vertical representations of levels $(\ell,\bell)=(0,-m)$ can be defined using the automorphism $\CS^2$. The corresponding modules are isomorphic to $(0,m)_{\vec v}$, and the action of the Drinfeld currents on the basis takes the form
\begin{align}\label{def_bvert_rep}
\begin{split}
&\rho_{\vec v}^{(0,-m)}(x^+(z))\ket{\vec 
v,\vec\lambda}\rangle=\gamma^{-m}\sum_{x\in R(\vec\lambda)}\delta(z\chi_x)\res_{z=\chi_x^{-1}}z^{-1}\CY_{\vec\lambda}(q_3^{-1}z^{-1})\ket{\vec v,\vec \lambda-x}\rangle,\\
&\rho_{\vec v}^{(0,-m)}(x^-(z))\ket{\vec v,\vec\lambda}\rangle =\sum_{x\in A(\vec\lambda)}\delta(z\chi_x)\res_{z=\chi_x^{-1}}\dfrac1{z\CY_{\vec\lambda}(z^{-1})}|\vec v,\vec\lambda+x\rangle\rangle,\\
&\rho_{\vec v}^{(0,-m)}(\psi^\pm(z))\ket{\vec v,\vec \lambda}\rangle=\gamma^{-m}\left[\Psi_{\lambda}(z^{-1})\right]_\pm\ket{\vec v,\vec\lambda}\rangle,\\
&\rho^{(0,-m)}_{\vec v}(a_k)|\vec v,\vec\l\rangle\rangle=-c_k\left(\sum_{x\in\vec\l}\chi_x^{-k}-\dfrac1{(1-q_1^{-k})(1-q_2^{-k})}\sum_{l=1}^mv_l^{-k}\right)|\vec v,\vec\l\rangle\rangle.
\end{split}
\end{align}
Since by definition $\rho^{(0,-m)}_{\vec v}(e)=\rho^{(0,m)}_{\vec v}(\CS^2\cdot e)$, the transformation matrix $\CM_{\CS^2}^{(0,m)}$ is actually trivial here, and $(p,q)$-branes of charges $(0,m)$ and $(0,-m)$ can be formally identified.

\subsubsection{Horizontal representations}
Representations with level $\ell\in\mathbb{Z}\setminus\{0\}$ are called horizontal representations, they are obtained as a tensor product of $|\ell|$ q-bosonic Fock modules. In this paper, we restrict ourselves to the case of a single q-boson, i.e. $|\ell|=1$. Then, all the modules $(\pm1,n)_u$ with $n\in\mathbb{Z}$ are isomorphic to the Fock space defined over the q-bosonic modes $\a_k$ with the commutation relations 
\begin{equation}\label{def_sk}
[\a_k,\a_l]=\s_k\d_{k+l},\quad\text{with}\quad \s_k=k\g^k(1-q_1^{k})(1-q_2^{k}).
\end{equation} 
Positive modes annihilate the Fock vacuum $\ket{\vac}$, while negative modes create excitations. We further introduce the dual state $\bra{\vac}$, annihilated by negative modes, and the normal ordering $:\cdots:$ is defined by moving the positive modes to the right. The horizontal representation will be defined using the vertex operators
\begin{equation}\label{def_eta_vphi}
\eta^\pm(z)=:\exp\left(\mp\sum_{k\neq0}\dfrac{z^{-k}}{k}\g^{\mp|k|/2}\a_k\right):,\quad\vphi^\pm(z)=\exp\left(\pm\sum_{k>0}\dfrac{z^{\mp k}}{k}(\g^k-\g^{-k})\a_{\pm k}\right).
\end{equation} 

In representation of levels $(\pm1,n)$ and weight $u$, the DIM algebra acts on the Fock modules as follows \cite{Feigin2009a}:\footnote{In contrast with the convention used in \cite{Bourgine2017b}, here we have rescaled the weights of representations $(\pm1,n)$ by sending $u\to\pm u\g^{\mp n}$.}
\begin{align}
\begin{split}\label{q-osc}
&\rho^{(1,n)}_u(a_k)=\dfrac{\g^k-\g^{-k}}{k}\a_k,\quad \rho^{(-1,n)}_u(a_k)=-\dfrac{\g^k-\g^{-k}}{k}\a_{-k},\quad \rho^{(\pm1,n)}_u(\hg)=\g^{\pm1},\\
&\rho^{(1,n)}_u(x^\pm(z))=u^{\pm1}z^{\mp n}\eta^\pm(z),\quad \rho^{(1,n)}_u(\psi^\pm(z))=\g^{\mp n}\vphi^\pm(z),\\
&\rho^{(-1,n)}_u(x^\pm(z))=-u^{\mp1}z^{\pm n}\eta^\mp(z^{-1}),\quad \rho^{(-1,n)}_u(\psi^\pm(z))=\g^{\mp n}\vphi^\mp(z^{-1}).
\end{split}
\end{align}
As the levels indicate, these representations are associated to a bound state of a single NS5-brane and $|n|$ D5-branes. The weight $u$ encodes the position of the $(p,q)$-brane along the axis '6' of the 56-plane. Due to the property $\rho_u^{(1,n)}(\CS^2\cdot e)=\rho_{u}^{(-1,-n)}(e)$ ($\forall e\in\DIM$), the transformation matrices $\CM_{\CS^2}^{(\pm1,n)}$ are trivial, and the weight $u$ remains invariant under $\CS^2$. As a result, $(p,q)$-branes of charges $(1,n)$ and $(-1,-n)$ can also be identified.

\subsubsection{Action of Miki's automorphism on the representations}
In this paper, we restrict ourselves to representations with levels $\ell,\bell\in\{-1,0,1\}$ that define a closed subset under the action of Miki's automorphism. To study this action, we need to introduce a new basis in the horizontal Fock space, in addition to the usual PBW basis. This is achieved using an isomorphism sending horizontal representations into representations acting on Macdonald symmetric polynomials $P_\l$ with the parameters $q=q_2$ and $t=q_1^{-1}$ \cite{Feigin2009a}. Under this isomorphism, the vacuum state $\ket{\vac}$ coincides with the constant $1$, and the oscillator modes are written in terms of the power sum symmetric polynomials $p_k$:
\begin{equation}
\a_{-k}\equiv(1-q_1^k)\g^{k/2}p_k,\quad \a_k\equiv k(1-q_2^k)\g^{k/2}\dfrac{\p}{\p p_k},\quad (k>0).
\end{equation} 
Under this identification, the zero mode of the vertex operator $\eta^+(z)$ reproduces the Macdonald operator. The latter is diagonalized by the Macdonald polynomials $P_\l$. It leads us to introduce the corresponding states $\ket{P_\l}$ in the Fock modules as an eigenbasis of the operator $\eta_0^+$ with the non-degenerate eigenvalues $E_\l$:\footnote{The two possible expressions for the function $\CYY(z)$ (given \ref{def_CYY} and \ref{def_S}) induce at the asymptotics $z\to\infty$ the relation 
\begin{equation}
\sum_{x\in A(\vec\l)}\chi_x-\sum_{x\in R(\vec\l)}q_3^{-1}\chi_x=\sum_lv_l-\s_1\g^{-1}\sum_{x\in\vec\l}\chi_x.
\end{equation}
which provides an alternative expression for $E_\l$.}
\begin{equation}\label{eigen_eta0}
\eta_0^+\ket{P_\l}=E_\l\ket{P_\l},\quad\text{with}\quad E_\l=\sum_{(i,j)\in A(\l)}q_1^{i-1}q_2^{j-1}-\sum_{(i,j)\in R(\l)}q_1^{i}q_2^{j}.
\end{equation}
However, this eigenvalue equation does not fix the normalization of the states. Instead of taking the standard norm of Macdonald polynomials, here we adjust it in such a way that the Pieiri rule with the elementary symmetric polynomial $e_1=p_1$ coincides precisely with the vertical action of $x_0^+$:\footnote{The original Pieiri rule for $e_1$ with the standard norm of Macdonald polynomials \cite{Macdonald} can be written as
\begin{equation}
e_1P_\l=\sum_{x\in A(\l)}\dfrac{\prod_{\superp{y\in R(\l)}{y<x}}(1-q_3^{-1}\chi_y/\chi_x)\prod_{\superp{y\in A(\l)}{y<x}}(1-q_3\chi_y/\chi_x)}{\prod_{\superp{y\in R(\l)}{y<x}}(1-\chi_y/\chi_x)\prod_{\superp{y\in A(\l)}{y<x}}(1-\chi_y/\chi_x)}\ P_{\l+x},
\end{equation} 
where the ordering on the boxes $x=(i,j)\in\l$ is defined such that $x<x'$ iff $i<i'$ or $i=i'$ and $j<j'$. However, here instead of the normalization $P_\l=m_\l+\cdots$ (with $m_\l$ the monomial symmetric function), we use $P_\l=n_\l m_\l+\cdots$ with
\begin{equation}
n_\l=\left(-\g^{1/2}\dfrac{(1-q_2^{-1})}{(1-q_3)}\right)^{|\l|}\prod_{\superp{x,y\in\l}{y<x}}S(\chi_x/\chi_y)\prod_{(i,j)\in\l}(1-q_1^{-i}q_2^{-j})\dfrac{1-q_1q_2^{1-j}}{1-q_2^{-j}}.
\end{equation}
We would have chosen the standard norm instead, the coefficients $n_\l$ would have appeared as the values of the matrices $\CM_\CS^{(\cdot,\cdot)}$ on the diagonal.}
\begin{equation}
\a_{-1}\ket{P_\l}=-\sum_{x\in A(\l)}\res_{z=\chi_x}\dfrac1{z\CY_\l(z)}\ket{P_{\l+x}}.
\end{equation}
In the same spirit, we also modified the norm of dual Macdonald polynomial $Q_\l$, and introduced the dual basis $\bra{P_\l}$ such that $\bra{P_\l}P_\mu\rangle=a_{\l}^{-1}\d_{\l,\mu}$ with the coefficients $a_{\l}$ given in \ref{scalar_vert}. This choice of normalization for the vertical and horizontal basis corresponds to a choice of framing factors for the topological vertex \cite{Taki2007}.

In order to compare with vertical representations, we also need the action of $\eta_0^-$ on the Macdonald states $\ket{P_\l}$. It can be obtained using the isomorphism $\s_V$ that sends the DIM algebra with parameters $q_1,q_2,q_3$ to the DIM algebra with inverse parameters $q_1^{-1},q_2^{-1},q_3^{-1}$. This isomorphism encodes the vertical reflection in the 56 plane \cite{Bourgine2017c}. It sends the mode $x_0^+$ to $x_0^-$, and as a consequence,
\begin{equation}
\eta_0^-\ket{P_\l}=E_\l^\ast\ket{P_\l},\quad\text{with}\quad E_\l^\ast=\sum_{(i,j)\in A(\l)}q_1^{-(i-1)}q_2^{-(j-1)}-\sum_{(i,j)\in R(\l)}q_1^{-i}q_2^{-j}.
\end{equation}
Finally, the action of $\a_1$ follows from the Pieiri rules applied to the dual Macdonald polynomial with our specific choice of normalization it gives
\begin{equation}\label{action_a1}
\a_1\ket{P_\l}=-\g^{-1}\sum_{x\in R(\l)}\res_{z=\chi_x}z^{-1}\CY_\l(q_3^{-1}z)\ket{P_{\l-x}}.
\end{equation} 

We now compare horizontal and vertical modules. It is seen in appendix \ref{AppA} that the DIM algebra is in fact generated by only four modes: $a_{\pm1}$ and $x_0^\pm$. Hence, it is sufficient to examine the action of these four modes in order to show the isomorphism \ref{prop_S} between vertical and horizontal modules. Comparing $\rho_v^{(0,\pm1)}(\CS\cdot e)$ and $\rho_u^{(\pm1,0)}(e)$ for $e\in\{a_{\pm1},x_0^\pm\}$, we deduce the existence of the isomorphisms $\CM_\CS^{(0,\pm1)}$ and $\CM_\CS^{(\pm1,0)}$ mapping vertical states $\dket{v,\l}$ to the Macdonald states $\ket{P_\l}$, and vice-versa:
\begin{equation}\label{CMCS}
\CM_\CS^{(0,1)}=\CM_\CS^{(0,-1)}=\sum_\l a_\l\ \ket{P_\l}\dbra{v,\l},\quad \CM_\CS^{(1,0)}=\CM_\CS^{(-1,0)}=\sum_\l a_\l\ \dket{v,\l}\bra{P_\l}.
\end{equation} 
Under this transformation, the vertical weights $v$ are mapped to horizontal weights $u=-\g v$,
\begin{equation}
(0,1)_v\to(-1,0)_{-\g v}\to(0,-1)_v\to(1,0)_{-\g v}\to (0,1)_v.
\end{equation} 
The presence of this extra factor is a matter of conventions, it is related to the fact that the weights of the vertical representation are usually defined as (minus) the roots of the Drinfeld polynomial that really correspond to $-\g v_l$ ($\ell=1,\cdots,m$).

Finally, we would like to consider the $\CS$-transformation of horizontal representations $(1,\pm1)$ and $(-1,\pm1)$. Using the properties 
\begin{equation}
\rho_u^{(1,n)}(x_n^+)=u\eta_0^+,\quad\text{and}\quad \rho_u^{(-1,n)}(x_n^-)=-u\eta_0^+,
\end{equation} 
it is easy to show that, the modules beeing isomorphic, if the matrices $\CM_\CS^{(\pm1,1)}$ and $\CM_\CS^{(1,\pm1)}$ exist, they must be diagonal in the basis $\ket{P_\l}$. Moreover, the identities $\CM_{\CS^2}^{(1,1)}=\CM_\CS^{(-1,1)}\CM_\CS^{(1,1)}$ (and similarly for the four other matrices) impose some relations among the eigenvalues of the different matrices $\CM_\CS$. At this stage, we can write down following the Ansatz:
\begin{equation}
\CM_\CS^{(1,1)}=\CM_\CS^{(-1,-1)}=\sum_\l a_\l d_\l\ \ket{P_\l}\bra{P_\l},\quad \CM_\CS^{(1,-1)}=\CM_\CS^{(-1,1)}=\sum_\l a_\l d_\l^{-1}\ \ket{P_\l}\bra{P_\l},
\end{equation} 
with $d_\l$ unknown coefficients. The horizontal weights are, once again, observed to be invariant under $\CS$-transformation.

The proof of the existence of the matrices $\CM_\CS$ between the representations $(\pm1,1)$ and $(\pm1,-1)$ is a little more involved. We consider only $\CM_{\CS}^{(1,1)}$, since the other matrices can be treated in the same way. The DIM algebra is generated by only four modes that we can choose to be $a_{\pm1}$ and $x_{\pm1}^\pm$. The previous arguments can be applied to both $x_{\pm1}^\pm$ using the isomorphism $\s_V$. Thus, it only remains to treat the case of $a_{\pm1}\propto\CS\cdot x_0^\mp$, for which we need to show that
\begin{equation}\label{MS11}
\rho_u^{(1,1)}(\CS\cdot x_0^\pm)=\CM_\CS^{(1,1)-1}\rho_{\tu}^{(-1,1)}(x_0^\pm)\CM_\CS^{(1,1)}\quad\Leftrightarrow\quad\CM_\CS^{(1,1)-1}\eta_{\pm1}^\pm\CM_\CS^{(1,1)}=\tu^{\mp1}\a_{\pm1}.
\end{equation}
For this purpose, we consider the commutator $\eta_{\pm1}^\pm=\pm\g^{\pm1/2}\s_1^{-1}[\a_{\pm1},\eta_0^\pm]$ in the Macdonald basis $\ket{P_\l}$. We will focus here on $\eta_1^+$, the treatment of $\eta_{-1}^-$ being similar. Comparing
\begin{equation}
\CM_\CS^{(1,1)-1}\eta_{1}^+\CM_\CS^{(1,1)}\ket{P_\l}=-\g^{-1/2}\s_1^{-1}\sum_{x\in R(\l)}\dfrac{d_\l}{d_{\l-x}}(E_\l-E_{\l-x})\res_{z=\chi_x}z^{-1}\CYY(q_3^{-1}z)\ket{P_{\l-x}}
\end{equation} 
with the action \ref{action_a1} of $\a_1$, we deduce that the constraint \ref{MS11} on $\eta_1^+$ is satisfied provided that the diagonal elements of the matrix $\CM_\CS^{(1,1)}$ equal
\begin{equation}
d_\l=d_{\vac}(-\g^{1/2}\tu^{-1})^{|\l|}\left(\prod_{(i,j)\in\l}q_1^{i-1}q_2^{j-1}\right)^{-1}.
\end{equation} 
The analysis of the constraint coming from $\eta_{-1}^-$ in \ref{MS11} gives the same value for $d_\l$. The choice of $d_\vac$ is let free, we set $d_\vac=1$ so that $\CM_{\CS}^{(1,1)}$ maps the vacuum state $\ket{\vac}$ to itself.

\section{$\CS$-transformation of intertwiners}
\subsection{A general approach to S-dual intertwiners}
\subsubsection{Definition of intertwiners and twisting}
\begin{figure}
\begin{center}
\begin{tikzpicture}
\draw[postaction={on each segment={mid arrow=black}}] (0,0) -- (1,0) -- (1.7,0.7);
\draw[postaction={on each segment={mid arrow=black}}] (1,-1) -- (1,0);
\node[above,scale=0.7] at (1,0) {$\Phi$};
\node[above,scale=0.7] at (0,0) {$(\ell_1,\bell_1)_{v_1}$};
\node[right,scale=0.7] at (1,-0.5) {$(\ell_2,\bell_2)_{v_2}$};
\node[above,scale=0.7] at (1.7,0.7) {$(\ell,\bell)_{v}$};
\end{tikzpicture}
\hspace{10mm}
\begin{tikzpicture}
\draw[postaction={on each segment={mid arrow=black}}] (-0.7,-0.7) -- (0,0) -- (0,1);
\draw[postaction={on each segment={mid arrow=black}}] (0,0) -- (1,0);
\node[left,scale=0.7] at (0,0) {$\Phi^{\ast}$};
\node[below,scale=0.7] at (1,0) {$(\ell_2,\bell_2)_{v_1}$};
\node[right,scale=0.7] at (0,0.5) {$(\ell_1,\bell_1)_{v_2}$};
\node[below,scale=0.7] at (-0.7,-0.7) {$(\ell,\bell)_{v}$};
\end{tikzpicture}
\end{center}
\caption{Representation of the intertwiners $\Phi$ and $\Phi^\ast$ as generalized topological vertices.}
\label{fig_INT}
\end{figure}
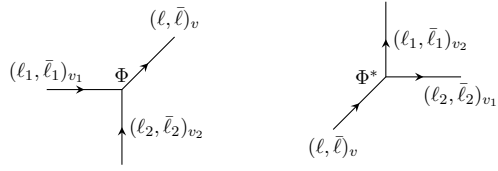

Trivalent intertwiners are the building blocks of the algebraic engineering of 5D $\CN=1$ gauge theories for which they play the role of the refined topological vertex. However, in contrast with topological strings, there exists several types of intertwiners in the DIM algebra, each parameterized by the level of the representations involved. In this way, it is possible to associate certain `higher rank' intertwiners to brane-webs involving more than one topological vertex (see \cite{Bourgine2017b}), leading to a more efficient method to compute topological strings amplitudes. In addition, it is also necessary to distinguish between the intertwiner $\Phi$, and the dual one $\Phi^\ast$. Indeed, the web of intertwiners should be oriented, the orientation corresponding to the mapping from one module to another. Intertwiners $\Phi$ and $\Phi^\ast$ correspond respectively to vertices with one and two outgoing arrows (see figure \ref{fig_INT}). Notice that $\Phi^\ast$ can be obtained from $\Phi$ by a reflection along the diagonal $x_5+x_6=0$ in the 56 plane (flipping also the arrows).

Algebraically, intertwiners are defined as the solution to the following equations:
\begin{align}
\begin{split}\label{AFS_lemmas}
&\rho_{v}^{(\ell,\bell)}(e)\Phi\mat{\ell_1}{\bell_1}{\ell_2}{\bell_2}=\Phi\mat{\ell_1}{\bell_1}{\ell_2}{\bell_2}\ \left(\rho_{v_1}^{(\ell_1,\bell_1)}\otimes\rho_{v_2}^{(\ell_2,\bell_2)}\ \D(e)\right),\\
&\left(\rho_{v_1}^{(\ell_1,\bell_1)}\otimes\rho_{v_2}^{(\ell_2,\bell_2)}\ \D'(e)\right)\Phi^\ast\mat{\ell_1}{\bell_1}{\ell_2}{\bell_2}=\Phi^\ast\mat{\ell_1}{\bell_1}{\ell_2}{\bell_2}\rho_{v}^{(\ell,\bell)}(e).
\end{split}
\end{align}
These equations do not determine the intertwiners uniquely, as it is still possible multiply them by a factor depending on the levels and weights (for instance). Whenever it is possible, this ambiguous factor will be set to one. Once this ambiguity is fixed, we expect that intertwiners involved in the correspondence with topological strings become unique. The relations \ref{AFS_lemmas} applied to the central charges $c$ and $\bc$ imply the conservation of the levels $\ell=\ell_1+\ell_2$ and $\bell=\bell_1+\bell_2$, corresponding to the charge conservation of the branes. Thus, intertwiners can be labeled by only two sets of levels $(\ell_1,\bell_1)$ and $(\ell_2,\bell_2)$. Obviously, intertwiners also depend on the weights of the representations, however we will not indicate them in order to lighten the notation. A constraint on these weights is also observed, it takes the form $\o^{(\ell,\bell)}(v)=\o^{(\ell_1,\bell_1)}(v_1)\o^{(\ell_2,\bell_2)}(v_2)$ with the representation-dependent functions of the weights $\o^{(\ell,\bell)}(v)$. Explicit calculations give the values $\o^{(\pm1,n)}(u)=u$ for horizontal representations, and $\o^{(0,\pm m)}(\vec v)=\prod_{l=1}^m(-\g v_l)$ for vertical representations.

There are two (somewhat equivalent) ways of defining the action of S-duality on intertwiners: either replacing the representations by the S-dual ones, or replacing the coproduct by its twisted version. The first point of view is relevant to string theory, since S-duality implies the rotation of $(p,q)$-branes that effectively replace representations by the dual ones. We will call \textit{rotated intertwiner} the intertwiner $\Phi\mat{-\bell_1}{\ell_1}{-\bell_2}{\ell_2}$ describing the topological vertex after rotation of the brane web. On the other hand, the replacement of the coproduct by its twisted version is more interesting from the integrability perspective. We will call \textit{S-dual intertwiner} the corresponding intertwiner, and denote it $\Phi_\CS$. This object is defined as the solution of the equations \ref{AFS_lemmas} with $\D$ replaced by $\D_\CS$:
\begin{align}
\begin{split}\label{AFS_lemmas_S}
&\rho_{v}^{(\ell,\bell)}(e)\Phi_{\CS}\mat{\ell_1}{\bell_1}{\ell_2}{\bell_2}=\Phi_{\CS}\mat{\ell_1}{\bell_1}{\ell_2}{\bell_2}\ \left(\rho_{v_1}^{(\ell_1,\bell_1)}\otimes\rho_{v_2}^{(\ell_2,\bell_2)}\ \D_\CS(e)\right),\\
&\left(\rho_{v_1}^{(\ell_1,\bell_1)}\otimes\rho_{v_2}^{(\ell_2,\bell_2)}\ \D_\CS'(e)\right)\Phi_\CS^{\ast}\mat{\ell_1}{\bell_1}{\ell_2}{\bell_2}=\Phi_{\CS}^\ast\mat{\ell_1}{\bell_1}{\ell_2}{\bell_2}\rho_{v}^{(\ell,\bell)}(e).
\end{split}
\end{align}
The relation between rotated and S-dual intertwiners involves the matrices $\CM_\CS$, it will be given below.

\subsubsection{Action of $\CS^2$ and inversion of intertwiners}
Before looking at the action of $\CS$, it is instructive to consider first $\CS^2$. There are two different ways to solve the equations \ref{AFS_lemmas} for $\Phi_{\CS^2}$ and $\Phi_{\CS^2}^\ast$ (i.e. with $\D$ and $\D'$ exchanged since $\D_{\CS^2}=\D'$ and $\CS^4=1$). The first possibility is to use the intertwining property of the universal R-matrix and replace $\D'$ with $\CR\D\CR^{-1}$. As a result, we find the solutions
\begin{equation}\label{sol1_SS}
\Phi_{\CS^2}\mat{\ell_1}{\bell_1}{\ell_2}{\bell_2}=\Phi\mat{\ell_1}{\bell_1}{\ell_2}{\bell_2}\CR\mat{\ell_1}{\bell_1}{\ell_2}{\bell_2}^{-1},\quad \Phi_{\CS^2}^\ast\mat{\ell_1}{\bell_1}{\ell_2}{\bell_2}=\CR\mat{\ell_1}{\bell_1}{\ell_2}{\bell_2}^{-1}\Phi^\ast\mat{\ell_1}{\bell_1}{\ell_2}{\bell_2},
\end{equation} 
where we have introduced the Lax matrix
\begin{equation}
\CR\mat{\ell_1}{\bell_1}{\ell_2}{\bell_2}=\left(\rho_{v_1}^{(\ell_1,\bell_1)}\otimes\rho_{v_2}^{(\ell_2,\bell_2)}\right)\CR.
\end{equation} 

A second solution can be found by introducing the matrices $\CM_{\CS^2}$ using the property \ref{prop_S}. A simple calculation gives
\begin{align}
\begin{split}\label{sol2_SS}
&\Phi_{\CS^2}\mat{\ell_1}{\bell_1}{\ell_2}{\bell_2}=\CM_{\CS^2}^{(\ell,\bell)-1}\Phi\mat{-\ell_1}{-\bell_1}{-\ell_2}{-\bell_2}\left(\CM_{\CS^2}^{(\ell_1,\bell_1)}\otimes\CM_{\CS^2}^{(\ell_2,\bell_2)}\right),\\
&\Phi_{\CS^2}^\ast\mat{\ell_1}{\bell_1}{\ell_2}{\bell_2}=\left(\CM_{\CS^2}^{(\ell_1,\bell_1)-1}\otimes\CM_{\CS^2}^{(\ell_2,\bell_2)-1}\right)\Phi^\ast\mat{-\ell_1}{-\bell_1}{-\ell_2}{-\bell_2}\CM_{\CS^2}^{(\ell,\bell)}.
\end{split}
\end{align}
This solution relates the two definitions of $\CS^2$-dual intertwiners. However, the main subtlety here is that this second solution may not coincide with the previous one, although we expect that they differ only by a normalization factor. In the following, as a point of reference for normalization factors, we consider \ref{sol2_SS} as the genuine solution.

Then, it is important to observe the following fact. Considering the two following products of intertwiners (taken in both intermediate modules $(\ell_1,\bell_1)_{v_1}$ and $(\ell_2,\bell_2)_{v_2}$) as endomorphisms of $(\ell,\bell)_v$,
\begin{equation}\label{invert_intw}
\CN\mat{\ell_1}{\bell_1}{\ell_2}{\bell_2}=\Phi\mat{\ell_1}{\bell_1}{\ell_2}{\bell_2}\Phi_{\CS^2}^\ast\mat{\ell_1}{\bell_1}{\ell_2}{\bell_2},\quad \CN^\ast\mat{\ell_1}{\bell_1}{\ell_2}{\bell_2}=\Phi_{\CS^2}\mat{\ell_1}{\bell_1}{\ell_2}{\bell_2}\Phi^\ast\mat{\ell_1}{\bell_1}{\ell_2}{\bell_2},
\end{equation} 
it is possible to show using the properties \ref{AFS_lemmas} that they commute with the action of the DIM algebra in the representation $(\ell,\bell)_v$. In the case of interest, namely vertical $(0,\pm m)$ and horizontal $(\pm1,n)$ representations, this property implies that $\CN$ and $\CN^\ast$ are constants (of course, depending on levels and weights). When these constants are finite, the dual intertwiner $\Phi^\ast_{\CS^2}$ provides the right inverse of $\Phi$, and similarly $\Phi_{\CS^2}$ gives the left inverse of $\Phi^\ast$. We would like to emphasize that these considerations, although very general, can also be used in practice to write down the inverse of an intertwiner using the solution \ref{sol2_SS} found previously. We will give below an explicit verification of the two identities \ref{invert_intw} in certain representations.

We would like to conclude this paragraph on the action of $\CS^2$ with a side remark. In fact, there is a third way to find a solution for $\Phi_{\CS^2}$ and $\Phi_{\CS^2}^\ast$, in which we introduce the permutation operator $\CP(a\otimes b)=(b\otimes a)\CP$. In this way, we find:
\begin{equation}\label{sol3_SS}
\Phi_{\CS^2}\mat{\ell_1}{\bell_1}{\ell_2}{\bell_2}=\Phi\mat{\ell_2}{\bell_2}{\ell_1}{\bell_1}\CP\mat{\ell_1}{\bell_1}{\ell_2}{\bell_2},\quad \Phi_{\CS^2}^\ast\mat{\ell_1}{\bell_1}{\ell_2}{\bell_2}=\CP\mat{\ell_2}{\bell_2}{\ell_1}{\bell_1}\Phi^\ast\mat{\ell_2}{\bell_2}{\ell_1}{\bell_1},
\end{equation} 
with
\begin{equation}
\CP\mat{\ell_1}{\bell_1}{\ell_2}{\bell_2}:(\ell_1,\bell_1)\otimes(\ell_2,\bell_2)\to(\ell_2,\bell_2)\otimes(\ell_1,\bell_1).
\end{equation}

\subsubsection{$\CS$-transformation of intertwiners}\label{sec_matrix}
Just like in the case of $\Phi_{\CS^2}$, there exists several ways of solving the twisted intertwining relations \ref{AFS_lemmas_S}. The simplest way, assuming the existence of the two-tensor $\CF_\CS$, is to replace $\D_{\CS}$ with $\CF_\CS\D\CF_\CS^{-1}$ (and $\D'_\CS$ with $\CF_\CS^\ast\D'\CF_\CS^{\ast-1}$). In this case, we find
\begin{equation}
\Phi_{\CS}\mat{\ell_1}{\bell_1}{\ell_2}{\bell_2}=\Phi\mat{\ell_1}{\bell_1}{\ell_2}{\bell_2}\CF_\CS\mat{\ell_1}{\bell_1}{\ell_2}{\bell_2}^{-1},\quad\Phi_{\CS}^\ast\mat{\ell_1}{\bell_1}{\ell_2}{\bell_2}=\CF_\CS^{\ast}\mat{\ell_1}{\bell_1}{\ell_2}{\bell_2}\Phi^\ast\mat{\ell_1}{\bell_1}{\ell_2}{\bell_2}.
\end{equation} 
with
\begin{equation}
\CF_\CS\mat{\ell_1}{\bell_1}{\ell_2}{\bell_2}=\rho^{(\ell_1,\bell_1)}_{v_1}\otimes\rho^{(\ell_2,\bell_2)}_{v_2}\ \CF_\CS,\quad \CF_\CS^{\ast}\mat{\ell_1}{\bell_1}{\ell_2}{\bell_2}=\rho^{(\ell_1,\bell_1)}_{v_1}\otimes\rho^{(\ell_2,\bell_2)}_{v_2}\ \CF_\CS^{\ast}.
\end{equation} 
This solution is not very useful since no expression is known for $\CF_\CS$ in general. Yet, it could be employed in a reverse way to find an expression for the specialization $\CF_\CS\mat{\ell_1}{\bell_1}{\ell_2}{\bell_2}$ from the knowledge of $\Phi_{\CS}$ and $\Phi$.

The second method employs the property \ref{prop_S}, it provides a relation between the S-dual intertwiner $\Phi_{\CS}$ and the rotated one,
\begin{align}\label{def_phi_CS}
\begin{split}
&\Phi_{\CS}\mat{\ell_1}{\bell_1}{\ell_2}{\bell_2}=\CM_\CS^{(\bell,-\ell)}\Phi\mat{\bell_1}{-\ell_1}{\bell_2}{-\ell_2}\left(\CM_\CS^{(\bell_1,-\ell_1)-1}\otimes\CM_\CS^{(\bell_2,-\ell_2)-1}\right),\\
&\Phi_{\CS}^\ast\mat{\ell_1}{\bell_1}{\ell_2}{\bell_2}=\left(\CM_\CS^{(\bell_1,-\ell_1)}\otimes\CM_\CS^{(\bell_2,-\ell_2)}\right)\Phi^\ast\mat{\bell_1}{-\ell_1}{\bell_2}{-\ell_2}\CM_\CS^{(\bell,-\ell)-1}.
\end{split}
\end{align}
Note however that in this formula, the rotation of intertwiners $\Phi$ and $\Phi^\ast$ is performed in the opposite direction, so that it corresponds the action of $\CS^{-1}$ on the representations.

Using the formula for the inversion of intertwiners \ref{invert_intw}, combined with the solution \ref{sol2_SS} for $\Phi_{\CS^2}$, the rotated intertwiner $\Phi\mat{-\bell_1}{\ell_1}{-\bell_2}{\ell_2}$ can be expressed in terms of the original one,
\begin{equation}\label{def_SPhi}
\Phi\mat{-\bell_1}{\ell_1}{-\bell_2}{\ell_2}=\CM_\CS^{(\ell,\bell)}\Phi\mat{\ell_1}{\bell_1}{\ell_2}{\bell_2}\bCF_\CS\mat{\ell_1}{\bell_1}{\ell_2}{\bell_2},\quad
\Phi^\ast\mat{-\bell_1}{\ell_1}{-\bell_2}{\ell_2}=\bCF_\CS^\ast\mat{\ell_1}{\bell_1}{\ell_2}{\bell_2}^{-1}\Phi^\ast\mat{\ell_1}{\bell_1}{\ell_2}{\bell_2}\CM_\CS^{(\ell,\bell)-1},
\end{equation} 
with
\begin{align}\label{expr_bCF}
\begin{split}
&\bCF_\CS\mat{\ell_1}{\bell_1}{\ell_2}{\bell_2}=\CN\mat{\ell_1}{\bell_1}{\ell_2}{\bell_2}^{-1}\ \left(\CM_{\CS^2}^{(\ell_1,\bell_1)-1}\otimes\CM_{\CS^2}^{(\ell_2,\bell_2)-1}\right)\Phi^\ast\mat{-\ell_1}{-\bell_1}{-\ell_2}{-\bell_2}\CM_\CS^{(-\bell,\ell)}\Phi\mat{-\bell_1}{\ell_1}{-\bell_2}{\ell_2},\\
&\bCF_\CS^\ast\mat{\ell_1}{\bell_1}{\ell_2}{\bell_2}^{-1}=\CN\mat{-\ell_1}{-\bell_1}{-\ell_2}{-\bell_2}^{-1}\Phi^\ast\mat{-\bell_1}{\ell_1}{-\bell_2}{\ell_2}\CM_\CS^{(-\bell,\ell)-1}\Phi\mat{-\ell_1}{-\bell_1}{-\ell_2}{-\bell_2}\left(\CM_{\CS^2}^{(\ell_1,\bell_1)}\otimes\CM_{\CS^2}^{(\ell_2,\bell_2)}\right).
\end{split}
\end{align}
In contrast with the transformation two-tensors $\CF_\CS\mat{\ell_1}{\bell_1}{\ell_2}{\bell_2}$ with which they are affiliated, the expression for the two-tensors $\bCF_\CS\mat{\ell_1}{\bell_1}{\ell_2}{\bell_2}$ can be written down explicitly, and we will provide several examples later on. Assuming that the solutions $\Phi_\CS$ obtained with the different methods are proportional, the two-tensors can be related as follows:
\begin{equation}\label{rep_FS}
\bCF_\CS\mat{\ell_1}{\bell_1}{\ell_2}{\bell_2}\propto\left(\CM_\CS^{(\ell_1,\bell_1)-1}\otimes\CM_\CS^{(\ell_2,\bell_2)-1}\right)\CF_\CS\mat{-\bell_1}{\ell_1}{-\bell_2}{\ell_2},\quad \bCF_\CS^\ast\mat{\ell_1}{\bell_1}{\ell_2}{\bell_2}\propto\left(\CM_\CS^{(\ell_1,\bell_1)-1}\otimes\CM_\CS^{(\ell_2,\bell_2)-1}\right)\CF_{\CS}^\ast\mat{-\bell_1}{\ell_1}{-\bell_2}{\ell_2}.
\end{equation} 

In fact, even though it can be made explicit, the exact expression for the two-tensors $\bCF_\CS$ will remain rather complicated. On the other hand, they satisfy simple covariance properties under the action of DIM algebra that can (and will) be exploited:
\begin{align}
\begin{split}\label{prop_bCF}
&\bCF_\CS\mat{\ell_1}{\bell_1}{\ell_2}{\bell_2}\left(\rho^{(-\bell_1,\ell_1)}_{\tv_1}\otimes\rho^{(-\bell_2,\ell_2)}_{\tv_2}\ \D(e)\right)=\left(\rho^{(\ell_1,\bell_1)}_{v_1}\otimes\rho^{(\ell_2,\bell_2)}_{v_2}\ \D(\CS\cdot e)\right)\bCF_\CS\mat{\ell_1}{\bell_1}{\ell_2}{\bell_2},\\
&\left(\rho^{(-\bell_1,\ell_1)}_{\tv_1}\otimes\rho_{\tv_2}^{(-\bell_2,\ell_2)}\ \D'(e)\right)\bCF_\CS^\ast\mat{\ell_1}{\bell_1}{\ell_2}{\bell_2}^{-1}=\bCF_\CS^\ast\mat{\ell_1}{\bell_1}{\ell_2}{\bell_2}^{-1}\left(\rho_{v_1}^{(\ell_1,\bell_1)}\otimes\rho_{v_2}^{(\ell_2,\bell_2)}\ \D'(\CS\cdot e)\right).
\end{split}
\end{align}
These relations have been established by a direct calculation.

\subsection{Example of intertwiners}
In this subsection, we provide the explicit expression for several types of intertwiners. These expressions will be used to check some of the identities obtained previously using Miki's automorphism. Moreover, they will also be employed in the next sections to investigate the action of S-duality on particular brane webs.

\subsubsection{Generalized AFS intertwiners}
The first intertwiners for the DIM algebra were introduced by Awata, Feigin and Shiraishi in \cite{Awata2011}, they correspond in our notation to $\Phi\mat{0}{1}{1}{n}$ and $\Phi^\ast\mat{0}{1}{1}{n}$. They are associated to the genuine refined topological vertex \cite{Awata2005,Iqbal2007} and describe the junction of a D5-brane and a NS5-brane, dressed by $n$-D5 brane, forming a bound state of one NS5-brane and $n+1$ D5-branes. These intertwiners were later generalized in \cite{Bourgine2017b} to vertical representations of higher level $m>0$ using a fusion technique. The new intertwiners involve the representations $(1,n+m)_{u'}$, $(0,m)_{\vec v}$ and $(1,n)_u$, and describe $m$ topological vertices with parallel preferred direction, and successively glued along a non-preferred direction. In this way, they represent a NS5-brane, dressed by $n$ D5-brane, meeting a stack of $m$ D5-brane (not necessarily overlayed), to form a bound state of one NS5-brane dressed by $n+m$ D5-branes. 

The generalized AFS intertwiners are conveniently presented as vectors of the vertical modules, with coefficients being vertex operators mapping the Fock spaces of the horizontal modules:
\begin{align}
\begin{split}\label{Intertwiners}
&\Phi\mat{0}{m}{1}{n}=\sum_{\vec\l}a_{\vec\l}\Phi_{\vec\l}\mat{0}{m}{1}{n}\langle\bra{\vec v,\vec\l},\quad \Phi^\ast\mat{0}{m}{1}{n}=\sum_{\vec\l}a_{\vec\l}\Phi_{\vec\l}^\ast\mat{0}{m}{1}{n}\ket{\vec v,\vec\l}\rangle,\\
&\Phi_{\vec\l}\mat{0}{m}{1}{n}=t_{\vec\l}\mat{0}{m}{1}{n}\ :\prod_{l=1}^m\Phi_\vac(v_l)\prod_{x\in\vec\l}\eta^+(\chi_x):,\quad \Phi_{\vec\l}^\ast\mat{0}{m}{1}{n}=t_{\vec\l}^\ast\mat{0}{m}{1}{n}\ :\prod_{l=1}^m\Phi_\vac^\ast(v_l)\prod_{x\in\vec\l}\eta^-(\chi_x):,
\end{split}
\end{align}
where $t_{\vec\l}$ and $t_{\vec\l}^\ast$ denote the normalization coefficients
\begin{equation}
t_{\vec\l}\mat{0}{m}{1}{n}=(u')^{|\vec\l|}\prod_{x\in\vec\l}\chi_x^{-n-m},\quad t_{\vec\l}^\ast\mat{0}{m}{1}{n}= u^{-|\vec\l|}\g^{-m|\vec\l|}\prod_{x\in\vec\l}\chi_x^{n}.
\end{equation}
The vertical components associated to the vacuum involve the operators $\Phi_\vac(v)$ and $\Phi_\vac^\ast(v)$ that defines a sort of Fermi sea. They can be built using a product over the boxes of an infinite, fully filled, Young diagram $\l_\infty=\{(i,j)\in\mathbb{Z}^{>0}\times\mathbb{Z}^{>0}\}$,
\begin{align}
\begin{split}\label{AFS_intw}
&\Phi_\vac(v)=:\prod_{x\in\vec\l_\infty}\eta^+(\chi_x)^{-1}:=:\exp\left(\sum_{k\neq0}\dfrac{(\g v)^{-k}}{\s_k}\g^{-|k|/2}\a_k\right):,\\
&\Phi_\vac^\ast(v)=:\prod_{x\in\vec\l_\infty}\eta^-(\chi_x)^{-1}:=:\exp\left(-\sum_{k\neq0}\dfrac{(\g v)^{-k}}{\s_k}\g^{|k|/2}\a_k\right):,
\end{split}
\end{align}
with the coefficients $\s_k$ given in \ref{def_sk}.

\subsubsection{New intertwiners $V\times H\leftrightarrow H$}\label{sec_new_I}
In fact, the generalized AFS intertwiner $\Phi$ presented above belong to a set of eight intertwiners obtained either by flipping the signs of the levels $(1,n)\to(-1,n)$, or $(0,m)\to(0,-m)$, or exchanging the representations $(0,\pm m)\leftrightarrow(\pm1,n)$. By convention, we denote the weight of representations $(\pm1,n)$ as $u$, $(0,\pm m)$ as $\vec v$, and $(\pm1,n+m)$ as $u'$. These intertwiners can also be decomposed over their vertical components, as in \ref{Intertwiners}, and the corresponding coefficients read
\begin{align}
\begin{split}
&\Phi_{\vec\l}\mat{0}{m}{-1}{n}=t_{\vec\l}\mat{0}{m}{-1}{n}:\prod_{l=1}^m\Phi_\vac^\ast(q_3^{-1}v_l^{-1})\prod_{x\in\vec\l}\eta^-(\chi_x^{-1}):,\quad t_{\vec\l}\mat{0}{m}{-1}{n}=u^{-|\vec\l|}\g^{-m|\vec\l|}\prod_{x\in\vec\l}\chi_x^{n},\\
&\Phi_{\vec\l}\mat{0}{-m}{1}{n}=t_{\vec\l}\mat{0}{-m}{1}{n}:\prod_{l=1}^m\Phi_\vac(q_3^{-1}v_l^{-1})^{-1}\prod_{x\in\vec\l}\eta^+(\chi_x^{-1})^{-1}:,\quad t_{\vec\l}\mat{0}{-m}{1}{n}=u^{-|\vec\l|}\prod_{x\in\vec\l}\chi_x^{-n},\\
&\Phi_{\vec\l}\mat{0}{-m}{-1}{n}=t_{\vec\l}\mat{0}{-m}{-1}{n}:\prod_{l=1}^m\Phi_\vac^\ast(v_l)^{-1}\prod_{x\in\vec\l}\eta^-(\chi_x)^{-1}:,\quad t_{\vec\l}\mat{0}{-m}{-1}{n}=(u')^{|\vec\l|}\g^{-m|\vec\l|}\prod_{x\in\vec\l}\chi_x^{-m+n},
\end{split}
\end{align}
and
\begin{align}
\begin{split}
&\Phi_{\vec\l}\mat{1}{n}{0}{m}=t_{\vec\l}\mat{1}{n}{0}{m}:\prod_{l=1}^m\Phi_\vac^\ast(v_l)^{-1}\prod_{x\in\vec\l}\eta^-(\chi_x)^{-1}:,\quad t_{\vec\l}\mat{1}{n}{0}{m}=(u')^{|\vec\l|}\g^{-m|\vec\l|}\prod_{x\in\vec\l}\chi_x^{-n-m},\\
&\Phi_{\vec\l}\mat{-1}{n}{0}{m}=t_{\vec\l}\mat{-1}{n}{0}{m}:\prod_{l=1}^m\Phi_\vac(q_3^{-1}v_l^{-1})^{-1}\prod_{x\in\vec\l}\eta^+(\chi_x^{-1})^{-1}:,\quad t_{\vec\l}\mat{-1}{n}{0}{m}=u^{-|\vec\l|}\prod_{x\in\vec\l}\chi_x^{n},\\
&\Phi_{\vec\l}\mat{1}{n}{0}{-m}=t_{\vec\l}\mat{1}{n}{0}{-m}:\prod_{l=1}^m\Phi_\vac^\ast(q_3^{-1}v_l^{-1})\prod_{x\in\vec\l}\eta^-(\chi_x^{-1}):,\quad t_{\vec\l}\mat{1}{n}{0}{-m}=u^{-|\vec\l|}\g^{-m|\vec\l|}\prod_{x\in\vec\l}\chi_x^{-n},\\
&\Phi_{\vec\l}\mat{-1}{n}{0}{-m}=t_{\vec\l}\mat{-1}{n}{0}{-m}:\prod_{l=1}^m\Phi_\vac(v_l)\prod_{x\in\vec\l}\eta^+(\chi_x):,\quad t_{\vec\l}\mat{-1}{n}{0}{-m}=(u')^{|\vec\l|}\prod_{x\in\vec\l}\chi_x^{-m+n}.
\end{split}
\end{align}
Note that the vertical coefficients of the intertwiner with opposite levels $(\ell,\bell)\to(-\ell,-\bell)$ can be obtained from the original ones by sending $\eta^\pm(z)\to\eta^\mp(z)^{-1}$, and $u\to u\g^{\pm m}$. In fact, this property follows from the action \ref{action_CSCS} of the automorphism $\CS^2$.

In the same way, it is possible to define eight dual intertwiners, including the one given in \ref{AFS_intw},
\begin{align}
\begin{split}
&\Phi_{\vec\l}^\ast\mat{0}{m}{-1}{n}=t_{\vec\l}^\ast\mat{0}{m}{-1}{n}:\prod_{l=1}^m\Phi_\vac(q_3^{-1}v_l^{-1})\prod_{x\in\vec\l}\eta^+(\chi_x^{-1}):,\quad t_{\vec\l}^\ast\mat{0}{m}{-1}{n}=(u')^{|\vec\l|}\prod_{x\in\vec\l}\chi_x^{-n-m},\\
&\Phi_{\vec\l}^\ast\mat{0}{-m}{1}{n}=t_{\vec\l}^\ast\mat{0}{-m}{1}{n}:\prod_{l=1}^m\Phi_\vac^\ast(q_3^{-1}v_l^{-1})^{-1}\prod_{x\in\vec\l}\eta^-(\chi_x^{-1})^{-1}:,\quad t_{\vec\l}^\ast\mat{0}{-m}{1}{n}=(u')^{|\vec\l|}\g^{-m|\vec\l|}\prod_{x\in\vec\l}\chi_x^{-m+n},\\
&\Phi_{\vec\l}^\ast\mat{0}{-m}{-1}{n}=t_{\vec\l}^\ast\mat{0}{-m}{-1}{n}:\prod_{l=1}^m\Phi_\vac(v_l)^{-1}\prod_{x\in\vec\l}\eta^+(\chi_x)^{-1}:,\quad t_{\vec\l}^\ast\mat{0}{-m}{-1}{n}=u^{-|\vec\l|}\prod_{x\in\vec\l}\chi_x^{-n},
\end{split}
\end{align}
and
\begin{align}
\begin{split}
&\Phi_{\vec\l}^\ast\mat{1}{n}{0}{m}=t_{\vec\l}^\ast\mat{1}{n}{0}{m}:\prod_{l=1}^m\Phi_\vac(v_l)^{-1}\prod_{x\in\vec\l}\eta^+(\chi_x)^{-1}:,\quad t_{\vec\l}^\ast\mat{1}{n}{0}{m}=u^{-|\vec\l|}\prod_{x\in\vec\l}\chi_x^{n},\\
&\Phi_{\vec\l}^\ast\mat{-1}{n}{0}{m}=t_{\vec\l}^\ast\mat{-1}{n}{0}{m}:\prod_{l=1}^m\Phi_\vac^\ast(q_3^{-1}v_l^{-1})^{-1}\prod_{x\in\vec\l}\eta^-(\chi_x^{-1})^{-1}:,\quad t_{\vec\l}^\ast\mat{-1}{n}{0}{m}=(u')^{|\vec\l|}\g^{-m|\vec\l|}\prod_{x\in\vec\l}\chi_x^{-(n+m)},\\
&\Phi_{\vec\l}^\ast\mat{1}{n}{0}{-m}=t_{\vec\l}^\ast\mat{1}{n}{0}{-m}:\prod_{l=1}^m\Phi_\vac(q_3^{-1}v_l^{-1})\prod_{x\in\vec\l}\eta^+(\chi_x^{-1}):,\quad t_{\vec\l}^\ast\mat{1}{n}{0}{-m}=(u')^{|\vec\l|}\prod_{x\in\vec\l}\chi_x^{-m+n},\\
&\Phi_{\vec\l}^\ast\mat{-1}{n}{0}{-m}=t_{\vec\l}^\ast\mat{-1}{n}{0}{-m}:\prod_{l=1}^m\Phi_\vac^\ast(v_l)\prod_{x\in\vec\l}\eta^-(\chi_x):,\quad t_{\vec\l}^\ast\mat{-1}{n}{0}{-m}=u^{-|\vec\l|}\g^{-m|\vec\l|}\prod_{x\in\vec\l}\chi_x^{-n}.
\end{split}
\end{align}

\paragraph{Action of $\CS^2$} These exact expressions give us the opportunity to verify some of the relations obtained using the action of $\CS^2$. For instance, combining the expressions of $\Phi_{\CS^2}$ given either in terms of the permutation operator \ref{sol3_SS}, or in terms of the matrices $\CM_{\CS^2}$ \ref{sol2_SS}, we deduce the relations
\begin{align}\label{SS_intw}
\begin{split}
&\Phi\mat{\ell_2}{\bell_2}{\ell_1}{\bell_1}=\CM_{\CS^2}^{(\ell,\bell)-1}\Phi\mat{-\ell_1}{-\bell_1}{-\ell_2}{-\bell_2}\ \left(\CM_{\CS^2}^{(\ell_1,\bell_1)}\otimes\CM_{\CS^2}^{(\ell_2,\bell_2)}\right)\CP\mat{\ell_2}{\bell_2}{\ell_1}{\bell_1},\\
&\Phi^\ast\mat{\ell_2}{\bell_2}{\ell_1}{\bell_1}=\CP\mat{\ell_1}{\bell_1}{\ell_2}{\bell_2}\left(\CM_{\CS^2}^{(\ell_1,\bell_1)-1}\otimes\CM_{\CS^2}^{(\ell_2,\bell_2)-1}\right)\ \Phi\mat{-\ell_1}{-\bell_1}{-\ell_2}{-\bell_2}\CM_{\CS^2}^{(\ell,\bell)},
\end{split}
\end{align}
where $\CP$ is the permutation operator in the appropriate representations, for instance
\begin{equation}
\CP\mat{0}{m}{1}{n}=\sum_{\vec\l,\mu}a_{\vec\l}a_\mu\left(\ket{P_\mu}\otimes\dket{\vec v,\vec\l}\right)\left(\dbra{\vec v,\vec\l}\otimes\bra{P_\mu}\right):(0,m)_{\vec v}\otimes(1,n)_u\to(1,n)_u\otimes(0,m)_{\vec v}.
\end{equation}
Projecting the relations \ref{SS_intw} on the vertical components, and taking into account that the matrices $\CM_{\CS^2}$ are trivial with our definition of horizontal and vertical representations, these relations indeed reduce to the following equalities between intertwiners' components, which can also be observed directly (similar equalities can be written for $\Phi^\ast$):
\begin{equation}
\Phi_{\vec\l}\mat{0}{m}{1}{n}=\Phi_{\vec\l}\mat{-1}{-n}{0}{-m},\quad \Phi_{\vec\l}\mat{0}{m}{-1}{n}=\Phi_{\vec\l}\mat{1}{-n}{0}{-m},\quad\Phi_{\vec\l}\mat{0}{-m}{1}{n}=\Phi_{\vec\l}\mat{-1}{-n}{0}{m},\quad \Phi_{\vec\l}\mat{0}{-m}{-1}{n}=\Phi_{\vec\l}\mat{-1}{-n}{0}{m}.
\end{equation} 
This action of $\CS^2$ on the intertwiners corresponds to a flip of the levels signs (or branes charges), together with an exchange of the vertical and horizontal representations (i.e. the two incoming branes in figure \ref{fig_INT}). Thus, it realizes the $180^\circ$ rotation of the brane web.

\paragraph{Inversion of intertwiners} It is also possible to check the inversion formula \ref{invert_intw} for intertwiners, using again the expression \ref{sol2_SS} of $\Phi_{\CS^2}$ involving the matrices $\CM_{\CS^2}$. For instance, we find for $\CN$:
\begin{equation}
\CN\mat{\ell_1}{\bell_1}{\ell_2}{\bell_2}=\Phi\mat{\ell_1}{\bell_1}{\ell_2}{\bell_2}\left(\CM_{\CS^2}^{(\ell_1,\bell_1)-1}\otimes\CM_{\CS^2}^{(\ell_2,\bell_2)-1}\right)\Phi^\ast\mat{-\ell_1}{-\bell_1}{-\ell_2}{-\bell_2}\CM_{\CS^2}^{(\ell,\bell)}.
\end{equation} 
In the present case, the matrices $\CM_{\CS^2}$ are trivial, and
\begin{equation}
\CN\mat{0}{\pm m}{1}{n}=\sum_{\vec\l}a_{\vec\l}(\vec v)\Phi_{\vec\l}\mat{0}{\pm m}{1}{n}\Phi_{\vec\l}^\ast\mat{0}{\mp m}{-1}{-n},\quad 
\CN\mat{0}{\pm m}{-1}{n}=\sum_{\vec\l}a_{\vec\l}(\vec v)\Phi_{\vec\l}\mat{0}{\pm m}{-1}{n}\Phi_{\vec\l}^\ast\mat{0}{\mp m}{1}{-n}.
\end{equation}
Since the product of intertwiners is done in both vertical and horizontal channels, the weights $u$, $u'$ and $\vec v$ must be the same for $\Phi$ and $\Phi^\ast$ in these formulas. However, due to the infinite dimensional nature of representations, it is necessary to introduce a regulator $\mu$ such that $u=\mu u^\ast$ where $u$ is the weight of representations $(\pm1,n)$ coupled to $\Phi$ and $u^\ast$ the weight of representations $(\pm1,-n)$ coupled to $\Phi^\ast$. Once normal-ordered, the product of vertical components, which is a priori a vertex operator, reduces to a simple constant. Observing further cancellations between this constant and the expression of the inverse norms $a_{\vec\l}$, we end up with
\begin{equation}
\CN\mat{0}{\pm m}{1}{n}=\phi(\mu^{\pm1})^{-m}\prod_{l,l'=1}^m\CG(v_l/(q_3v_{l'}))^{-1},\quad \CN\mat{0}{\pm m}{-1}{n}=\phi(\mu^{\mp1})^{-m}\prod_{l,l'=1}^m\CG(v_l/v_{l'})^{-1}
\end{equation} 
where $\phi(z)$ is the Euler function, $\phi(z)^{-1}=\sum_{\l}z^{|\l|}$.

\paragraph{Conjugation} We would like to conclude the analysis of this set of intertwiners with a short remark. It is possible to relate further the vertical components of intertwiners using the conjugation operation $\dagger$ defined on Fock space operators by 
$(\a_k)^\dagger=\a_{-k}$,
\begin{equation}
(\eta^\pm(z))^\dagger=:\eta^\pm(z^{-1})^{-1}:,\quad (\vphi^\pm(z))^\dagger=\vphi^\mp(z^{-1})^{-1},\quad \Phi_\vac(v)^\dagger=:\Phi(q_3^{-1}v^{-1})^{-1}:,\quad \Phi_\vac^\ast(v)^\dagger=:\Phi^\ast(q_3^{-1}v^{-1})^{-1}:,
\end{equation}
which implies, for instance,
\begin{equation}
\Phi_{\vec\l}\mat{0}{m}{1}{n}^\dagger=\Phi_{\vec\l}\mat{0}{-m}{1}{n+m},\quad \Phi_{\vec\l}\mat{0}{m}{-1}{n}^\dagger=\Phi_{\vec\l}\mat{0}{-m}{-1}{n+m}.
\end{equation}
In the first identity, we have assumed that the weights transforms as $(u')^\dagger=u^{-1}$, and $u^\dagger=(u')^{-1}$ in the second one. This operation of conjugation exchanges the two horizontal spaces $(\pm1,n)_u\leftrightarrow(\pm1,n+m)_{u'}$, while also flipping the sign of the vertical level $(0,m)\to(0,-m)$. It can be seen as flipping the horizontal (and oblique) arrows associated to the (D5-dressed) NS5-branes in the representation of figure \ref{fig_INT}, at the cost of replacing the vertical representation by its $\CS^2$-dual.

\subsubsection{New intertwiners $H\times H\leftrightarrow V$}\label{sec_new_II}
Until now, we have only discussed a certain type of intertwiners that couples a horizontal representation to the tensor product of a horizontal and a vertical representation. However, another type of intertwiners appears in the study of Lax matrices presented in the next section. These intertwiners couple a vertical representation to a tensor product of two horizontal representations. They can be constructed following two different methods. In appendix \ref{AppC}, they have been obtained by solving directly the intertwining relations \ref{AFS_lemmas}. Since this method is rather technical, we present here another derivation based on the reflector state (and its dual) belonging to the tensor product of two Fock modules,
\begin{equation}\label{def_reflector}
\ketS{\O}=\exp\left(\sum_{k>0}\dfrac1{\s_k}\ \a_{-k}\otimes\a_{-k}\right)\left(\ket{\vac}\otimes\ket{\vac}\right),\quad \braS{\O}=\left(\bra{\vac}\otimes\bra{\vac}\right)\exp\left(\sum_{k>0}\dfrac1{\s_k}\ \a_{k}\otimes\a_{k}\right).
\end{equation}
Such states were introduced in \cite{Bourgine2017c} under the name \textit{horizontal reflection states}, their essential property is the `reflection' of the action of q-oscillator modes:
\begin{equation}
\left(\a_k\otimes1\right)\ketS{\O}=\left(1\otimes(\a_k)^\dagger\right)\ketS{\O},\quad \braS{\O}\left(\a_k\otimes1\right)=\braS{\O}\left(1\otimes(\a_k)^\dagger\right),
\end{equation}
where $(\a_k)^\dagger=\a_{-k}$.

The construction of the dual intertwiner is slightly easier. For definiteness, the first horizontal representation will carry the levels $(-1,n)$ and weight $u$, while the second horizontal representation will bear the levels $(1,m)$ and weight $u'$. They form a vertical representation of levels $(0,m+n)$ (with $m+n>0$ for simplicity) and weight $\vec v$ such that $uu'=\prod_l(-\g v_l)$,
\begin{equation}
\Phi^\ast\mat{-1}{n}{1}{m}:(0,n+m)_{\vec v}\to (-1,n)_u\otimes(1,m)_{u'}.
\end{equation} 
This intertwiner can be obtained as the reflection of a generalized AFS intertwiner acting in the first and third tensor space,
\begin{equation}
\Phi^\ast\mat{-1}{n}{1}{m}=\Phi\mat{0}{m+n}{1}{-n}_{13}\left(1\otimes\ketS{\O}\right),\quad\text{with}\quad \Phi\mat{0}{m+n}{1}{-n}:(0,n+m)_{\vec v}\otimes(1,-n)_{u^{-1}}\to (1,m)_{u'}.
\end{equation} 
Note that the weight in the left horizontal space for the auxiliary intertwiner has been inverted, in agreement with the weights conservation relation. Decomposing on the vertical components, we find
\begin{equation}
\Phi^\ast\mat{-1}{n}{1}{m}=\sum_{\vec\l}a_{\vec\l}(\vec v)\ \ketS{\vec v,\vec\l}\langle\bra{\vec v,\vec\l},\quad\text{with}\quad\ketS{\vec v,\vec\l}=\left(1\otimes \Phi_{\vec\l}\mat{0}{n+m}{1}{-n}\right)\ketS{\O}.
\end{equation} 
This expression indeed coincides with the one obtained in appendix \ref{AppC}, the coefficient $t_{\vec\l}\mat{0}{m+n}{1}{-n}$ reproducing the norm $n_{\vec\l}^\ast$ of the states $\ketS{\vec v,\vec\l}$. Geometrically, the reflector $\ketS{\O}$ effectively reverse the arrow on the leg corresponding to the representation $(1,-n)$, and flip the sign of the levels to produce the representation $(-1,n)_u$.

The same construction holds for the intertwiner
\begin{equation}
\Phi\mat{-1}{n}{1}{m}:(-1,n)_u\otimes(1,m)_{u'}\to (0,n+m)_{\vec v},
\end{equation}
using the dual reflector and an auxiliary intertwiner $\Phi^\ast$,
\begin{equation}
\Phi\mat{-1}{n}{1}{m}=\left(1\otimes\braS{\O}\right)\Phi^\ast\mat{0}{m+n}{1}{-n}_{23},\quad \Phi^\ast\mat{0}{m+n}{1}{-n}:(1,m)_{u'}\to(0,n+m)_{\vec v}\otimes(1,-n)_{u^{-1}}.
\end{equation}
The vertical decomposition reads
\begin{equation}
\Phi\mat{-1}{n}{1}{m}=\sum_{\vec\l} a_{\vec\l}\ket{\vec v,\vec\l}\rangle\braS{\vec v,\vec\l},\quad\text{with}\quad\braS{\vec v,\vec\l}=\braS{\O}\left(1\otimes\Phi_{\vec\l}^\ast\mat{0}{m+n}{1}{-n}\right),
\end{equation} 
it reproduces the expression found in appendix \ref{AppC} with $t_{\vec\l}^\ast\mat{0}{m+n}{1}{-n}$ identified with the norm $n_{\vec\l}$.

\section{$\CS$-transformation of Lax matrices}
\subsection{Lax matrices}\label{sec_Lax}
Coupling intertwiners is simply realized by taking a product, either a product of operators for horizontal representations, or a scalar product for vertical representations. In effect, intertwiners $\Phi$ and $\Phi^\ast$ can be coupled in three different ways, depending on the choice of the common leg/representation, which leads to different algebraic objects. In this section, we discuss only the case of a coupling along the common representation $(\ell,\bell)$,
\begin{equation}
\Phi^\ast\mat{\ell_1^\ast}{\bell_1^\ast}{\ell_2^\ast}{\bell_2^\ast}\Phi\mat{\ell_1}{\bell_1}{\ell_2}{\bell_2}:(\ell_1,\bell_1)_{v_1}\otimes (\ell_2,\bell_2)_{v_2}\to(\ell_1^\ast,\bell_1^\ast)_{v_1^\ast}\otimes (\ell_2^\ast,\bell_2^\ast)_{v_2^\ast}.
\end{equation}
Couplings along the other two legs is the subject of the next section. Moreover, for simplicity, we restrict ourselves to the case $(\ell_i,\bell_i)=(\ell_i^\ast,\bell_i^\ast)$ ($i=1,2$), and denote
\begin{equation}
\CL\mat{\ell_1}{\bell_1}{\ell_2}{\bell_2}=\Phi^\ast\mat{\ell_1}{\bell_1}{\ell_2}{\bell_2}\Phi\mat{\ell_1}{\bell_1}{\ell_2}{\bell_2}.
\end{equation} 
Due to the intertwining properties \ref{AFS_lemmas} of $\Phi$ and $\Phi^\ast$, the operator $\CL$ intertwines between the coproduct $\D$ and its opposite $\D'$,
\begin{equation}\label{covar_Lax}
\left(\rho^{(\ell_1,\bell_1)}_{v_1^\ast}\otimes\rho^{(\ell_2,\bell_2)}_{v_2^\ast}\ \D'(e)\right)\CL\mat{\ell_1}{\bell_1}{\ell_2}{\bell_2}=\CL\mat{\ell_1}{\bell_1}{\ell_2}{\bell_2}\left(\rho_{v_1}^{(\ell_1,\bell_1)}\otimes\rho_{v_2}^{(\ell_2,\bell_2)}\ \D(e)\right).
\end{equation}
The rotation property of intertwiners \ref{def_SPhi} implies a similar property for $\CL$:
\begin{equation}\label{trans_Lax}
\CL\mat{-\bell_1}{\ell_1}{-\bell_2}{\ell_2}=\bCF_\CS^\ast\mat{\ell_1}{\bell_1}{\ell_2}{\bell_2}^{-1}\CL\mat{\ell_1}{\bell_1}{\ell_2}{\bell_2}\bCF_\CS\mat{\ell_1}{\bell_1}{\ell_2}{\bell_2}.
\end{equation}

The property \ref{covar_Lax} is also satisfied by the Lax matrix $\CR\mat{\ell_1}{\bell_1}{\ell_2}{\bell_2}$ which is, by definition, the evaluation of the universal R-matrix in the representations $(\ell_1,\bell_1)_{v_1}\otimes(\ell_2,\bell_2)_{v_2}$. However, this intertwining property alone does not fully determine the Lax matrix, since the universal R-matrix is further constraint to obey two extra relations in \ref{def_R}. In fact, the solution of the universal equation $\D'=\CR\D\CR^{-1}$ is unique, up to a factor $\CA$ belonging to the centralizer of $\D(\DIM)$ in $\DIM\otimes\DIM$ \cite{Chari1994}. In practice, it has been observed that $\CL$ is indeed proportional to $\CR$, with a factor depending on levels and weights:
\begin{equation}
\CL\mat{\ell_1}{\bell_1}{\ell_2}{\bell_2}=\CA\mat{\ell_1}{\bell_1}{\ell_2}{\bell_2}\left(\rho^{(\ell_1,\bell_1)}\otimes\rho^{(\ell_2,\bell_2)}\ \CR\right).
\end{equation} 
This factor is responsible for the anomaly term observed in \cite{Awata2016b} for the Yang-Baxter equation. Its expressions is known in a number of cases, for instance $\CA\mat{0}{m}{1}{n}=\Zol(\vec v)^{-1}$ within our conventions \cite{Bourgine2019}. With a slight abuse of terminology, we will also refer to $\CL$ as an (ill-normalized) Lax matrix.

The same formulas can be written for the $\CS$-twisted coproduct, thus defining the quantities $\CL_\CS$ and $\CA_\CS$. Using the formula \ref{def_phi_CS} to express the $\CS$-dual intertwiners in terms of rotated intertwiners, we find the relation
\begin{equation}\label{CL_dual}
\CL_\CS\mat{-\bell_1}{\ell_1}{-\bell_2}{\ell_2}=\left(\CM_\CS^{(\ell_1,\bell_1)}\otimes\CM_\CS^{(\ell_2,\bell_2)}\right)\CL\mat{\ell_1}{\bell_1}{\ell_2}{\bell_2}\left(\CM_\CS^{(\ell_1,\bell_1)-1}\otimes\CM_\CS^{(\ell_2,\bell_2)-1}\right).
\end{equation} 
Assuming a proper mapping of the vacuum states, this relation, together with the rotation property \ref{trans_Lax}, implies the equality between the vacuum expectation values of the Lax matrices:
\begin{equation}
\la\CL_\CS\mat{-\bell_1}{\ell_1}{-\bell_2}{\ell_2}\ra=\la\CL\mat{\ell_1}{\bell_1}{\ell_2}{\bell_2}\ra=\la\CL\mat{-\bell_1}{\ell_1}{-\bell_2}{\ell_2}\ra.
\end{equation}
This is the equality that we will check on the two examples below. We shall see that they reproduce the S-duality relations observed among the instanton partition functions. The key to show this equality is the study of the vacuum transformation. Unfortunately, this task turned out too difficult for the transformations $\bCF_\CS$ and $\bCF_\CS^\ast$ in \ref{trans_Lax}, and we will have to rely on some indirect arguments.

\subsection{Example I: resolved conifold}
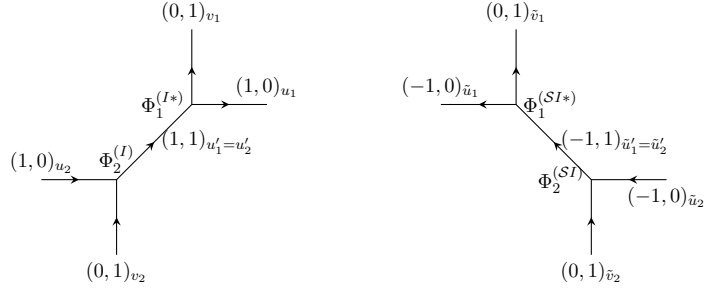
\begin{figure}
\begin{center}
\begin{tikzpicture}
\draw[postaction={on each segment={mid arrow=black}}] (-1,0) -- (0,0) -- (1,1)--(1,2);
\draw[postaction={on each segment={mid arrow=black}}] (0,-1) -- (0,0);
\draw[postaction={on each segment={mid arrow=black}}] (1,1) -- (2,1);
\node[above,scale=0.7] at (0,0) {$\Phi_2^{(I)}$};
\node[left,scale=0.7] at (1,1) {$\Phi_1^{(I\ast)}$};
\node[above,scale=0.7] at (-1,0) {$(1,0)_{u_2}$};
\node[below,scale=0.7] at (0,-1) {$(0,1)_{v_2}$};
\node[right,scale=0.7] at (0.5,0.5) {$(1,1)_{u'_1=u'_2}$};
\node[above,scale=0.7] at (1,2) {$(0,1)_{v_1}$};
\node[above,scale=0.7] at (2,1) {$(1,0)_{u_1}$};
\end{tikzpicture}
\hspace{10mm}
\begin{tikzpicture}
\draw[postaction={on each segment={mid arrow=black}}] (2,0) -- (1,0) -- (0,1)--(-1,1);
\draw[postaction={on each segment={mid arrow=black}}] (0,1) -- (0,2);
\draw[postaction={on each segment={mid arrow=black}}] (1,-1) -- (1,0);
\node[left,scale=0.7] at (1,0) {$\Phi_2^{(\CS I)}$};
\node[right,scale=0.7] at (0,1) {$\Phi_1^{(\CS I\ast)}$};
\node[below,scale=0.7] at (2,0) {$(-1,0)_{\tu_2}$};
\node[below,scale=0.7] at (1,-1) {$(0,1)_{\tv_2}$};
\node[right,scale=0.7] at (0.5,0.5) {$(-1,1)_{\tu'_1=\tu'_2}$};
\node[above,scale=0.7] at (0,2) {$(0,1)_{\tv_1}$};
\node[above,scale=0.7] at (-1,1) {$(-1,0)_{\tu_1}$};
\end{tikzpicture}
\end{center}
\caption{Representation of the Lax matrix $\CL^{(I)}$ and $\CL^{(\CS I)}$.}
\label{fig_LI}
\end{figure}

Our first example is the toric diagram corresponding to the resolved conifold, it is represented on figure \ref{fig_LI}. The diagram on the left defines the Lax matrix
\begin{equation}
\CL^{(I)}=\CL\mat{0}{1}{1}{0}:(0,1)_{v_2}\times(1,0)_{u_2}\to(0,1)_{v_1}\times(1,0)_{u_1},
\end{equation} 
which involves the horizontal representation $(1,1)_{u'}$ in the intermediate channel. We have introduced here, and in figure \ref{fig_LI}, a shortcut notation for the Lax matrix and its intertwiners $\Phi^{(I)}=\Phi\mat{0}{1}{1}{0}$ and $\Phi^{(I\ast)}=\Phi^\ast\mat{0}{1}{1}{0}$. This Lax matrix is related to an S-dual Lax matrix through the formula \ref{CL_dual}. Since the matrices $\CM_\CS$ involved map the vertical vacuum to the horizontal one, and vice-versa, their vacuum expectation values are the same:
\begin{equation}
\left\langle\CL^{(I)}\right\rangle=\left(\langle\bra{v_1,\vac}\otimes\bra{\vac}\right)\CL^{(I)}\left(\ket{v_2,\vac}\rangle\otimes\ket{\vac}\right)=\left(\bra{\vac}\otimes\langle\bra{\tv_1,\vac}\right)\CL_\CS\mat{-1}{0}{0}{1}\left(\ket{\vac}\otimes\ket{\tv_2,\vac}\rangle\right)=\left\langle\CL_\CS\mat{-1}{0}{0}{1}\right\rangle.
\end{equation}
In addition, we would like to undo the rotation, using \ref{trans_Lax} and show that $\la\CL^{(I)}\ra$ is equal to the v.e.v. of the rotated Lax matrix $\CL^{(\CS I)}=\CL\mat{-1}{0}{0}{1}$ represented on figure \ref{fig_LI} (right). This Lax matrix involves the rotated intertwiners $\Phi^{(\CS I)}=\Phi\mat{-1}{0}{0}{1}$ and $\Phi^{(\CS I\ast)}=\Phi^\ast\mat{-1}{0}{0}{1}$.

We will denote the weights of the original Lax matrix $\CL^{(I)}$ as $u_i$, $v_i$ and $u'_i=-\g u_iv_i$ with $i=1,2$, and the weights of the rotated matrix $\CL^{(\CS I)}$ with a tilde: $\tu_i$, $\tv_i$ and $\tu'_i=-\g\tu_i\tv_i$. Using the known expressions for intertwiners, both Lax matrices can be decomposed over their vertical components,
\begin{equation}
\CL^{(I)}=\sum_{\l_1,\l_2}a_{\l_1}a_{\l_2}\CL^{(I)}_{\l_1,\l_2}\ \dket{v_1,\l_1}\dbra{v_2,\l_2},\quad
\CL^{(\CS I)}=\sum_{\l_1,\l_2}a_{\l_1}a_{\l_2}\CL^{(\CS I)}_{\l_1,\l_2}\ \dket{\tv_1,\l_1}\dbra{\tv_2,\l_2},
\end{equation} 
which are, in turn, expressed as vertex operators in the Fock space:
\begin{align}
\begin{split}\label{Lax_I}
&\CL^{(I)}_{\l_1,\l_2}=(\g u_1)^{-|\l_1|}(-\g u_2v_2)^{|\l_2|}\dfrac{N(\l_2,\l_1|\g v_2/v_1)}{\CG(v_2/(\g v_1))}\prod_{x\in\l_2}\chi_x^{-1}\ :\Phi_\vac^\ast(v_1)\Phi_\vac(v_2)\prod_{x\in\l_1}\eta^-(\chi_x)\prod_{x\in\l_2}\eta^+(\chi_x):,\\
&\CL^{(\CS I)}_{\l_1,\l_2}=(-\tu_1\tv_1)^{|\l_1|}(\tu_2)^{-|\l_2|}\dfrac{N(\l_1,\l_2|\g \tv_1/\tv_2)}{\CG(\tv_1/(\g\tv_2))}\prod_{x\in\l_1}\chi_x^{-1}\ :\Phi_\vac^\ast(\tv_1)^\dagger\Phi_\vac(\tv_2)^\dagger\prod_{x\in\l_1}\eta^-(\chi_x)^\dagger\prod_{x\in\l_2}\eta^+(\chi_x)^\dagger:.
\end{split}
\end{align}
The expression for the Nekrasov factors $N(\l_1,\l_2|r)$ and the function $\CG(z)$ can be found in appendix \ref{AppB}. Projecting on the vacuum states, we deduce the respective v.e.v.
\begin{equation}
\la\CL^{(I)}\ra=\CG(v_2/(\g v_1))^{-1},\quad \la\CL^{(\CS I)}\ra=\CG(\tv_1/(\g\tv_2))^{-1}.
\end{equation} 
The two v.e.v. are equal, provided that the ratios $v_2/v_1$ and $\tv_1/\tv_2$ coincide. We will see below that this is indeed the case. Note that the anomaly factor is irrelevant here since $\CA\mat{0}{1}{1}{0}=\CG(q_3^{-1})^{-1}$ is a weight-independent constant.

\paragraph{Transformation of the vacuum} The Lax matrices $\CL^{(I)}$ and $\CL^{(\CS I)}$ are related through the formula \ref{trans_Lax}, explicitly
\begin{equation}
\CL^{(\CS I)}=\bCF_\CS^{(I\ast)-1}\CL^{(I)}\bCF_\CS^{(I)},\quad\text{with}\quad \bCF_\CS^{(I)}=\bCF_\CS\mat{0}{1}{1}{0},\quad \bCF_\CS^{(I\ast)}=\bCF_\CS^\ast\mat{0}{1}{1}{0}.
\end{equation} 
Thus, the equality of the v.e.v. would follow if $\bCF_\CS^{(I)}$ and $\bCF_\CS^{(I\ast)-1}$ map the product of vacuum states in the representations $I$ to the equivalent product of vacuum states for the representations $\CS I$. Since the treatment of $\bCF_\CS^{(I\ast)}$ is identical, we will focus here only on $\bCF_\CS^{(I)}$. The general expression for the two tensors $\bCF$ is given in \ref{expr_bCF}, it reduces in our case to 
\begin{equation}
\bCF_\CS^{(I)}=\CN\mat{0}{1}{1}{0}^{-1}\sum_{\l_1,\l_2}a_{\l_1}a_{\l_2}\Phi_{\l_1}^\ast\mat{0}{-1}{-1}{0}\CM_\CS^{(-1,1)}\Phi_{\l_2}\mat{-1}{0}{0}{1}\ \dket{v_2,\l_1}\dbra{\tv_2,\l_2}.
\end{equation} 
The factor $\CN\mat{0}{1}{1}{0}$ appearing in this formula is independent of the weights and can be neglected. In general, such factors depend only on half of the weights (say $u_2,v_2,...$ but not $u_1,v_1,...$), and cannot generate a non-trivial dependence in the gauge theory parameters that would correspond to ratios of opposite weights (e.g. $u_1/u_2$, or $v_1/v_2$). Thus, the two tensor $\bCF_\CS^{(I)}$  maps the product of vacuum states in representations $(-1,0)_{\tu_2}\otimes(0,1)_{\tv_2}$ to the state
\begin{equation}
\ket{\CS I\vac}=\bCF_\CS^{(I)}\left(\ket{\vac}\otimes\dket{\tilde{v}_2,\vac}\right)=\sum_{\l}a_{\l}\ \dket{v_2,\l}\otimes\Phi_{\l}^\ast\mat{0}{-1}{-1}{0}\CM_\CS^{(-1,1)}\Phi_{\vac}\mat{-1}{0}{0}{1}\ket{\vac}.
\end{equation}
We would like to identify this state with  $\ket{v_2,\vac}\rangle\otimes\ket{\vac}$ in the module $(0,1)_{v_2}\otimes(1,0)_{u_2}$. Unfortunately, this expression seems too complicated for a direct approach, and we will rely on an indirect argument.

Our argument follows from the comparison of the states' transformation properties under the action of the DIM algebra. To lighten the notations, the index $2$ for the weights will be dropped in this calculation. Taking the coproduct \ref{Drinfeld_coproduct} of Drinfeld currents in the proper representations, it is possible to show that\footnote{These relations should be read as follows. In the first line, we examine only the action of positive modes, they all vanish except for the mode $k=0$ that is diagonal with eigenvalue $u^{-1}$. Similarly, in the second equation, we examine only the modes strictly positive, they are all diagonal.}
\begin{align}\label{char_vac_I}
\begin{split}
&\left(\rho_v^{(0,1)}\otimes\rho_u^{(1,0)}\ \D(x^-_{k\geq0})\right)\left(\dket{v,\vac}\otimes\ket{\vac}\right)=\d_{k,0}\ 
u^{-1}\left(\dket{v,\vac}\otimes\ket{\vac}\right),\\
&\left(\rho_v^{(0,1)}\otimes\rho_u^{(1,0)}\ \D(a_{k>0})\right)\left(\dket{v,\vac}\otimes\ket{\vac}\right)=-\dfrac1k(\g^k-\g^{-k})(v\g^{1/2})^k\left(\dket{v,\vac}\otimes\ket{\vac}\right).
\end{split}
\end{align}
We recognize here the transformation properties of the coherent state $\ketA{v,\vac}$ constructed in the Fock module in appendix \ref{AppC}. In fact, these properties characterize the state uniquely, up to an overall normalization factor. Indeed, the action of $a_{k>0}$ corresponds in the module $(1,1)_{u'_1}$ to the action of the positive modes $\a_k$. This permits to identify states up to a possible shift of the vacuum state $\ket{\vac}$ (i.e. the unique state annihilated by all the modes $\a_{k>0}$). This degree of freedom is further fixed by the action of $x_0^-$, leaving only the possibility of a different norm. Note that this identification also determines the relation between the weights, since
\begin{equation}
\rho_{u'}^{(1,1)}(x_0^-)\ketA{v,\vac}=-\dfrac{\g v}{u'}\ketA{v,\vac},
\end{equation} 
which implies $u'=-\g uv$ as required by the weights conservation relation.\footnote{Naively, we could expect that the tensor product of modules $(0,1)_{v}\otimes(1,0)_{u}$ is isomorphic to the module $(1,1)_{u'}$. However, the intertwiner $\Phi^{(I\ast)}$ has only a left inverse and nothing guarantees, a priori, the existence of this isomorphism.}

On the other hand, the characterization of the state $\ket{\CS I\vac}$ follows indirectly from the covariance property \ref{prop_bCF} of $\bCF_\CS^{(I)}$. This property implies that, for any element $e\in\DIM$,
\begin{equation}
\left(\rho_v^{(0,1)}\otimes\rho_u^{(1,0)}\ \D(\CS\cdot e)\right)\ket{\CS I\vac}=\bCF_\CS^{(I)}\left(\rho_{\tilde{u}}^{(-1,0)}\otimes\rho_{\tilde{v}}^{(0,1)}\ \D(e)\right)\left(\ket{\vac}\otimes\dket{\tilde{v},\vac}\right).
\end{equation}
As a result, if the state $\ket{\vac}\otimes\dket{\tilde{v},\vac}$ in the module $(-1,0)_{\tu}\times(0,1)_\tv$ is annihilated by $\D(e)$ for some operator $e$, the state $\ket{\CS I\vac}$ will be annihilated by $\D(\CS\cdot e)$. Similarly, if $\ket{\vac}\otimes\dket{\tilde{v},\vac}$ is an eigenstate of $\D(e)$, so is the state $\ket{\CS I\vac}$ for $\D(\CS\cdot e)$, with the same eigenvalue. Using the characterization of the state $\ket{\vac}\otimes\dket{\tilde{v},\vac}$ obtained in appendix \refOld{AppD1}, it is possible to show that\footnote{The action of $a_k$ follows from the re-expansion of the current $\psi^+(z)$, for which the modes act as
\begin{equation}
\left(\rho_v^{(0,1)}\otimes\rho_u^{(1,0)}\ \D(\psi^+_{k})\right)\ket{\CS I\vac}=\left(\g\d_{k,0}-(\g-\g^{-1})(-\tilde{u}\g^{-2})^k\right)\ket{\CS I\vac}.
\end{equation}}
\begin{align}
\begin{split}
&\left(\rho_v^{(0,1)}\otimes\rho_u^{(1,0)}\ \D(x^-_{k\geq0})\right)\ket{\CS I\vac}=\d_{k,0}\ (-\g^{3/2}\tilde{v})^{-1}\ket{\CS I\vac},\\
&\left(\rho_v^{(0,1)}\otimes\rho_u^{(1,0)}\ \D(a_{k>0})\right)\ket{\CS I\vac}=-\dfrac1k(\g^k-\g^{-k})(-\tu \g^{-1})^k\ket{\CS I\vac}.
\end{split}
\end{align}
Comparing this result with the characterization \ref{char_vac_I} of $\ket{v,\vac}\rangle\otimes\ket{\vac}$, we deduce that the two states coincide, up to their norm. Unfortunately, the determination of the relative norm remains difficult, since the projection
\begin{equation}
\left(\dbra{v,\vac}\otimes\bra{\vac}\right)\ket{\CS I\vac}=\bra{\vac}\Phi_{\vac}^\ast\mat{0}{-1}{-1}{0}\CM_\CS^{(-1,1)}\Phi_{\vac}\mat{-1}{0}{0}{1}\ket{\vac}
\end{equation} 
is hard to evaluate. Yet, just like the coefficients $\CN$, this normalization coefficient can only depend on the weights $u_1$, $v_1$ and $\tu_1$, $\tv_1$, and do not produce any dependence on the gauge theory parameters that correspond to $u_1/u_2$ and $v_1/v_2$.

\paragraph{Weights} The comparison between actions of DIM algebra on the two states provides also the relationship between the weights $\tu_i=-\g^{3/2}v_i$ and $\tv_i=-\g^{-3/2}u_i$. Surprising, it does not coincide with the transformation of the weights for vertical $(0,1)$ and horizontal $(1,0)$ representations taken separately, in which case we would have $\tu_i=-\g v_i$ and $\tv_i=-\g^{-1}u_i$. It is thus essential to consider the $\CS$-transformation of the whole tensor product $(0,1)\otimes(1,0)$ to find the proper weight mapping.\footnote{In fact, Miki's automorphism is not quite unique (see appendix \ref{AppA}), and in this particular case it is possible to redefine it such that the two weights transformation coincide. However, this will no longer be the case for the two other examples we will treat, so that we decided to keep the simpler definition \ref{Miki_init} for $\CS$.} Note that in both cases, the weights of the representation in the intermediate channel $u'=-\g u_iv_i$ becomes $\tu'=-\g\tv_i\tu_i$, in agreement with the $\CS$-transformation of module $(1,1)_{u'}$. Physically speaking, it seems that the presence of the extra branes requires to adjust the position of the branes after rotation with an extra shift of $\pm(\e_1+\e_2)/4$ in order to observe the invariance of the amplitude. It would be interesting to investigate this phenomenon with more care, for instance by studying the effect of the branes on the graviphoton field responsible for the Omega-background.

Eventually, the exact exponent of the factor $\g$ in the weights transformation is not relevant for the gauge theory quantity that depends only on the ratio $v_2/v_1$. Since, according to this transformation, $v_2/v_1=\tu_2/\tu_1=\tv_1/\tv_2$, the v.e.v. of the Lax matrices $\la\CL^{(I)}\ra$ and $\la\CL^{(\CS I)}\ra$ do coincide.

\subsection{Example II: pure $U(1)$ gauge theory}
\begin{figure}
\begin{center}
\begin{tikzpicture}
\draw[postaction={on each segment={mid arrow=black}}] (-0.7,0.7) -- (0,0) -- (2,0)--(2.7,-0.7);
\draw[postaction={on each segment={mid arrow=black}}] (0,-1) -- (0,0);
\draw[postaction={on each segment={mid arrow=black}}] (2,0) -- (2,1);
\node[left,scale=0.7] at (0,0) {$\Phi_2^{(II)}$};
\node[right,scale=0.7] at (2,0) {$\Phi_1^{(II\ast)}$};
\node[below,scale=0.7] at (0,-1) {$(0,1)_{v_2}$};
\node[above,scale=0.7] at (-0.7,0.7) {$(1,-1)_{u_2}$};
\node[above,scale=0.7] at (2,1) {$(0,1)_{v_1}$};
\node[below,scale=0.7] at (1,0) {$(1,0)_{u_1'=u_2'}$};
\node[below,scale=0.7] at (2.7,-0.7) {$(1,-1)_{u_1}$};
\end{tikzpicture}
\hspace{10mm}
\begin{tikzpicture}
\draw[postaction={on each segment={mid arrow=black}}] (-0.7,-0.7) -- (0,0) -- (0,2)--(0.7,2.7);
\draw[postaction={on each segment={mid arrow=black}}] (1,0) -- (0,0);
\draw[postaction={on each segment={mid arrow=black}}] (0,2) -- (-1,2);
\node[left,scale=0.7] at (0,0) {$\Phi_2^{(\CS II)}$};
\node[right,scale=0.7] at (0,2) {$\Phi_1^{(\CS II\ast)}$};
\node[right,scale=0.7] at (0,1) {$(0,1)_{\tv_1=\tv_2}$};
\node[above,scale=0.7] at (0.7,2.7) {$(1,1)_{\tu'_1}$};
\node[below,scale=0.7] at (-1,2) {$(-1,0)_{\tu_1}$};
\node[below,scale=0.7] at (-0.7,-0.7) {$(1,1)_{\tu_2'}$};
\node[below,scale=0.7] at (1,0) {$(-1,0)_{\tu_2}$};
\end{tikzpicture}
\end{center}
\caption{Representation of the S-dual Lax matrices obtained as $\Phi^{(II\ast)}\Phi^{(II)}$ and $\Phi^{(\CS II\ast)}\cdot\Phi^{(\CS II)}$.}
\label{fig_S_II}
\end{figure}
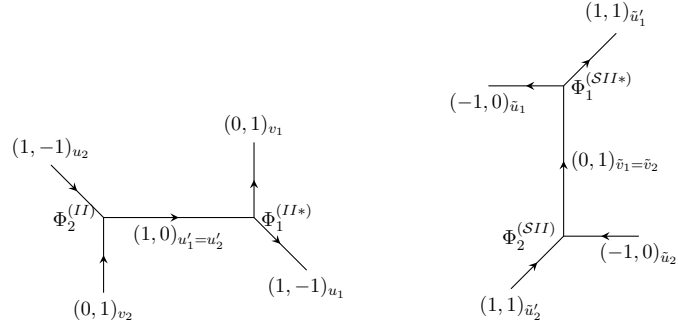

We would like to repeat the previous analysis in the case of the brane-web describing the pure $U(1)$ gauge theory (see figure \ref{fig_S_II}). In fact, this diagram has already been considered from the algebraic point of view by Awata and Kanno in \cite{Awata2009a}. In their paper, they investigated the invariance of the partition function under the choice of preferred direction. Changing the preferred direction effectively rotates the brane-web, although there is an important difference between their computation and ours: in our case, topological vertices are also rotated, which requires the use of a different intertwiner.

We introduce again a shortcut notation for the Lax matrices $\CL^{(II)}=\CL\mat{0}{1}{1}{-1}$, $\CL^{(\CS II)}=\CL\mat{-1}{0}{1}{1}$ and their associated intertwiners $\Phi^{(II)}=\Phi\mat{0}{1}{1}{-1}$, $\Phi^{(II\ast)}=\Phi^\ast\mat{0}{1}{1}{-1}$ and $\Phi^{(\CS II)}=\Phi\mat{-1}{0}{1}{1}$, $\Phi^{(\CS II\ast)}=\Phi^\ast\mat{-1}{0}{1}{1}$. The convention for the labeling of weights can be seen on figure \ref{fig_S_II}. Since the matrices $\CM_\CS$ are diagonal, the equality between the v.e.v. of the Lax matrix $\CL^{(II)}$ and the $\CS$-dual one boils down to an equality between
\begin{equation}
\la\CL^{(II)}\ra=\left(\dbra{v_1,\vac}\otimes\bra{\vac}\right)\CL\mat{0}{1}{1}{-1}\left(\dket{v_2,\vac}\otimes\ket{\vac}\right)\quad\text{and}\quad
\la\CL^{(\CS II)}\ra=\left(\bra{\vac}\otimes\bra{\vac}\right)\CL\mat{-1}{0}{1}{1}\left(\ket{\vac}\otimes\ket{\vac}\right).
\end{equation} 

Once again, the Lax matrix $\CL^{(II)}$ can be decomposed over its vertical components,
\begin{align}
\begin{split}
&\CL^{(II)}=\sum_{\l_1,\l_2}a_{\l_1}a_{\l_2}\CL^{(II)}_{\l_1,\l_2}\ \dket{v_1,\l_1}\dbra{v_2,\l_2}:(0,1)_{v_2}\otimes(1,-1)_{u_2}\to(0,1)_{v_1}\otimes (1,-1)_{u_1}\\
&\text{with}\quad \CL^{(II)}_{\l_1,\l_2}=(\g u_1)^{-|\l_1|}(u'_2)^{|\l_2|}\prod_{x\in\l_1}\chi_x^{-1}\dfrac{N(\l_2,\l_1|\g v_2/v_1)}{\CG(v_2/(\g v_1))}:\Phi_\vac^\ast(v_1)\Phi_\vac(v_2)\prod_{x\in\l_1}\eta^-(\chi_x)\prod_{x\in\l_2}\eta^+(\chi_x):,
\end{split}
\end{align}
and the weights obey the conservation relations $u'_i=-\g u_iv_i$ ($i=1,2$) and $u'_1=u'_2$. The v.e.v. is easily computed from this expression,
\begin{equation}
\la\CL^{(II)}\ra=\bra{\vac}\CL_{\vac,\vac}^{(II)}\ket{\vac}=\CG(v_2/(\g v_1))^{-1}.
\end{equation} 
On the other hand, the Lax matrix $\CL^{(\CS II)}$ is purely horizontal:
\begin{align}
\begin{split}
\CL^{(\CS II)}&=\sum_\l a_\l\ \ketS{\tv,\l}\braS{\tv,\l}:(-1,0)_{\tu_2}\times(1,1)_{\tu'_2}\to(-1,0)_{\tu_1}\times(1,1)_{\tu'_1},\\
&=\sum_\l a_\l (\g^{-1}\tu_1'\tu_2)^{|\l|}\prod_{x\in\l}\chi_x^{-1}\ \left(1\otimes:\Phi_\vac(\tv)\prod_{x\in\l}\eta^+(\chi_x):\right)\ketS{\O}\braS{\O}\left(1\otimes:\Phi_\vac^\ast(\tv)\prod_{x\in\l}\eta^-(\chi_x):\right),
\end{split}
\end{align}
with the weights satisfying the relations $\tv_i=-\g^{-1}\tu_i\tu'_i$ and $\tv_1=\tv_2=\tv$. The computation of the v.e.v. follows from the normal-ordering of operators, 
\begin{equation}
\la\CL^{(II)}\ra=\sum_\l\left(\dfrac{\tu'_1}{\g \tu'_2}\right)^{|\l|}\Zv(\l).
\end{equation}
We will show below that the weights transformation implies that $\qf=\tu'_1/(\g\tu'_2)=v_2/(\g v_1)$. Thus, the equality between the v.e.v. of the Lax matrices becomes
\begin{equation}\label{rel_S_U1}
\CG(\qf)^{-1}=\sum_\l\qf^{|\l|}\Zv(\l),
\end{equation}
which is indeed the identity arising from the application of S-duality to the $U(1)$ gauge theory. It is derived in \cite{Awata2009a} using a change of preferred direction, and a proof can be found in \cite{Macdonald,Haiman2002} (see also \cite{Nakajima2003}).

\paragraph{Transformation of the vacuum} We have seen previously that, since $\CL^{(\CS II)}=\bCF_\CS^{(II\ast)-1}\CL^{(II)}\bCF_\CS^{(II)}$, the equality between v.e.v. of Lax matrices would follow from the transformation of the vacua under the two-tensors $\bCF_\CS^{(II)}=\bCF_\CS\mat{0}{1}{1}{-1}$ and $\bCF_\CS^{(II\ast)}=\bCF_\CS^\ast\mat{0}{1}{1}{-1}$. The treatment of $\bCF_\CS^{(II)}$ and $\bCF_\CS^{(II\ast)}$ is similar, and we will discuss only the first one. The expression of this two-tensor follows from \ref{expr_bCF}:\footnote{In this expression, the products over $x=(i,j)\in\l$ involve the variables $\chi_x=\tv q_1^{i-1}q_2^{j-1}$ while the products over $x\in\mu$ involve $\chi_x=vq_1^{i-1}q_2^{j-1}$. Note also that we have dropped the index `2' for the weights to lighten the notations.}
\begin{equation}\small
\bCF_\CS^{(II)}=\CN\mat{0}{1}{1}{-1}^{-1}\sum_{\l,\mu}a_{\l} a_{\mu}\dfrac{(\g^{-1}\tu)^{|\l|}}{(-u)^{|\mu|}}\prod_{x\in\mu}\chi_x^{-1}\left(\dket{v,\mu}\otimes:\Phi_\vac(v)^{-1}\prod_{x\in\mu}\eta^+(\chi_x)^{-1}:\ket{P_\l}\right)\braS{\O}\left(1\otimes:\Phi_\vac^\ast(\tv)\prod_{x\in\l}\eta^-(\chi_x):\right).
\end{equation}
Unfortunately, its action on the vacuum state remains fairly complicated,
\begin{equation}
\ket{\CS II\vac}=\bCF_\CS^{(II)}\left(\ket{\vac}\otimes\ket{\vac}\right)=\CN\mat{0}{1}{1}{-1}^{-1}\sum_{\l,\mu}a_{\l} a_{\mu}\dfrac{(\g^{-1}\tu)^{|\l|}}{(-u)^{|\mu|}}\prod_{x\in\mu}\chi_x^{-1}\left(\dket{v,\mu}\otimes:\Phi_\vac(v)^{-1}\prod_{x\in\mu}\eta^+(\chi_x)^{-1}:\ket{P_\l}\right),
\end{equation} 
and we need to employ again an indirect method. Note however that, this time, the coefficient $\CN$ can be fixed by computing the overlap
\begin{equation}\label{overlap}
\left(\dbra{v,\vac}\otimes\bra{\vac}\right)\ket{\CS II\vac}=\CN\mat{0}{1}{1}{-1}^{-1}\sum_{\l}a_{\l}(\g^{-1}\tu)^{|\l|}\bra{\vac}e^{-\sum_{k>0}\frac{(\g^{3/2}v)^{-k}}{\s_k}\a_k}\ket{P_\l}.
\end{equation} 

Hence, we need to compare the action of the DIM algebra on the states $\dket{v,\vac}\otimes\ket{\vac}$ and $\ket{\CS II\vac}$. The characterization of the first one is easily deduced using the coproduct taken in the appropriate representations:
\begin{align}\label{char_var_II}
\begin{split}
&\left(\rho_v^{(0,1)}\otimes\rho_u^{(1,-1)}\ \D(x_{k>0}^-)\right)\dket{v,\vac}\otimes\ket{\vac}=\d_{k,1}\ u^{-1}\ \dket{v,\vac}\otimes\ket{\vac},\\
&\left(\rho_v^{(0,1)}\otimes\rho_u^{(1,-1)}\ \D(a_{k>0})\right)\dket{v,\vac}\otimes\ket{\vac}=-\dfrac1k(\g^k-\g^{-k})(\g^{1/2}v)^k\ \dket{v,\vac}\otimes\ket{\vac}.
\end{split}
\end{align}
These transformation properties coincide with those of the coherent state $\ketA{v,\vac}$ (defined in appendix \ref{AppC}) belonging to the Fock module $(1,0)_{-\g uv}$. The action of DIM on the state $\ket{\CS II\vac}$ follows again from the property \ref{prop_bCF} obeyed by $\bCF_\CS^{(II)}$ that implies
\begin{equation}\label{prop_CSII_vac}
\left(\rho_v^{(0,1)}\otimes\rho_u^{(1,-1)}\ \D(\CS\cdot e)\right)\ket{\CS II\vac}=\bCF_\CS^{(II)}\left(\rho_{\tu}^{(-1,0)}\otimes\rho_{\tu'}^{(1,1)}\ \D(e)\right)\left(\ket{\vac}\otimes\ket{\vac}\right).
\end{equation}
The results obtained for the state $\ket{\vac}\otimes\ket{\vac}$ in the appendix \ref{AppD2} imply
\begin{align}
\begin{split}
&\left(\rho_v^{(0,1)}\otimes\rho_u^{(1,-1)}\ \D(x_{k>0}^-)\right)\ket{\CS II\vac}=\d_{k,1}\ (\g^{1/2}\tu')^{-1}\ \ket{\CS II\vac},\\
&\left(\rho_v^{(0,1)}\otimes\rho_u^{(1,-1)}\ \D(a_{k>0})\right)\ket{\CS II\vac}=(1-q_3)(-\tu\g^{-1})^k\ \ket{\CS II\vac}.
\end{split}
\end{align}
It is readily observed that the states are characterized by the same action for the elements $x_k^-$ and $a_k$ (with $k>0$) of the DIM algebra. The intertwiner is again invertible and defines an isomorphism between the tensor module $(0,1)_v\otimes(1,-1)_u$ and the module $(1,0)_{u'}$ in the intermediate channel. In this module, states are uniquely characterized by the action of $a_k$ and $x_k^-$ for $k>0$, up to a possible normalization factor. Thus, the vacuum state is indeed mapped to the vacuum state under $\bCF_\CS^{(II)}$. The same is true for $\bCF_\CS^{(II)\ast}$. Moreover, comparing the action of DIM, we also deduce the weights transformation $\tu_i=-\g^{3/2}v_i$ and $\tu'_i=\g^{-1/2}u_i$ (for $i=1,2$). This transformation still differs by factors $\g^{\pm1/2}$ from the weight transformation of the modules considered individually. However, these factors disappear in the ratio $v_2/v_1=\tu'_1/\tu'_2$ entering in \ref{rel_S_U1}.

\section{$\CT$-operators and the $U(2)$ self-dual diagram}
\subsection{$\CT$-operators}
In this section, we discuss yet another type of algebraic object, obtained by coupling the intertwiner $\Phi\mat{\ell_1}{\bell_1}{\ell_2}{\bell_2}$ to the dual one $\Phi^\ast\mat{\ell_1^\ast}{\bell_1^\ast}{\ell_2^\ast}{\bell_2^\ast}$ through the legs bearing the representation $(\ell_i,\bell_i)=(\ell_i^\ast,\bell_i^\ast)$ with either $i=1$ or $i=2$. We call this type of objects \textit{$\CT$-operators} in reference to those constructed from linear quiver gauge theories and identified with the Baxter $\CT$-operator of an underlying integrable system \cite{Awata2016,Bourgine2017b}.\footnote{Actually, this term is slightly abused here, since the underlying integrable system is a chain of length two, with boundary operators applied on each site. Yet, it makes perfect sense for linear quivers of higher rank \cite{Bourgine2017b}.} These operators will be denoted $\CT^{(i)}=\Phi\cdot_i\Phi^\ast$ where the index $i$ in the product $\cdot_i$ refers to the coupling channel.

Since the two representations $(\ell_1,\bell_1)$ and $(\ell_2,\bell_2)$ now play a different role, the covariance property under the action of the DIM algebra might seem to be lost. Yet, the co-associativity of the coproduct, namely the property $(\D\otimes1)\D=(1\otimes\D)\D$, ensures that the operators $\CT^{(i)}$ obey the following properties:
\begin{align}
\begin{split}\label{covar_T}
&\left(\rho^{(\ell,\bell)}\otimes\rho^{(\ell_2^\ast,\bell_2^\ast)}\ \D'(e)\right)\CT^{(1)}=\CT^{(1)}\left(\rho^{(\ell_2,\bell_2)}\otimes\rho^{(\ell^\ast,\bell^\ast)}\ \D'(e)\right),\\
&\left(\rho^{(\ell,\bell)}\otimes\rho^{(\ell_1^\ast,\bell_1^\ast)}\ \D(e)\right)\CT^{(2)}=\CT^{(2)}\left(\rho^{(\ell_1,\bell_1)}\otimes\rho^{(\ell^\ast,\bell^\ast)}\ \D(e)\right).
\end{split}
\end{align}
In contrast with the covariance property \ref{covar_Lax} of Lax matrices, here the left and right hand sides involve the same coproduct (either $\D$ or $\D'$). For simplicity, we will again restrict ourselves to the case where levels on both sides coincide, i.e. $(\ell,\bell)=(\ell^\ast,\bell^\ast)$ and $(\ell_i,\bell_i)=(\ell_i^\ast,\bell_i^\ast)$, and denote the corresponding operators
$\CT^{(i)}\mat{\ell_1}{\bell_1}{\ell_2}{\bell_2}$.

The same kind of operators can be constructed using the S-dual intertwiners associated to the twisted coproduct. They obey the covariance properties \ref{covar_T} with $\D$, $\D'$ replaced by $\D_\CS$, $\D_\CS'$, and will be denoted $\CT_\CS^{(i)}$. The relation \ref{def_phi_CS} between S-dual and rotated intertwiners extends to $\CT$-operators in the form
\begin{align}
\begin{split}\label{rotate_CT}
&\CT_\CS^{(1)}\mat{\ell_1}{\bell_1}{\ell_2}{\bell_2}=\left(\CM_\CS^{(\bell,-\ell)}\otimes\CM_\CS^{(\bell_2,-\ell_2)}\right)\CT^{(1)}\mat{\bell_1}{-\ell_1}{\bell_2}{-\ell_2}\left(\CM_\CS^{(\bell_2,-\ell_2)-1}\CM_\CS^{(\bell,-\ell)-1}\right),\\
&\CT_\CS^{(2)}\mat{\ell_1}{\bell_1}{\ell_2}{\bell_2}=\left(\CM_\CS^{(\bell,-\ell)}\otimes\CM_\CS^{(\bell_1,-\ell_1)}\right)\CT^{(2)}\mat{\bell_1}{-\ell_1}{\bell_2}{-\ell_2}\left(\CM_\CS^{(\bell_1,-\ell_1)-1}\CM_\CS^{(\bell,-\ell)-1}\right).
\end{split}
\end{align}

The main difficulty, working with $\CT$-operators instead of Lax matrices, is the lack of a rotation formula like \ref{trans_Lax} relating the rotated $\CT$-operators to the original ones by a two-tensor transformation. Yet, we would like to give here an argument to justify the existence of such tensors. In the next subsection, we will provide a concrete example for which the vacuum properties support our proposal. This argument is based on the fact that after, multiplication by a permutation, the $\CT$-operators satisfy the covariance property \ref{covar_Lax} of a Lax matrix. As a result, we expect that these permuted objects coincide with a Lax matrix up to a normalization (or anomaly) factor, i.e.
\begin{equation}
\CL\mat{\ell}{\bell}{\ell_2}{\bell_2}\propto\CT^{(1)}\mat{\ell_1}{\bell_1}{\ell_2}{\bell_2}\CP\mat{\ell}{\bell}{\ell_2}{\bell_2}\propto\CP\mat{\ell_1}{\bell_1}{\ell}{\bell}\CT^{(2)}\mat{\ell_1}{\bell_1}{\ell_2}{\bell_2}.
\end{equation}
This motivates the introduction of the two-tensors $\bCF^{(i)}$ and $\bCF^{(i)\ast}$, presumably differing from $\bCF$ and $\bCF^\ast$ only by a permutation and a normalization, and such that
\begin{equation}\label{trans_CT}
\CT^{(i)}\mat{-\bell_1}{\ell_1}{-\bell_2}{\ell_2}=\bCF_\CS^{(i)\ast}\mat{\ell_1}{\bell_1}{\ell_2}{\bell_2}^{-1}\CT^{(i)}\mat{\ell_1}{\bell_1}{\ell_2}{\bell_2}\bCF_\CS^{(i)}\mat{\ell_1}{\bell_1}{\ell_2}{\bell_2}.
\end{equation}
We will not use any explicit formula for these tensors, but only their covariance property that should follow from the ones obeyed by $\bCF$ and $\bCF^\ast$ in \ref{prop_bCF}. For instance, in the case of $\CT^{(1)}$, we expect that\footnote{We have used the Ansatz
\begin{equation}
\bCF_\CS^{(1)}\mat{\ell_1}{\bell_1}{\ell_2}{\bell_2}\propto \CP\mat{\ell}{\bell}{\ell_2}{\bell_2}\bCF_\CS\mat{\ell}{\bell}{\ell_2}{\bell_2}\CP\mat{-\bell_2}{\ell_2}{-\bell}{\ell},\quad \bCF_\CS^{(1)\ast}\mat{\ell_1}{\bell_1}{\ell_2}{\bell_2}\propto\bCF_\CS^\ast\mat{\ell}{\bell}{\ell_2}{\bell_2}
\end{equation}}
\begin{align}
\begin{split}\label{prop_bCF1}
&\bCF_\CS^{(1)}\mat{\ell_1}{\bell_1}{\ell_2}{\bell_2}\left(\rho^{(-\bell_2,\ell_2)}_{\tv_2}\otimes\rho^{(-\bell,\ell)}_{\tv}\ \D'(e)\right)=\left(\rho^{(\ell_2,\bell_2)}_{v_2}\otimes\rho^{(\ell,\bell)}_{v}\ \D'(\CS\cdot e)\right)\bCF_\CS^{(1)}\mat{\ell_1}{\bell_1}{\ell_2}{\bell_2},\\
&\left(\rho^{(-\bell,\ell)}_{\tv}\otimes\rho_{\tv_2}^{(-\bell_2,\ell_2)}\ \D'(e)\right)\bCF_\CS^{(1)\ast}\mat{\ell_1}{\bell_1}{\ell_2}{\bell_2}^{-1}=\bCF_\CS^{(1)\ast}\mat{\ell_1}{\bell_1}{\ell_2}{\bell_2}^{-1}\left(\rho_{v}^{(\ell,\bell)}\otimes\rho_{v_2}^{(\ell_2,\bell_2)}\ \D'(\CS\cdot e)\right).
\end{split}
\end{align}
These covariance properties are in agreement with the characterization \ref{covar_T} of the $\CT$-operator $\CT^{(1)}$. They imply certain relations between the vacuum and its rotation that will be checked below by direct computation on a specific example.

\subsection{Example: pure $U(2)$ gauge theory}
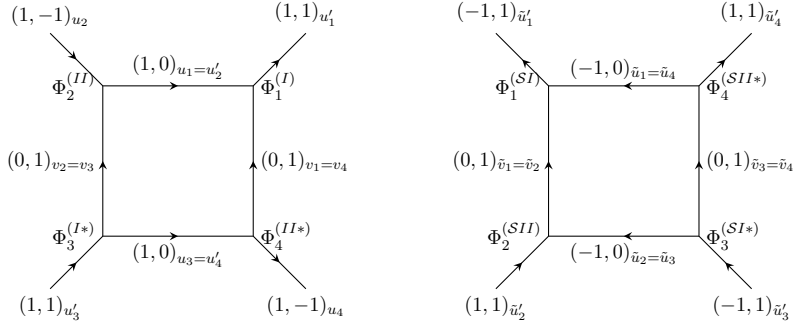
\begin{figure}
\begin{center}
\begin{tikzpicture}
\draw[postaction={on each segment={mid arrow=black}}] (-0.7,-0.7) -- (0,0) -- (2,0)--(2.7,-0.7);
\draw[postaction={on each segment={mid arrow=black}}] (-0.7,2.7) -- (0,2) -- (2,2)--(2.7,2.7);
\draw[postaction={on each segment={mid arrow=black}}] (0,0) -- (0,2);
\draw[postaction={on each segment={mid arrow=black}}] (2,0) -- (2,2);
\node[left,scale=0.7] at (0,0) {$\Phi_3^{(I\ast)}$};
\node[left,scale=0.7] at (0,2) {$\Phi_2^{(II)}$};
\node[right,scale=0.7] at (2,2) {$\Phi_1^{(I)}$};
\node[right,scale=0.7] at (2,0) {$\Phi_4^{(II\ast)}$};
\node[below,scale=0.7] at (-0.7,-0.7) {$(1,1)_{u_3'}$};
\node[below,scale=0.7] at (2.7,-0.7) {$(1,-1)_{u_4}$};
\node[above,scale=0.7] at (-0.7,2.7) {$(1,-1)_{u_2}$};
\node[above,scale=0.7] at (2.7,2.7) {$(1,1)_{u_1'}$};
\node[above,scale=0.7] at (1,2) {$(1,0)_{u_1=u_2'}$};
\node[below,scale=0.7] at (1,0) {$(1,0)_{u_3=u_4'}$};
\node[left,scale=0.7] at (0,1) {$(0,1)_{v_2=v_3}$};
\node[right,scale=0.7] at (2,1) {$(0,1)_{v_1=v_4}$};
\end{tikzpicture}
\hspace{10mm}
\begin{tikzpicture}
\draw[postaction={on each segment={mid arrow=black}}] (-0.7,-0.7) -- (0,0) -- (0,2)--(-0.7,2.7);
\draw[postaction={on each segment={mid arrow=black}}] (2.7,-0.7) -- (2,0) -- (2,2)--(2.7,2.7);
\draw[postaction={on each segment={mid arrow=black}}] (2,0) -- (0,0);
\draw[postaction={on each segment={mid arrow=black}}] (2,2) -- (0,2);
\node[left,scale=0.7] at (0,0) {$\Phi_2^{(\CS II)}$};
\node[left,scale=0.7] at (0,2) {$\Phi_1^{(\CS I)}$};
\node[right,scale=0.7] at (2,2) {$\Phi_4^{(\CS II\ast)}$};
\node[right,scale=0.7] at (2,0) {$\Phi_3^{(\CS I\ast)}$};
\node[below,scale=0.7] at (-0.7,-0.7) {$(1,1)_{\tu_2'}$};
\node[below,scale=0.7] at (2.7,-0.7) {$(-1,1)_{\tu_3'}$};
\node[above,scale=0.7] at (-0.7,2.7) {$(-1,1)_{\tu_1'}$};
\node[above,scale=0.7] at (2.7,2.7) {$(1,1)_{\tu_4'}$};
\node[above,scale=0.7] at (1,2) {$(-1,0)_{\tu_1=\tu_4}$};
\node[below,scale=0.7] at (1,0) {$(-1,0)_{\tu_2=\tu_3}$};
\node[left,scale=0.7] at (0,1) {$(0,1)_{\tv_1=\tv_2}$};
\node[right,scale=0.7] at (2,1) {$(0,1)_{\tv_3=\tv_4}$};
\end{tikzpicture}
\end{center}
\caption{Representation of the two dual $U(2)$ brane-webs.}
\label{fig_U2}
\end{figure}

\begin{figure}
\begin{center}
\begin{tikzpicture}
\draw[postaction={on each segment={mid arrow=black}}] (-0.7,-0.7) -- (0,0) -- (0.7,-0.7);
\draw[postaction={on each segment={mid arrow=black}}] (-0.7,2.7) -- (0,2) -- (0.7,2.7);
\draw[postaction={on each segment={mid arrow=black}}] (0,0) -- (0,2);
\node[right,scale=0.7] at (0,0) {$\Phi^{\ast}\mat{0}{2}{1}{-1}$};
\node[right,scale=0.7] at (0,2) {$\Phi\mat{0}{2}{1}{-1}$};
\node[below,scale=0.7] at (-0.7,-0.7) {$(1,1)_{u_3'}$};
\node[below,scale=0.7] at (0.7,-0.7) {$(1,-1)_{u_4}$};
\node[above,scale=0.7] at (-0.7,2.7) {$(1,-1)_{u_2}$};
\node[above,scale=0.7] at (0.7,2.7) {$(1,1)_{u_1'}$};
\node[right,scale=0.7] at (0,1) {$(0,2)_{\vec v}$};
\end{tikzpicture}
\hspace{10mm}
\begin{tikzpicture}
\draw[postaction={on each segment={mid arrow=black}}] (-0.7,-0.7) -- (0,0) -- (0,2)--(-0.7,2.7);
\draw[postaction={on each segment={mid arrow=black}}] (0.7,-0.7) -- (0,0);
\draw[postaction={on each segment={mid arrow=black}}] (0,2)--(0.7,2.7);
\node[right,scale=0.7] at (0,0) {$\Phi\mat{-1}{1}{1}{1}$};
\node[right,scale=0.7] at (0,2) {$\Phi^\ast\mat{-1}{1}{1}{1}$};
\node[below,scale=0.7] at (-0.7,-0.7) {$(1,1)_{\tu_2'}$};
\node[below,scale=0.7] at (0.7,-0.7) {$(-1,1)_{\tu_3'}$};
\node[above,scale=0.7] at (-0.7,2.7) {$(-1,1)_{\tu_1'}$};
\node[above,scale=0.7] at (0.7,2.7) {$(1,1)_{\tu_4'}$};
\node[left,scale=0.7] at (0,1) {$(0,2)_{\vec\tv}$};
\end{tikzpicture}
\end{center}
\caption{Folded version of the two dual $U(2)$ diagrams.}
\label{fig_U2_folded}
\end{figure}
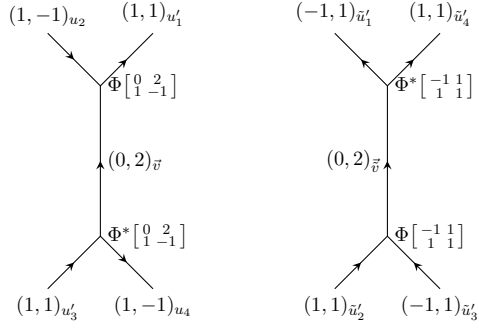

We would like to treat the $\CT$-operator associated to the pure $U(2)$ gauge theory as an example. Including the anomaly factors, this operator is given by
\begin{equation}
\CT=\CA(\vec v)\CT^{(1)}\mat{0}{2}{1}{-1}:(1,-1)_{u_2}\times(1,1)_{u'_3}\to(1,1)_{u'_1}\times(1,-1)_{u_4},
\end{equation} 
where intertwiners are internally coupled through the vertical representation $(0,2)_{\vec v}$ (see figure \ref{fig_U2_folded} left). Since we restricted ourselves to levels $\pm1$ in this paper, we need to decompose the $\CT$-operator in terms of four intertwiners instead of two, exploiting in reverse the fusion method developed in \cite{Bourgine2017b}. As a result, the operator $\CT$ is written in terms of four generalized AFS intertwiners following the gluing rules of the brane-web for pure $U(2)$ $\CN=1$ SYM,
\begin{equation}\label{def_CT}
\CT=\sum_{\l_1,\l_2}a_{\l_1}a_{\l_2}\ \Phi_{\l_1}^{(I)}\Phi_{\l_2}^{(II)}\otimes\Phi_{\l_1}^{(II\ast)}\Phi_{\l_2}^{(I\ast)}.
\end{equation}
It turns out that these intertwiners coincide with those employed in the two previous examples, and that we denoted with the shortcut notations $\Phi^{(I)}$, $\Phi^{(II)}$,... The weights associated to each representation are indicated on the figure \ref{fig_U2}: each intertwiner $\Phi_i$ depends on the weights $u_i$, $v_i$ and $u'_i$ constraint by the conservation relation $u'_i=-\g u_iv_i$ (for $i=1\cdots4$). In addition, the two vertical couplings impose the relations $v_1=v_4$ and $v_2=v_3$, and the two horizontal couplings the relations $u_1=u'_2$ and $u_3=u'_4$. Introducing the expressions of intertwiners found previously, and normal-ordering the vertex operators in the two horizontal channels, the $\CT$-operator takes the explicit form\footnote{We have also used the property
\begin{equation}
\dfrac{a_{\vec\l}}{a_{\l_1}a_{\l_2}}=\dfrac{(-\g v_1)^{-|\l_2|}(-\g v_2)^{-|\l_1|}\prod_{x\in\vec\l}\chi_x}{N(\l_1,\l_2|v_1/v_2)N(\l_2,\l_1|v_2/v_1)}.
\end{equation}}
\begin{equation}
\CT=\CG(v_2/(q_3v_1))\CG(v_2/v_1)\sum_{\vec\l}\left(\dfrac{u_1'}{q_3u'_3}\right)^{|\vec\l|}\Zv(\vec v,\vec\l)\ :\Phi_\vac(v_1)\Phi_\vac(v_2)\prod_{x\in\vec\l}\eta^+(\chi_x):\otimes :\Phi_\vac^\ast(v_1)\Phi_\vac^\ast(v_2)\prod_{x\in\vec\l}\eta^-(\chi_x):,
\end{equation}
with $\vec v=(v_1,v_2)$ and $\vec\l=(\l_1,\l_2)$. As already mentionned, the v.e.v. of this operator reproduces the instanton partition function of the gauge theory. The latter depends on the gauge coupling $\qf=u'_1/u'_3$ and the ratio $\rf=v_2/v_1$ of (exponentiated) Coulomb branch v.e.v.,
\begin{equation}
\la\CT\ra=\left(\bra{\vac}\otimes\bra{\vac}\right)\CT\left(\ket{\vac}\otimes\ket{\vac}\right)=\CG(\rf)\CG(q_3^{-1}\rf)\sum_{\vec\l}(\qf/q_3)^{|\vec\l|}\Zv(\vec v,\vec\l).
\end{equation} 

The formula \ref{rotate_CT} relates the S-dual $\CT$-operators (associated to the $\CS$-twisted coproduct) with the rotated ones. Since the matrices $\CM_\CS^{(1,\pm1)}$ involved in this formula map the vacuum state to itself in the Fock modules, the v.e.v. of the two $\CT$-operators should coincide. The rotated brane-web diagram is represented on figure \ref{fig_U2} (right), it involves the dual intertwiners $\Phi^{(\CS I)}$, $\Phi^{(\CS II)}$,... studied in the previous examples. The weights labeling has also been represented on this figure: the intertwiner $\Phi_i$ now depends on the weights $\tu_i$, $\tu_i'$ and $\tv_i$ obeying the conservation relation:
\begin{equation}
\tu'_1=-\g\tu_1\tv_1,\quad \tv_2=-\g^{-1}\tu_2\tu'_2,\quad \tu'_3=-\g\tu_3\tv_3,\quad \tv_4=-\g^{-1}\tu_4\tu'_4.
\end{equation}
In addition, horizontal and vertical couplings impose $\tu_1=\tu_4$, $\tu_2=\tu_3$, $\tv_1=\tv_2$ and $\tv_3=\tv_4$. These coupled intertwiners produce the operator\footnote{Here we have introduced the permutation
\begin{equation}
\CP\mat{1}{1}{-1}{1}=\sum_{\l,\mu}a_\l a_\mu\ \left(\ket{P_\l}\otimes\ket{P_\mu}\right)\left(\bra{P_\mu}\otimes\bra{P_\l}\right),
\end{equation} 
in order to produce an operator $\CT^\CS:(1,1)_{\tu'_2}\times(-1,1)_{\tu_3'}\to(-1,1)_{\tu'_1}\times(1,1)_{\tu_4'}$ in agreement with the brane-web diagram and the covariance property of $\CT$-operators.}
\begin{equation}\label{def_CTS}
\CT^\CS=\sum_{\l_1\,\l_2}a_{\l_1}a_{\l_2}\left(\Phi_{\l_1}^{(\CS I)}\otimes1\right)\ketS{\tv_4,\l_2}\braS{\tv_2,\l_1}\left(\Phi_{\l_2}^{(\CS I\ast)}\otimes1\right)\CP\mat{1}{1}{-1}{1}.
\end{equation}
By definition, $\CT^\CS\propto\CT^{(1)}\mat{-2}{0}{1}{1}$, up to a normalization factor. Using the expression of the intertwiners obtained previously, and exploiting the reflection property of the state $\ketS{\O}$, this operator writes
\begin{align}
\begin{split}
\CT^\CS=\CG(\tv_1/(q_3\tv_3))\CG(\tv_1/\tv_3)\sum_{\vec\l}\left(\dfrac{\tu_3'}{q_3\tu'_1}\right)^{|\vec\l|}\Zv(\vec\tv,\vec\l)\ &\left(1\otimes :\Phi_\vac(\tv_1)\Phi_\vac(\tv_3)\prod_{x\in\vec\l}\eta^+(\chi_x):\right)\ketS{\O}\\
&\braS{\O}\left(:\Phi_\vac^\ast(\tv_1)\Phi_\vac^\ast(\tv_2)\prod_{x\in\vec\l}\eta^-(\chi_x):\otimes1\right),
\end{split}
\end{align}
with $\vec\tv=(\tv_1,\tv_3)$ (the weight $\tv_3$ being associated to the second Young diagram $\l_2$). Once again, the v.e.v. reproduces the instanton partition function of the gauge theory,\footnote{Note that $\Zv((v_1,v_2),(\l_1,\l_2))=\Zv((v_2,v_1),(\l_2,\l_1))$, and the instanton part is invariant under the replacement $\tilde{\rf}\to\tilde{\rf}^{-1}$.}
\begin{equation}
\la\CT^\CS\ra=\left(\bra{\vac}\otimes\bra{\vac}\right)\CT^\CS\left(\ket{\vac}\otimes\ket{\vac}\right)=\CG(\tilde{\rf})\CG(q_3^{-1}\tilde{\rf})\sum_{\vec\l}(\tilde{\qf}/q_3)^{|\vec\l|}\Zv(\vec\tv,\vec\l).
\end{equation} 
However, the gauge coupling now corresponds to $\tilde{\qf}=\tu_3'/\tu'_1$ and the ratio of Coulomb branch v.e.v. is given by $\tilde{\rf}=\tv_1/\tv_3$. We will see in the next paragraph that the transformation of weights gives $\tilde{\qf}=\qf^{-1}$ and (presumably) $\tilde{\rf}=\qf\rf$. The invariance of the $U(2)$ partition function under this replacement of the parameters $(\qf,\rf)\to(\qf^{-1},\qf\rf)$ is called \textit{slicing invariance}, it is a consequence of the fiber-base duality for the toric diagram \cite{Ito2012,Mitev2014}. From the gauge theory point of view, this invariance follows from the Weyl reflection of the enhanced global $E_1$ symmetry at the UV fixed point. As we have seen here, it arises in our formalism as the equality between the v.e.v. of $\CT$-operators associated to the two coalgebraic structures.

\paragraph{S-transformation of the vacuum} In the previous subsection, we have assumed that the rotated $\CT$-operator $\CT^\CS$ can be obtained from $\CT$ using the two-tensors
\begin{equation}
\CT^\CS=\bCF_\CS^{(1)\ast}\mat{0}{2}{1}{-1}^{-1}\CT\bCF_\CS^{(1)}\mat{0}{2}{1}{-1}.
\end{equation} 
If so, the equality between the v.e.v. of $\CT$-operators can be seen as a consequence of the mapping by $\bCF_\CS^{(1)}$ of the vacuum state $\ket{0}=\ket{\vac}\otimes\ket{\vac}$ for $\CT^\CS$ to the vacuum state for $\CT$ which turns out to be also $\ket{0}$, and a similar mapping for the dual vacuum $\bra{0}=\bra{\vac}\otimes\bra{\vac}$ under $\bCF_\CS^{(1)\ast}$. In order to test this hypothesis, we will compare the action of several DIM algebra elements on the vacuum states and on the transformed states
\begin{equation}\label{def_CSvac}
\ket{\CS\vac}=\bCF_\CS^{(1)}\mat{0}{2}{1}{-1}\ket{0},\quad \bra{\CS\vac}=\bra{0}\bCF_\CS^{(1)\ast}\mat{0}{2}{1}{-1}^{-1},
\end{equation}
belonging to the modules $(1,-1)_{u_2}\otimes(1,1)_{u'_3}$ and $(1,1)_{u'_1}\otimes(1,-1)_{u_4}$ respectively. The action on the vacuum can be obtained by a direct calculation, it reads
\begin{align}\label{char_var_III}
\begin{split}
&\left(\rho_{u_2}^{(1,-1)}\otimes\rho_{u'_3}^{(1,1)}\ \D'(x_{k>0}^+)\right)\ket{0}=\d_{k,1}u'_3\ket{0},\quad\left(\rho_{u_2}^{(1,-1)}\otimes\rho_{u'_3}^{(1,1)}\ \D'(x_{k>0}^-)\right)\ket{0}=\d_{k,1} u_2^{-1}\ket{0},\\
&\bra{0}\left(\rho_{u'_1}^{(1,1)}\otimes\rho_{u_4}^{(1,-1)}\ \D'(x_{k<0}^+)\right)=\d_{k,-1}u_4\bra{0},\quad\bra{0}\left(\rho_{u'_1}^{(1,1)}\otimes\rho_{u_4}^{(1,-1)}\ \D'(x_{k<0}^-)\right)=\d_{k,-1} (u'_1)^{-1}\bra{0},\\
&\left(\rho_{u_2}^{(1,-1)}\otimes\rho_{u'_3}^{(1,1)}\ \D'(\psi^+(z))\right)\ket{0}=\ket{0},\quad \bra{0}\left(\rho_{u'_1}^{(1,1)}\otimes\rho_{u_4}^{(1,-1)}\ \D'(\psi^-(z))\right)=\bra{0},\\
\end{split}
\end{align}
On the other hand, the action of these elements on the states \ref{def_CSvac} follows once again from the covariance property of the two-tensors,
\begin{align}
\begin{split}
&\left(\rho_{u_2}^{(1,-1)}\otimes\rho_{u'_3}^{(1,1)}\ \D'(\CS\cdot e)\right)\ket{\CS\vac}=\bCF_\CS^{(1)}\mat{0}{2}{1}{-1}\left(\rho_{\tu'_2}^{(1,1)}\otimes\rho_{\tu'_3}^{(-1,1)}\ \D'(e)\right)\left(\ket{\vac}\otimes\ket{\vac}\right),\\
&\bra{\CS\vac}\left(\rho_{u'_1}^{(1,1)}\otimes\rho_{u_4}^{(1,-1)}\ \D'(\CS\cdot e)\right)=\left(\bra{\vac}\otimes\bra{\vac}\right)\left(\rho_{\tu'_1}^{(-1,1)}\otimes\rho_{\tu'_4}^{(1,1)}\ \D'(e)\right)\bCF_\CS^{(1)\ast}\mat{0}{2}{1}{-1}^{-1},
\end{split}
\end{align}
for $e\in\DIM$. Using the results obtained in appendix \ref{AppD3}, we find the same action on the states $\ket{\CS\vac}$ and $\bra{\CS\vac}$ for the modes $x_{k>0}^\pm$ and $\psi_k^+$ (resp. $x_{k<0}^\pm$ and $\psi_{-k}^-$) as in \ref{char_var_III}, provided that we identify the weights as follows:
\begin{equation}
\tu_1'=\g u'_1,\quad \tu'_2=\g^{-1}u_2,\quad \tu'_3=\g u'_3,\quad \tu'_4=\g^{-1}u_4.
\end{equation} 
This remarkable agreement between the action of DIM on the different states support the rotation formula for the $\CT$-operators. Moreover, the weights identification implies $\tilde{\qf}=\qf^{-1}$, which is indeed required for the slicing invariance. Unfortunately this identification is not sufficient to deduce the relation between the two other gauge parameters $\rf$ and $\tilde{\rf}$. However, we expect that the transformation of the module $(0,2)_{\vec v}$ in the intermediate channel relates the vertical and horizontal weights as $\tu_1=\g^\a v_1$ and $\tu_2=\g^\a v_2$ for some power $\a$ depending on the surrounding branes configuration. This unknown factor would cancel in the ratio $\rf=v_2/v_1=\tu_2/\tu_1$, hence reproducing the relation $\tilde{\rf}=\qf\rf$.

\begin{figure}
\begin{center}
\begin{tikzpicture}[scale=0.9]
\draw[postaction={on each segment={mid arrow=black}}] (-0.7,-0.7) -- (0,0) -- (2,0)--(2.7,-0.7);
\draw[postaction={on each segment={mid arrow=black}}] (0,0) -- (0,1);
\draw[postaction={on each segment={mid arrow=black}}] (2,0) -- (2,1);
\draw [black] plot [only marks, mark=square*] coordinates {(2.7,-0.7) (-0.7,-0.7)};
\node[left,scale=0.7] at (0,0) {$\Phi_3^{(I\ast)}$};
\node[right,scale=0.7] at (2,0) {$\Phi_4^{(II\ast)}$};
\node[below,scale=0.7] at (-0.7,-0.7) {$(1,1)_{u_3'}$};
\node[below,scale=0.7] at (2.7,-0.7) {$(1,-1)_{u_4}$};
\node[below,scale=0.7] at (1,0) {$(1,0)_{u_3=u_4'}$};
\node[left,scale=0.7] at (0,1) {$(0,1)_{v_3}$};
\node[right,scale=0.7] at (2,1) {$(0,1)_{v_4}$};
\end{tikzpicture}
\hspace{10mm}
\begin{tikzpicture}[scale=0.9]
\draw[postaction={on each segment={mid arrow=black}}] (-0.7,2.7) -- (0,2) -- (2,2)--(2.7,2.7);
\draw[postaction={on each segment={mid arrow=black}}] (0,1) -- (0,2);
\draw[postaction={on each segment={mid arrow=black}}] (2,1) -- (2,2);
\draw [black] plot [only marks, mark=square*] coordinates {(-0.7,2.7) (2.7,2.7)};
\node[left,scale=0.7] at (0,2) {$\Phi_2^{(II)}$};
\node[right,scale=0.7] at (2,2) {$\Phi_1^{(I)}$};
\node[above,scale=0.7] at (-0.7,2.7) {$(1,-1)_{u_2}$};
\node[above,scale=0.7] at (2.7,2.7) {$(1,1)_{u_1'}$};
\node[above,scale=0.7] at (1,2) {$(1,0)_{u_1=u_2'}$};
\node[left,scale=0.7] at (0,1) {$(0,1)_{v_2}$};
\node[right,scale=0.7] at (2,1) {$(0,1)_{v_1}$};
\end{tikzpicture}
\hspace{10mm}
\begin{tikzpicture}[scale=0.9]
\draw[postaction={on each segment={mid arrow=black}}] (2.7,-0.7) -- (2,0) -- (2,2)--(2.7,2.7);
\draw[postaction={on each segment={mid arrow=black}}] (2,0) -- (1,0);
\draw[postaction={on each segment={mid arrow=black}}] (2,2) -- (1,2);
\draw [black] plot [only marks, mark=square*] coordinates {(2.7,-0.7) (2.7,2.7)};
\node[right,scale=0.7] at (2,2) {$\Phi_4^{(\CS II\ast)}$};
\node[right,scale=0.7] at (2,0) {$\Phi_3^{(\CS I\ast)}$};
\node[below,scale=0.7] at (2.7,-0.7) {$(-1,1)_{\tu_3'}$};
\node[above,scale=0.7] at (2.7,2.7) {$(1,1)_{\tu_4'}$};
\node[above,scale=0.7] at (1,2) {$(-1,0)_{\tu_4}$};
\node[below,scale=0.7] at (1,0) {$(-1,0)_{\tu_3}$};
\node[right,scale=0.7] at (2,1) {$(0,1)_{\tv_3=\tv_4}$};
\end{tikzpicture}
\hspace{10mm}
\begin{tikzpicture}[scale=0.9]
\draw[postaction={on each segment={mid arrow=black}}] (-0.7,-0.7) -- (0,0) -- (0,2)--(-0.7,2.7);
\draw[postaction={on each segment={mid arrow=black}}] (1,0) -- (0,0);
\draw[postaction={on each segment={mid arrow=black}}] (1,2) -- (0,2);
\draw [black] plot [only marks, mark=square*] coordinates {(-0.7,2.7) (-0.7,-0.7)};
\node[left,scale=0.7] at (0,0) {$\Phi_2^{(\CS II)}$};
\node[left,scale=0.7] at (0,2) {$\Phi_1^{(\CS I)}$};
\node[below,scale=0.7] at (-0.7,-0.7) {$(1,1)_{\tu_2'}$};
\node[above,scale=0.7] at (-0.7,2.7) {$(-1,1)_{\tu_1'}$};
\node[above,scale=0.7] at (1,2) {$(-1,0)_{\tu_1}$};
\node[below,scale=0.7] at (1,0) {$(-1,0)_{\tu_2}$};
\node[left,scale=0.7] at (0,1) {$(0,1)_{\tv_1=\tv_2}$};
\end{tikzpicture}
\end{center}
\caption{Representation of the Gaiotto states $\dket{G,\vec v}$, $\dbra{G,\vec v}$ and the S-dual states (the black squares represent projections on the vacuum state).}
\label{fig_U2_Gaiotto}
\end{figure}
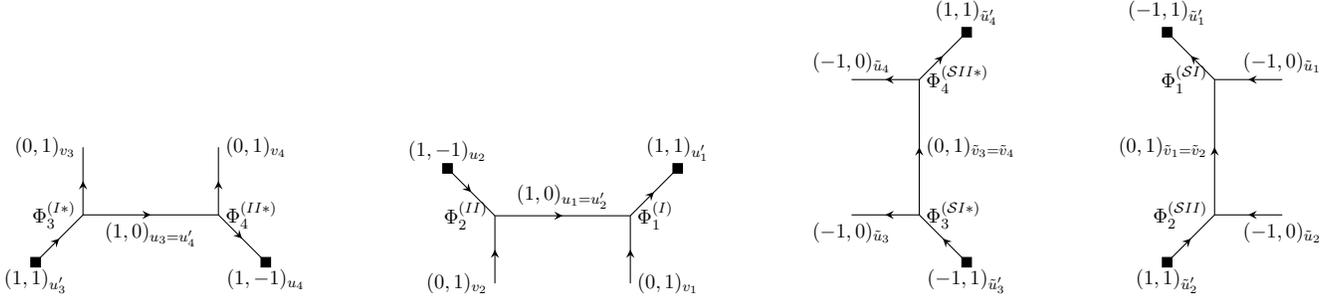

\paragraph{Gaiotto states} We would like to conclude this study of the $U(2)$ gauge theory with a short remark on Gaitto states. These states were originally defined as Whittaker states for the Virasoro algebra \cite{Gaiotto2009,Marshakov2009,Kanno2011,Kanno2012}, but this definition can be extended to q-Virasoro and q-$W_m$ algebras \cite{Awata2009,Awata2010,Taki2014,Awata2011a,Feigin2011}. In fact, they can also be defined in the vertical modules $(0,m)$ of the DIM algebra, seen as a product of q-$W_m$ with a q-Heisenberg algebra. Defined in this way, these states coincide with the horizontal v.e.v. of generalized AFS intertwiners \cite{Bourgine2017b}. In the case $m=2$ relevant to $U(2)$ gauge theories and q-Virasoro algebra, they read
\begin{align}
\begin{split}
&\dket{G}=\sum_{\l_1,\l_2} a_{\l_1}a_{\l_2}\bra{\vac}\Phi_{\l_1}^{(II\ast)}\Phi_{\l_2}^{(I\ast)}\ket{\vac}\ \dket{v_4,\l_1}\otimes\dket{v_3,\l_2},\\
&\dbra{G}=\sum_{\l_1,\l_2} a_{\l_1}a_{\l_2}\bra{\vac}\Phi_{\l_1}^{(I)}\Phi_{\l_2}^{(II)}\ket{\vac}\ \dbra{v_1,\l_1}\otimes\dbra{v_2,\l_2}.
\end{split}
\end{align}
As can be seen by comparing with expression \ref{def_CT} of the $\CT$-operator, the scalar product of two Gaiotto states reproduces the instanton partition function of the underlying gauge theory, i.e. $\la\CT\ra=\dbra{G}G\rangle\rangle$. In terms of brane-web diagram, these states can be represented as a cutting of the $U(2)$ diagram in half along a horizontal line, and vacuum states inserted at the endpoint (see figure \ref{fig_U2_Gaiotto} left). It was shown in \cite{Bourgine2016} that the q-Virasoro Whittaker condition follows from the intertwiners covariance properties under the action of the DIM algebra, that imposes
\begin{equation}
\left(\rho_{v_4}^{(0,1)}\otimes \rho_{v_3}^{(0,1)}\ \D(x^-_k)\right)\dket{G}=\left\{
\begin{array}{cc}
\cdots & (k<-1)\\
(u'_3)^{-1}\dket{G} & (k=-1)\\
0 & (k=0)\\
-\g^{-2}u_4^{-1}\dket{G}& (k=1)\\
\cdots & (k>1)\\
\end{array}\right.,
\end{equation} 
and a similar relation for $x_k^+$ and $\dbra{G}$. Incidentally, this condition also implies the following relations for the S-dual modes $y_k^\pm=\CS\cdot x_k^\pm$ and $b_k=\CS\cdot a_k$:
\begin{align}\label{act_SGaiotto}
\begin{split}
&\left(\rho_{v_4}^{(0,1)}\otimes \rho_{v_3}^{(0,1)}\ \D(y^+_{k<0})\right)\dket{G}=-\d_{k,-1}(\g u'_3)^{-1}\dket{G},\\
&\left(\rho_{v_4}^{(0,1)}\otimes \rho_{v_3}^{(0,1)}\ \D(y^-_{k<0})\right)\dket{G}=-\d_{k,-1}(\g u_4)^{-1}\dket{G},\\
&\left(\rho_{v_4}^{(0,1)}\otimes \rho_{v_3}^{(0,1)}\ \D(b_{k<0})\right)\dket{G}=0.
\end{split}
\end{align}

The 90$^\circ$ rotation of the brane-web diagrams associated to Gaiotto states are represented on figure \ref{fig_U2_Gaiotto} (right). In the algebraic language, they define the states $\ketS{GS}$ and $\braS{GS}$ belonging to the tensor product of two Fock modules $(-1,0)$,
\begin{equation}
\ketS{GS}=\CP\mat{-1}{0}{-1}{0}\left(1\otimes\left(\left(1\otimes\bra{\vac}\right)\Phi_4^{(SII\ast)}\right)\right)\Phi_3^{(SI\ast)}\ket{\vac},\quad \braS{GS}=\bra{\vac}\Phi_1^{(SI)}\left(1\otimes\left(\Phi_2^{(SII)}\left(1\otimes\ket{\vac}\right)\right)\right).
\end{equation} 
Using the known expression for these intertwiners, and the reflection property of the state $\ketS{\O}$, we can write these states as 
\begin{align}
\begin{split}
&\ketS{GS}=\sum_\l\left(\dfrac{\tu'_3}{\g\tu_4}\right)^{|\l|}\Zv(\l)\prod_{x\in\l}\chi_x^{-1}\ \CO_\l(\tv_3)^\dagger\ket{0}\in(-1,0)_{\tu_4}\otimes(-1,0)_{\tu_3},\\
&\braS{GS}=\sum_\l \left(\dfrac{\tu_2}{\g\tu'_1}\right)^{|\l|}\Zv(\l)\prod_{x\in\l}\chi_x\ \bra{0}\CO_\l(\tv_1)^\dagger\in(-1,0)_{\tu_1}\otimes(-1,0)_{\tu_2},
\end{split}
\end{align}
where $\CO_\l^\dagger$ is the conjugation of the following operator acting on the tensor product of two Fock spaces:
\begin{equation}
\CO_\l(v)=:\Phi_\vac(v)\prod_{x\in\l}\eta^+(\chi_x):\otimes:\Phi_\vac^\ast(v)\prod_{x\in\l}\eta^-(\chi_x):.
\end{equation}
The scalar product of these states reproduces the v.e.v. of the rotated Lax matrix, namely $\la\CT^\CS\ra=\braS{GS}\!\!\ketS{GS}$.\footnote{To show this fact, we can use the normal-ordering property for operators $\CO_\l^\dagger$:
\begin{equation}
\CO_{\l_1}(v_1)^\dagger\CO_{\l_2}(v_2)^\dagger=(-q_3v_1)^{-|\l_2|}(-v_2)^{|\l_1|}\prod_{x\in\l_1}\chi_x^{-1}\prod_{x\in\l_2}\chi_x\dfrac{\CG(v_1/v_2)\CG(v_1/(q_3v_2))}{N(\l_1,\l_2|v_1/v_2)N(\l_2,\l_1|v_2/v_1)}:\CO_{\l_1}(v_1)^\dagger\CO_{\l_2}(v_2)^\dagger:,
\end{equation}}

We would like to compare our rotated Gaiotto states with the states introduced by Kimura and Pestun in their formalism of quiver W-algebra \cite{Kimura2015,Kimura2016}. In fact, our problem contains two different q-$W_2$ algebra. The first one comes from the vertical module $(0,2)$, it is associated to the two D5-branes in the intermediate channel, and is involved in the q-AGT correspondence \cite{Alday2010,Awata2009,Awata2010}. This is the q-$W_2$ algebra acting on the genuine Gaiotto states $\dket{G}$ and $\dbra{G}$. The second q-$W_2$ algebra comes from the horizontal modules $(-1,1)\otimes(1,1)$ or $(-1,0)\otimes(-1,0)$, it is associated to the two NS5-branes, and acts on the rotated states $\ketS{GS}$ and $\braS{GS}$. This is the algebra corresponding to Kimura\&Pestun's quiver W-algebra. Indeed, following \cite{Mironov2016,Bourgine2017b}, we can rewrite the operators acting on the product of two Fock spaces in terms of the modes
\begin{equation}
k\b_k=-(\g^k-\g^{-k})\left(\g^{|k|/2}\a_k\otimes 1+\g^{-|k|/2}1\otimes\a_k\right),\quad ks_k=\dfrac{\g^{-|k|/2}}{1-q_2^{-k}}\a_k\otimes1-\dfrac{\g^{|k|2}}{1-q_2^{-k}}1\otimes\a_k.
\end{equation} 
The modes $\b_k=\rho^{(-1,0)}_{\tu_4}\otimes\rho_{\tu_3}^{(-1,0)}\ \D(a_{-k})$ describe the Cartan sector of the horizontal representation $(-2,0)$. By definition, they form a q-Heisenberg subalgebra. On the other hand, the modes $s_k$ define the q-Virasoro stress energy tensor in \cite{Shiraishi1995} (with $q=q_2$ and $t=q_1^{-1}$ here), they obey the commutation relations
\begin{equation}
[s_k,s_l]=-\dfrac1k\dfrac{1-q_1^{k}}{1-q_2^{-k}}(1+q_3^{k})\d_{k+l}.
\end{equation} 
Since the modes $\b_k$ and $s_k$ commute, Heisenberg and q-Virasoro components can be decoupled. The operators $\CO_\l$ can be written in terms of the vertex operator $\Sf(z)$ (introduced in \cite{Kimura2015}) by exploiting the property
\begin{equation}
\eta^+(z)\otimes\eta^-(z)=:\Sf(z)^{-1}\Sf(q_2z):,\quad \Sf(z)=\exp\left(\sum_{k\neq 0}z^{k}s_{-k}\right),
\end{equation} 
where the normal ordering of the modes $\a_k$ naturally extends to $s_k$. We have
\begin{equation}
\CO_\l(v)=:\prod_{x\in\l}\dfrac{\Sf(q_2\chi_x)}{\Sf(\chi_x)}\prod_{x\in\l_\infty}\dfrac{\Sf(\chi_x)}{\Sf(q_2\chi_x)}:=:\prod_{i=1}^{\infty}\Sf(vq_1^{i-1}q_2^{\l_i}):.
\end{equation} 
The Kimura-Pestun construction of the partition function proceeds from another decomposition, namely $\CZ=\langle0\ket{\CZ}$ with
\begin{align}
\begin{split}
\ket{\CZ}&=\CG(\tilde{\rf})\CG(\tilde{\rf}q_3^{-1})\sum_{\l_1,\l_2}(\tilde{\qf} q_3^{-1})^{|\l_1|+|\l_2|}\Zv(\vec{\tv},\vec\l):\CO_{\l_1}(\tv_1)\CO_{\l_2}(\tv_3):\ket{0}\\
&=\dfrac{\CG(\tilde{\rf}q_3^{-1})}{\CG(\tilde{\rf}^{-1})}\sum_{\l_1,\l_3}\prod_{\superp{l=1,3}{i=1\cdots\infty}}^{\to}\Sf(\tv_lq_1^{i-1}q_2^{\l_i^{(l)}})\ket{0}.
\end{split}
\end{align}
In the second line, the ordered product produces the extra factors of $\CG$ and $\Zv$ of the first line upon normal ordering, provided we introduce also the proper zero modes in $\Sf(z)$ \cite{Kimura2015}. Finally, note that since the rotated state $\ketS{GS}$ is constructed upon the modes $s_k$, it is annihilated by the positive modes $\b_{k>0}$, i.e. by the action of the Cartan modes $a_{-k}=\CS^{-1}\cdot b_{-k}$, which can be seen as part of a S-dual Whittaker condition \ref{act_SGaiotto}.

\section{Perspectives}
As we have seen in the three examples treated in this paper, the S-duality relations between $\CN=1$ gauge theories can be understood at the algebraic level in terms of twisted co-algebraic structure. The key ingredient is the equivalence \ref{prop_S} between the action of Miki's automorphism, and the replacement of representations $(\ell,\bell)\to(-\bell,\ell)$ that renders the rotation of branes. Despite the lack of a general statement, we were able to show this relation in the cases we needed to treat our three examples. A second important ingredient was the existence of two-tensors that translates the rotation of topological vertices in the language of intertwiners. This rotation map by two-tensors was then naturally extended to Lax matrices, and used to show the equality between their v.e.v. by studying the vacuum transformation. Unfortunately, a similar transformation for T-matrices could only be conjectured, but it was seen to provide the correct vacuum transformation property in the example of pure $U(2)$ gauge theory. In all these examples, the equality between the v.e.v. of operators associated to the two coproducts $\D$ and $\D_\CS$ reproduced the known S-duality relation for the underlying gauge theory. This is a strong indication that the formalism we introduced here is correct, even though several mathematical development would be required to make it fully rigorous. Eventually, it seems that a general proof of S-duality relations could be achieved by this method.

One of the strength of the algebraic engineering of 5D $\CN=1$ gauge theories is the possibility to extend the results to higher rank gauge groups, and higher rank (linear) quivers, without much effort. This kind of generalization seems also feasible here, but it would require a deeper study of horizontal representations $(m,0)$ with higher level $|m|>1$, supposedly S-dual to vertical representations $(0,m)$. The construction of new intertwiners coupled to these higher representations would also be needed. This is certainly the next natural step for this study. Another important, yet relatively easy, application for our formalism is the treatment of gauge theories with fundamental flavors. In fact, the simplest examples of $U(2)$ gauge theories with such matter fields can already be treated using the intertwiners constructed in this paper.

The application of this method to D-type quivers is a more challenging problem. The algebraic engineering of these gauge theories has been presented in \cite{Bourgine2017c}, it involves the introduction of reflection states in vertical modules. By the studying the action of Miki's automorphism on these states, one should be able to obtain the algebraic realization of the S-dual theory, a gauge theory with gauge group of type SO. Very interestingly, this approach could solve the longstanding problem of finding a expansion formula for the instanton partition function of these gauge theories with Young diagrams replaced by a more general combinatorial structure \cite{Nekrasov2004,Nakamura2014,Nakamura2015}.

The S-duality transformation of qq-characters is another interesting open problem. These objects are generating functions of Wilson loops \cite{Kim2016}, they encode a set of constraints on the partition function called \textit{non-perturbative Schwinger-Dyson equations} \cite{NPS,Nekrasov2015,Nekrasov2016b,Jeong2017,Jeong2018}. These constraints can be understood in terms of the action of DIM algebra \cite{Bourgine2015c,Bourgine2016}, and the qq-character obtained by insertion of an algebra element in the v.e.v. of $\CT$-operators, i.e. formally $\chi(z)\sim\la(\rho\otimes\rho\ \D(x^+(z)))\CT\ra$. It seems natural to expect that this object is transformed into $\la(\rho\otimes\rho\ \D_\CS(x^+(z)))\CT_\CS\ra$ under the action of Miki's automorphism. Yet, the proper transformation still remains to be worked out in details. Besides, the S-transformation of Wilson loops obtained in \cite{Agarwal2018} should also be reproduced. We hope to address this issue in the near future.

Very recently, the algebraic engineering has been extended to 4D $\CN=2$ quiver gauge theories in \cite{Bourgine2018a}. At first sight, the automorphism $\CS$ seems to be lost in the degenerate version of DIM algebra used in the 4D case. However, a more involved realization of S-duality might still be found. This is actually suggested by the correspondence observed in \cite{Nedelin2017} between the conformal blocks of the ordinary Virasoro algebra and the d-Virasoro algebra (a.k.a. a degenerate limit of q-Virasoro). Indeed, the former appears in the vertical representation of the degenerate DIM algebra (in the case of $U(2)$ gauge groups), while the former has been identified with the degenerate version of Kimura-Pestun's quiver W-algebra (in the $A_1$ case) obtained from the horizontal representation. Thus, a relation seems to exist between vertical and horizontal representations of the degenerate DIM algebra, but this problem certainly deserves a deeper study.

Finally, integrable aspects have not been discussed in this paper. Quantum integrable systems are usually built upon a quantum group symmetry algebra, but this construction also applies to quantum toroidal algebra (that are affine quantum groups) \cite{Feigin2015,Feigin2016,Feigin2016a}. In this way, it is possible to build a quantum integrable system using one of the universal R-matrices, either $\CR$ or $\CR_\CS$, of the DIM algebra. The presence of Miki's automorphism is a new feature of these integrable models, it should generate some important dualities among such systems. Moreover, the R-matrix is interpreted as a scattering factor in 2D integrable quantum field theories \cite{Zamolodchikov1978}, and it would be interesting to develop a similar interpretation for the two-tensor $\CF_\CS$ associated to Miki's automorphism twist. At a general level, combining S-duality and integrability is an enthralling perspective, and we hope to be able to push forward this discussion in a near future.

\section*{Acknowledgments} I would like to thank Omar Foda, Sasha Garbali, Joonho Kim, Antonio Sciarappa and Paul Zinn-Justin for discussions, and more particularly Yutaka Matsuo for valuable advice, and Elli Pomoni for useful comments on the final version of the manuscript. I am very grateful to Melbourne University, CEA-Saclay and DESY for their warm hospitality at the various stages of preparation of this article.

\appendix
\section{More on Miki's automorphism}\label{AppA}
This appendix presents a detailed discussion of Miki's automorphism. Specifically, we first derive the transformation of the Drinfeld currents' modes following \cite{Miki2007}. Then, we examine the action of these modes in the vertical representation $(0,1)$, and observe that their action defines a horizontal representation, leading to an equivalence between the two. Finally, we study the action of grading elements, and how they interplay with Miki's automorphism.

The first step is to introduce some notation. We denote by $y^\pm(z)=\CS\cdot x^\pm(z)$ and $\xi^\pm(z)=\CS\cdot\psi^\pm(z)$ the image of the Drinfeld currents under Miki's automorphism. Just like the original currents in \ref{com_ak}, the S-dual currents can be decomposed in terms of modes 
\begin{equation}
y^\pm(z)=\sum_{k\in\mathbb{Z}}z^{-k}y_k^\pm,\quad \xi^\pm(z)=\sum_{k\geq0}z^{\mp k}\xi_{\pm k}^\pm=\xi_0^\pm\exp\left(\pm\sum_{k>0}z^{\mp k}b_{\pm k}\right),
\end{equation}
with $y_k^\pm=\CS\cdot x_k^\pm$, $b_k=\CS\cdot a_k$ and $\xi^\pm_{\pm k}=\CS\cdot\psi^\pm_{\pm k}$. Note also that, due to the property \ref{action_CSCS}, the action of Miki's automorphism on these modes reads $\CS\cdot y_k^\pm=-x_{-k}^\mp$, $\CS\cdot b_k=-a_{-k}$ and $\CS\cdot\xi^\pm_{\pm k}=\psi^\mp_{\mp k}$. Finally, since $\CS$ is an automorphism, these new modes satisfy the q-commutation relations of the DIM algebra, and in particular the relations \ref{com_ak} that reads here
\begin{align}
\begin{split}\label{com_bk}
&[b_k,b_l]=-(\g^{k\bc}-\g^{-k\bc})c_k\d_{k+l},\quad [b_k,y_l^\pm]=\pm\g^{\pm |k|\bc/2}c_k y_{l+k}^\pm,\\
&[y^+_k,y^-_l]=\left\{
\begin{array}{l}
\k\g^{-(k-l)\bc/2}\xi^+_{k+l},\quad k+l>0\\
\k\g^{-(k-l)\bc/2}\xi_0^+-\k\g^{(k-l)\bc/2}\xi_0^-,\quad k+l=0\\
-\k\g^{(k-l)\bc/2}\xi^-_{k+l},\quad k+l<0.\\
\end{array}
\right.
\end{split}
\end{align}

\subsection{Derivation of Miki's transformation}
The expression for the S-dual modes $y_k^\pm$, $\xi_{\pm k}^\pm$ in terms of the original ones can be obtained from the transformation \ref{Miki_init} of the four modes $a_{\pm1}$, $x_0^\pm$ (and the two central charges $c$ and $\bc$) using the algebraic relations \ref{com_ak}. This construction is based on the observation that the second commutation relation in \ref{com_ak} allows us to reconstruct recursively the modes $x_k^\pm$ from $x_0^\pm$:
\begin{equation}\label{rel_xk_x0}
x^\pm_k=(\pm\g^{\pm c/2})^kc_1^{-k}\left(\text{ad}_{a_1}\right)^k x_0^\pm,\quad x^\pm_{-k}=(\pm\g^{\pm c/2})^kc_1^{-k}\left(\text{ad}_{a_{-1}}\right)^k x_0^\pm,\quad (k>0),
\end{equation}
where $\text{ad}_XY=[X,Y]$ denotes the adjoint action, and $c_1,\ \s_1$ are the specialization of the coefficients $c_k,\ \s_k$ defined in \ref{exp_g} and \ref{q-osc} at $k=1$. Note that $c_k/\s_k=(\g^k-\g^{-k})/k^2$ and $\k=\s_1^2/c_1$. Once the modes $x_k^\pm$ are known, it is possible to find the modes $\psi_{\pm k}^\pm$ using the third relation in \ref{com_ak}, 
\begin{equation}\label{rel_psik}
\psi_{\pm k}^\pm=\k^{-1}\hg^{\pm(k-2)/2}[x_{\pm1}^\pm,x_{\pm(k-1)}^\mp].
\end{equation}

We will start by deriving the expression of the modes $y_k^\pm$. Applying Miki's automorphism to the relations \ref{rel_xk_x0}, we find:
\begin{equation}
y^\pm_k=-(\pm)^{k}\g^{\mp k\bc/2}(\g-\g^{-1})^{k-1}c_1^{-k}\left(\text{ad}_{x_0^+}\right)^k a_{\mp1},\quad y^\pm_{-k}=-(\pm)^{k}\g^{\mp k\bc/2}(\g-\g^{-1})^{k-1}c_1^{-k}\left(\text{ad}_{x_0^-}\right)^k a_{\mp1}.
\end{equation}
The modes $a_{\pm1}$ can be eliminated using the commutation relation \ref{com_ak} once again, and we end up with
\begin{equation}\label{Miki_y}
y^\pm_k=(\pm)^k\g^{-(c\pm k\bc)/2}\s_1^{-(k-1)}\left(\text{ad}_{x_0^+}\right)^{k-1} x_{\mp1}^+,\quad y^\pm_{-k}=-(\pm)^k \g^{(c\mp k\bc)/2} \s_1^{-(k-1)}\left(\text{ad}_{x_0^-}\right)^{k-1} x^-_{\mp1}.
\end{equation} 
This is indeed the expression given by Miki in \cite{Miki2007}. It is readily observed that the generators $y_k^\pm$ have degrees $(\bd,-d)=(k,\mp1)$ (for $k\in\mathbb{Z}$).

The modes $\xi_{\pm k}^\pm$ can be obtained by applying Miki's automorphism to the relation \ref{rel_psik}, using the expressions of the modes $y_k^\pm$ obtained previously (here $k>1$),
\begin{equation}\label{Miki_xi}
\xi_{\pm k}^\pm=-(\mp)^k(\g-\g^{-1})\s_1^{-(k-1)}\g^{\mp c}\text{ad}_{x_{\mp1}^\pm}\left(\text{ad}_{x_0^\pm}\right)^{k-2}x_{\pm1}^{\pm},\quad \xi_{\pm1}^\pm=\pm\g^{\mp c}(\g-\g^{-1})x_0^\pm,\quad \xi_0^\pm=\g^{\mp c},
\end{equation} 
which coincide again with Miki's expressions \cite{Miki2007}. The modes $\xi_{\pm k}^\pm$ have degrees $(\bd,-d)=(\pm k,0)$.

\paragraph{Remark} An alternative expression for the modes $y_k^\pm$ and $\xi_{\pm k}^\pm$ can be obtained from the Baker-Campbell-Hausdorff formula,
\begin{equation}\label{BCH}
e^{\text{ad}_X}Y=e^{X}Ye^{-X}\implies \left(\text{ad}_X\right)^kY=\sum_{l=0}^k(-1)^{l}\left(\superp{k}{l}\right)\ X^{k-l}YX^l.
\end{equation} 

\paragraph{Dictionary with Miki's notations} In order to compare with the results of Miki \cite{Miki2007}, it is necessary to translate the notations as follows (from 'Miki'$\to$'Ours'):
\begin{align}
\begin{split}
&\g^2\to q_1,\quad (q\g)^{-2}\to q_2,\quad q^2\to q_3,\quad q\to\g,\quad C\to\hg=\g^c,\quad \mathcal{C}\to \psi_0^+=\g^{-\bc},\\
&(q-q^{-1})(\g^k-\g^{-k})a_k\to a_k,\quad (\g-\g^{-1})C^{-k/2}X_k^\pm\to\pm x_k^\pm,\quad \Phi_{\pm k}^\pm\to\g^{\pm\bc}\psi_{\pm k}^\pm.
\end{split}
\end{align}

\subsection{Choice of preferred direction}
We consider here a vertical representation of levels $(0,1)$ and examine the action of the dual modes $b_k$, $y^\pm_k$ and $\xi_k^\pm$. First, we observe that the modes $b_k$ form an Heisenberg sub-algebra,
\begin{equation}
[\rho_v^{(0,1)}(b_k),\rho_v^{(0,1)}(b_l)]=-(\g^k-\g^{-k})c_k\d_{k+l}.
\end{equation} 
In fact, the commutation relations between the modes $y_k^\pm$ and $b_k$ reproduce those of the modes $x_k^\pm$ and $a_k$ in a horizontal representation $(-1,0)$. As a consequence, it is possible to express them in terms of the q-oscillator modes $\a_k$ satisfying \ref{q-osc},
\begin{equation}
\rho^{(0,1)}_v(y^\pm(z))=-u^{\mp1}:\exp\left(\pm\sum_{k\neq0}\dfrac{z^{k}}{k}\g^{\pm|k|/2}\a_k\right):,\quad \rho^{(0,1)}_v(\xi^\pm(z))=\exp\left(\mp\sum_{k>0}\dfrac{z^{\mp k}}{k}(\g^k-\g^{-k})\a_{\mp k}\right).
\end{equation} 
Thus, we have found a hidden horizontal representation inside the vertical one, but it remains to identify the `horizontal states' in the vertical module. Let's start with the Fock vacuum $\ket{\vac}$, by definition it is annihilated by the positive modes $\a_{k>0}$, or, equivalently, by the modes $\rho^{(0,1)}_v(\xi^-_{-k<0})$. In Miki's relations \ref{Miki_xi}, these modes are expressed in terms of $x_k^-$, and therefore they annihilate the vertical vacuum $\dket{v,\vac}$. The same argument shows that $\rho^{(0,1)}_v(y^\pm_{-k<0})$ also annihilates the state $\dket{v,\vac}$, which is consistent with the Fock representation of the currents $y^\pm(z)$:
\begin{equation}
\rho^{(0,1)}_v(y^\pm(z))\ket{\vac}=-u^{\mp1}\exp\left(\mp\sum_{k>0}\dfrac{z^{-k}}{k}\g^{\pm|k|/2}\a_{-k}\right)\ket{\vac}.
\end{equation} 
Indeed, the r.h.s. contains only negative powers of $z$, which implies that the action of negative modes does vanish. Thus, we can identify the Fock vacuum with the vertical vacuum, up to a normalization factor that will be fixed later.

In order to identify the other states in the Fock space, we consider the S-duality relations \ref{Miki_init}, and more precisely $a_1=-(\g-\g^{-1})y_0^-$ and $b_1=(\g-\g^{-1})x_0^+$. Comparing the vertical action of these modes, with the action of oscillators in the Macdonald basis,
\begin{align}
\begin{split}
&\rho^{(0,1)}_v(a_1)\dket{v,\l}=-(\g-\g^{-1})\g vE_\l\dket{v,\l},\quad\rho_v^{(0,1)}(x_0^+)\dket{v,\l}=\sum_{x\in A(\l)}\res_{z=\chi_x}\dfrac1{z\CY_\l(z)}\dket{v,\l+x},\\
&\rho_v^{(0,1)}(y_0^-)\ket{P_\l}=-uE_\l\ket{P_\l},\quad \rho_v^{(0,1)}(b_1)\ket{P_\l}=(\g-\g^{-1})\sum_{x\in A(\l)}\res_{z=\chi_x}\dfrac1{z\CY_\l(z)}\ket{P_{\l+x}},
\end{split}
\end{align}
we observe that states can be simply identified as $\dket{v,\l}\sim\ket{P_\l}$ (since $\ket{P_\l}$ is defined as the non-degenerate eigenbasis of $\eta_0^+$), and that the horizontal weight is given by $u=-\g v$. The comparison of the actions of $b_1$ and $x_0^+$ further fixes the normalization factor to one. Generalizing our argument, we establish that the representations $(0,\pm1)$ and $(\mp1,0)$ are in fact equivalent, up to a rotation of the generators sending $x^\pm(z)$, $\psi^\pm(z)$ to $y^\pm(z)$, $\xi^\pm(z)$. Somehow, this choice of the Drinfeld currents can also be seen as a choice of preferred direction for the topological vertex.

In addition, Fock modules also contain a set of coherent states $\ketA{\vec w,\vec\mu}$, $\ketB{\vec w,\vec\mu}$ that will be constructed in appendix \ref{AppC} (note that we have labeled these states with the m-tuple Young diagrams $\vec\mu$ and the weight vector $\vec w$ here to avoid confusions). For instance, the states $\ketB{\vec w,\vec\mu}$ are characterized by the relation
\begin{equation}
\rho_v^{(0,1)}(\xi^-(\g^{1/2}z))\ket{\vec w,\vec\mu}_B=\g^{-2m}\left[\Psi_{\vec\mu}(z)\right]_-\ket{\vec w,\vec\mu}_B.
\end{equation}
In particular, since $\xi_{-1}^-\propto x_0^-$, they are eigenstates of $x_0^-$ in the vertical representation,
\begin{equation}
\rho^{(0,1)}(x_0^-)\ket{\vec w,\vec\mu}_B=\left(\g^{-1/2}\s_1\sum_{x\in\vec\mu}\chi_x^{-1}-\g^{-3/2}\sum_{l=1}^mw_l^{-1}\right)\ket{\vec w,\vec\mu}_B.
\end{equation}
They can be seen as a sort of Gaiotto states for a vertical module of lower level \cite{Bourgine2016}.

\subsection{Gradings and uniqueness of Miki's automorphism}\label{AppA3}
It turns out that Miki's automorphism is not unique due to the presence of two other automorphisms. These automorphisms, denoted $\t_\o$ and $\bt_\bo$, have been defined in \ref{def_gradings} using the grading operators $d$ and $\bd$. Their string theory realization can be understood by examination of their composition with representations. Indeed, comparing \ref{prop_gradings} with the explicit form of vertical/horizontal representations, we realize that the composed representations can be obtained alternatively from a simple shift of the weights: $\rho_{\vec v}^{(0,\pm m)}\circ\t_\o=\rho_{\o^{\mp1}\vec v}^{(0,\pm m)}$ and $\rho_u^{(\pm1,n)}\circ\t_\o\simeq\rho_{\o^{\mp n}u}^{(\pm1,n)}$ (up to a rescaling of the q-bosonic modes $\a_k$). Since the weights encode the position of the branes, we deduce that the action of $\t_\o$ corresponds to an overall translation along the NS5-direction (thus shifting the D5-positions corresponding to vertical weights). Similarly, the composition with $\bt_\bo$ leaves vertical representations invariant (up to a rescaling of the states, i.e. a change of basis) $\rho_{\vec v}^{(0,\pm m)}\circ\bt_\bo\simeq\rho_{\vec v}^{(0,\pm m)}$
but shifts the horizontal weights $\rho_u^{(\pm1,n)}\circ\bt_\bo=\rho_{\bo^{\pm1}u}^{(\pm1,n)}$. We deduce that this automorphism encodes the translation along the D5-direction, shifting NS5-branes positions. Note also that the two automorphisms $\t_\o$ and $\bt_\bo$ commute, as translations should do.

In this subsection, we investigate the interplay between these translations and the rotations of the brane-web realized by Miki's automorphism $\CS$. The $\CS$-transformation of the gradings follows from the invariance requirement of the commutation relations \ref{act_gradings} defining their action on Drinfeld currents: $(d,\bd)\to(-\bd,d)$. As a result, if an element $e$ has gradings $(d_e,\bd_e)$ under $(d,\bd)$, then the rotated element has the gradings $(\bd_e,-d_e)$. In fact, we have already observed this behavior for the modes $x_k^\pm$, $\psi_k^\pm$ and $y_k^\pm$, $\xi_k^\pm$ above. We also deduce the following properties:
\begin{equation}
\CS\circ\t_\o=\bt_{(\CS\cdot\o)^{-1}}\circ\CS,\quad \CS\circ\bt_{\bo}=\t_{\CS\cdot\bo}\circ\CS,\quad \o,\bo\in\mathbb{C}[c,\bc].
\end{equation} 
We recover here the fact that the translation axis is rotated by $\CS$, so that the translation $\bt_\o$ along the NS5-direction becomes a translation along (minus) the D5-direction $\bt_{(\CS\cdot\o)^{-1}}$, and the translation along the D5-direction becomes a translation along the NS5-direction. Note that here we consider a case slightly more general, allowing the shift parameters $\o$ and $\bo$ to be functions of the central charges $(c,\bc)$, so that the translation may now depend on the branes charges. This extra freedom is important in order to explore the possibility to twist Miki's automorphism.

We define the twisted Miki's automorphism by composition with the grading automorphisms, $\CS_{\o\bo}=\CS\circ\t_\o\circ\bt_{\bo}$. The requirement that $\CS_{\o\bo}$ is of order four imposes a restriction on the parameters $\o$ and $\bo$:
\begin{equation}
\o(\CS^2\cdot\o)^{-1}=(\CS^3\cdot\bo)(\CS\cdot\bo)^{-1}.
\end{equation} 
In order to maintain the property $\D_{\CS^2}=\D'$, we may further want to require that $\CS_{\o\bo}^2=\CS^2$, in which case we also have $\o=\CS^3\cdot\bo=(\CS\cdot\bo)^{-1}$. Taking the general Ansatz $\o=\o_0\g^{\mu c+\nu\bc}$, $\bo=\bo_0\g^{\bar\mu c+\bar\nu\bc}$ with $\o_0,\mu,\nu,\bar\o_0,\bar\mu,\bar\nu\in\mathbb{C}$, the condition $\CS_{\o\bo}^4=1$ implies $\bar\mu=\nu$ and $\bar\nu=-\mu$. The extra condition on $\CS_{\o\bo}^2$ further requires $\o_0=\bo_0=\pm1$. Thus, it is possible to choose another definition for Miki's automorphism in this paper, taking for instance $\CS_{\o\bo}$ with $\o=\pm\g^{\mu c+\nu\bc}$ and $\bo=\pm\g^{\nu c-\mu\bc}$ so that
\begin{equation}
S_{\o\bo}: a_1\to\pm(\g-\g^{-1})\g^{-\nu c+\mu\bc}x_0^+\to -a_{-1}\to\mp(\g-\g^{-1})\g^{\nu c-\mu\bc}x_0^-\to a_1.
\end{equation}

The main interest for deforming the definition of Miki's automorphism lies the possibility to define a different coproduct twist. Indeed, due to the presence of the two terms $c\otimes d$ and $d\otimes c$ in coproducts of the form
\begin{equation}
\D(cd)=cd\otimes1+1\otimes cd+c\otimes d+d\otimes c,
\end{equation} 
the twisting of $\D$ by $\CS$ and $\CS_{\o\bo}$ are different (here we have chosen the sign $\o_0=+1$ for simplicity):
\begin{equation}
\D_{\CS_{\o\bo}}=\left(\t_{\CS\cdot\bo_{(2)}}\bt_{(\CS\o_{(2)})^{-1}}\otimes\t_{\CS\cdot\bo_{(1)}}\bt_{(\CS\o_{(1)})^{-1}}\right)\D_{\CS},
\end{equation} 
where $\o_{(i)}=\g^{\mu c_{(i)}+\nu\bc_{(i)}}$, $c_{(1)}=c\otimes1$ and $c_{(2)}=1\otimes c$ (and the same for $\bc_{(i)}$, $\bo_{(i)}$,...).

\section{Building blocks of instanton partition functions}\label{AppB}
We present in this section a short reminder on the various building blocks used to construct instanton partition functions of 5D $\CN=1$ gauge theories. These building blocks are associated to the field content of the theory, with contributions coming from (anti)-fundamental matter fields, Chern-Simons terms, vector gauge multiplets and bifundamental fields:
\begin{align}\label{def_bblocks}
\begin{split}
&\Zf(\vec v,\vec\l|\vec m)=\prod_{f=1}^{n_f}\prod_{x\in\vec\l}(1-\chi_x/(q_3m_f)),\quad \Zaf(\vec v,\vec\l|\vec m)=\prod_{f=1}^{n_f}\prod_{x\in\vec\l}(1-m_f/\chi_x),\quad \ZCS(\vec v,\vec\l|\k)=\prod_{x\in\vec\l}(\chi_x)^\k\\
&\Zv(\vec v,\vec\l)=\prod_{l,l'=1}^m\dfrac1{N(v_l,\l^{(l)},v_{l'},\l^{(l')})},\quad \Zbf(\vec v,\vec\l,\vec v',\vec\l'|\mu)=\prod_{l=1}^m\prod_{l'=1}^{m'}N(v_l,\l^{(l)},\mu v_{l'}',\l^{(l')\prime}).
\end{split}
\end{align}
The last two building blocks are themselves written using the Nekrasov factor
\begin{equation}\label{Nek_factors}
N(v,\l|v',\l')=\prod_{\superp{x\in\l}{y\in\l'}}S\left(\dfrac{\chi_x}{\chi_y}\right)\times\prod_{x\in\l}\left(1-\dfrac{\chi_x}{q_3 v'}\right)\times\prod_{x\in\l'}\left(1-\dfrac{v}{\chi_x}\right),
\end{equation} 
with the function $S(z)$ defined in \ref{def_S}. In addition, perturbative (one-loop) contributions involve the function $\CG(z)$ defined for $|q_1|,|q_2|<1$ as
\begin{equation}\label{def_CG}
\CG(z)=\exp\left(-\sum_{k=1}^\infty\dfrac1k\dfrac{z^k}{(1-q_1^{k})(1-q_2^{k})}\right)=\prod_{i,j=1}^\infty\left(1-zq_1^{i-1}q_2^{j-1}\right).
\end{equation}
For instance, the one-loop vector contribution writes
\begin{equation}
\Zol(\vec v)=\prod_{l,l'=1}^m\CG(v_l/(q_3v_{l'})).
\end{equation} 

Note that the Nekrasov factor is invariant under the rescaling of the weights, and we can alternatively denote $N(\l,\l'|v/v')=N(v,\l|v',\l')$. Consequently, the rank $m=1$ vector contribution does not depend on the weight, and we will simply denote it $\Zv(\l)$. Alternatively, it is also possible to write down
\begin{equation}
\Zv(\l)=\prod_{x\in\l}\left(1-q_1^{l(x)+1}q_2^{-a(x)}\right)^{-1}\left(1-q_1^{-l(x)}q_2^{a(x)+1}\right)^{-1}
\end{equation} 
where $a(x)=\l_i-j$ and $l(x)=\tilde{\l}_j-i$ denote the arm and leg length (respectively) associated to the box $x=(i,j)\in\l$.

\section{Construction of the intertwiners $V\leftrightarrow H\times H$}\label{AppC}
In this appendix, we construct the intertwiners $\Phi=\Phi\mat{-1}{n}{1}{m}$, and $\Phi^{\ast}=\Phi^\ast\mat{-1}{n}{1}{m}$. By convention the horizontal representations $(-1,n)$ (resp. $(1,m)$) have weight $u$ (resp. $u'$), and the vertical representations $(0,n+m)$ have the usual vector of weights $\vec v$. These intertwiners are defined by the equations \ref{AFS_lemmas} that, in the present case, take the form
\begin{equation}\label{AFS_VHH}
\rho_{\vec v}^{(0,n+m)}(e)\Phi=\Phi\left(\rho_{u}^{(-1,n)}\otimes\rho_{u'}^{(1,m)}\ \D(e)\right),\quad \left(\rho_{u}^{(-1,n)}\otimes\rho_{u'}^{(1,m)}\ \D'(e)\right)\Phi^{\ast}=\Phi^{\ast}\rho_{\vec v}^{(0,n+m)}(e).
\end{equation}
We restrict ourselves to the case $n+m>0$.

\subsection{New operators in the Cartan}
For this construction, it is useful to introduce new operators in the Cartan sector, defined in terms of the modes $a_k$ as follows:
\begin{equation}
\U^\pm(z)=\exp\left(\sum_{k>0}\dfrac{(\g z)^{\mp k}}{\g^k-\g^{-k}}a_{\pm k}\right)\implies \psi^\pm(z)=\psi^\pm_0 \U^\pm(q_3^{-1}z)\U^\pm(z)^{-1}.
\end{equation}
The vertical and horizontal representations of these operators follows from those of the modes $a_k$, namely
\begin{align}
\begin{split}
&\rho^{(0,m)}_{\vec v}(\U^+(z))|\vec v,\vec\l\rangle\rangle=\left[\CY_{\vec\l}(z)\right]_+|\vec v,\vec\l\rangle\rangle,\quad
\rho^{(0,m)}_{\vec v}(\U^-(z))|\vec v,\vec\l\rangle\rangle=\prod_{l=1}^m(-z/v_l)\ \left[\CY_{\vec\l}(z)\right]_-|\vec v,\vec\l\rangle\rangle,\\
&\rho^{(1,n)}(\U^\pm(z))=\u^\pm(z),\quad \rho^{(-1,n)}(\U^\pm(z))=\u^\mp(q_3^{-1}z^{-1})^{-1},\quad \u^\pm(z)=\exp\left(\sum_{k>0}\dfrac{(\g z)^{\mp k}}{k}\a_{\pm k}\right).
\end{split}
\end{align}
The new vertex operators $\u^\pm(z)$ acting in the horizontal Fock modules satisfy the normal-ordering property
\begin{equation}
\u^+(z)\u^-(w)=S(\g w/z):\u^+(z)\u^-(w):.
\end{equation}
Note also that they can be used to decompose the vertex operators $\eta^\pm(z)$ and $\vphi^\pm(z)$:
\begin{equation}
\vphi^\pm(z)=\u^\pm(q_3^{-1}z)\u^\pm(z)^{-1},\quad \eta^+(z)=\u^-(\g^{-3/2}z)\u^+(\g^{-1/2}z)^{-1},\quad \eta^-(z)=\u^-(\g^{-1/2}z)^{-1}\u^+(\g^{-3/2}z).
\end{equation} 

\subsection{Coherent states in the Fock modules}
The main ingredient in the construction of our intertwiners is a set of coherent states belonging to the Fock modules. These states are parameterized by a vector of weights $\vec v$, and an $m$-tuple Young diagram, they are defined as follows:\footnote{Note that
\begin{equation}
\u^+(z)\u_\vac^-(v)=\left[1-v/(\g^{1/2}z)\right]_+:\u^+(z)\u_\vac^-(v):,\quad \u^+(z^{-1})\tup_\vac^-(v)=\left[(1-z/(\g^{5/2}v))^{-1}\right]_- :\u^+(z^{-1})\tup_\vac^-(v):
\end{equation}}
\begin{align}
\begin{split}
&\ketA{\vec v,\vec\l}=\prod_{x\in\vec\l}\u^-(\g^{-3/2}\chi_x)\prod_{l=1}^{m}\u_\vac^-(v_l)\ket{\vac},\quad \u_\vac^-(v)=\exp\left(-\sum_{k>0}\dfrac{(\g^{1/2}v)^k}{\s_k}\a_{-k}\right),\\
&\ketB{\vec v,\vec\l}=\prod_{x\in\vec\l}\u^-(\g^{-3/2}\chi_x^{-1})^{-1}\prod_{l=1}^{m}\tilde{\u}_\vac^-(v_l)\ket{\vac},\quad \tilde{\u}_\vac^-(v)=\exp\left(\sum_{k>0}\dfrac{(\g^{3/2}v)^{-k}}{\s_k}\a_{-k}\right).
\end{split}
\end{align}
In fact, these states can also be defined using the generalized AFS intertwiners,
\begin{align}
\begin{split}\label{def_coh_st}
&\ketA{\vec v,\vec\l}=\Phi\mat{0}{m}{1}{-m}\left(\dket{\vec v,\vec\l}\otimes\ket{\vac}\right)=\Phi_{\vec\l}\mat{0}{m}{1}{-m}\ket{\vac},\\
&\ketB{\vec v,\vec\l}=\Phi\mat{0}{-m}{1}{0}\left(\dket{\vec v,\vec\l}\otimes\ket{\vac}\right)=\Phi_{\vec\l}\mat{0}{-m}{1}{0}\ket{\vac},
\end{split}
\end{align}
where, in order to fix the norms such that to $\ketA{\vec v,\vec\l}=\ket{\vac}+\cdots$ (and $\ketB{\vec v,\vec\l}=\ket{\vac}+\cdots$), the weight $u'$ (resp. $u$) have been set to one.

Due to the intertwining property \ref{AFS_lemmas}, DIM's horizontal action action on such states can be written
\begin{align}
\begin{split}
&\rho_{u'=1}^{(1,0)}(e)\ketA{\vec v,\vec\l}=\Phi\mat{0}{m}{1}{-m}\left(\rho^{(0,m)}_{\vec v}\otimes\rho^{(1,-m)}_u\ \D(e)\right)\left(\dket{\vec v,\vec\l}\otimes\ket{\vac}\right),\\
&\rho_{u'}^{(1,-m)}(e)\ketB{\vec v,\vec\l}=\Phi\mat{0}{-m}{1}{0}\left(\rho^{(0,-m)}_{\vec v}\otimes\rho^{(1,0)}_{u=1}\ \D(e)\right)\left(\dket{\vec v,\vec\l}\otimes\ket{\vac}\right).
\end{split}
\end{align}
Since the action of DIM's algebra on the r.h.s. is already known, these identities can be used to characterize the algebraic properties of the coherent states. In particular, if the r.h.s. is annihilated by the action of an element of DIM algebra, so is the r.h.s.. Similarly, if the r.h.s. is an eigenstate, then it is also the case of the coherent state. As a result, we deduce the following properties:
\begin{align}
\begin{split}
&\a_{k>0}\ket{\vec v,\vec\l}_{A}=\g^{-k/2}\left[\dfrac{\s_k}{k}\sum_{x\in\vec\l}\chi_x^{k}-\sum_{l=1}^m(\g v_l)^{k}\right]\ket{\vec v,\vec\l}_{A},\\
&\a_{k>0}\ket{\vec v,\vec\l}_{B}=-\g^{-k/2}\left[\dfrac{\s_k}{k}\sum_{x\in\vec\l}\chi_x^{-k}-\sum_{l=1}^m(\g v_l)^{-k}\right]\ket{\vec v,\vec\l}_{B}.
\end{split}
\end{align}
Thus, these coherent states are eigenstates of the positive modes $\a_k$, these relations characterize them uniquely, up to their norm, and a possible linear shift by the vacuum state. These relations also imply:
\begin{equation}\label{prop_coh_st}
\vphi^+(\g^{-1/2}z)\ket{\vec v,\vec\l}_A=\left[\PsiY(z)\right]_+\ket{\vec v,\vec\l}_A,\quad \vphi^+(\g^{-1/2}z^{-1})\ket{\vec v,\vec\l}_B=\g^{-2m}\left[\PsiY(z)\right]_-\ket{\vec v,\vec\l}_B,\\
\end{equation} 
and further
\begin{align}\label{act_eta_coh}
\begin{split}
&\eta^+(z)\ket{\vec v,\vec\l}_A=\left[\CYY(z)^{-1}\right]_+\ \u^-(\g^{-3/2}z)\ket{\vec v,\vec\l}_A,\\
&\eta^-(\g^{-1}z)\ket{\vec v,\vec\l}_A=\left[\CYY(q_3^{-1}z)\right]_+\ \u^-(\g^{-3/2}z)^{-1}\ket{\vec v,\vec\l}_A,\\
&\eta^+(z^{-1})\ket{\vec v,\vec\l}_B=\prod_{l=1}^{n+m}\left(\dfrac{-z}{q_3v_l}\right)\left[\CYY(q_3^{-1}z)\right]_-\ \u^-(\g^{-3/2}z^{-1})\ket{\vec v,\vec\l}_B,\\
&\eta^-(\g^{-1}z^{-1})\ket{\vec v,\vec\l}_B=\prod_{l=1}^{n+m}\left(\dfrac{-v_l}{z}\right)\left[\CYY(z)^{-1}\right]_-\ \u^-(\g^{-3/2}z^{-1})^{-1}\ket{\vec v,\vec\l}_B.
\end{split}
\end{align}
Note that the action of algebra elements on the states $\ket{\vec v,\vec\l}_A$ involves operators expanded as series in $z^{-1}$ (in agreement with the radial ordering that assumes $|z|>|\chi_x|$), so that the function in the r.h.s. must also be expanded in inverse powers of $z$, which justifies the notation $\left[\cdots\right]_+$ employed. Similarly, the action on the states $\ket{\vec v,\vec\l}_A$ involves operators expanded as series in $z$, and the r.h.s. involves $\left[\cdots\right]_-$ (in agreement with $|z^{-1}|>|\chi_x^{-1}|$).

Dual states can also be introduced,
\begin{align}
\begin{split}
&\braA{\vec v,\vec\l}=\bra{\vac}\prod_{l=1}^{n+m}\u_\vac^+(v_l)\prod_{x\in\vec\l}\u^+(\g^{-3/2}\chi_x),\quad \u_\vac^+(v)=\exp\left(-\sum_{k>0}\dfrac{(\g^{1/2}v)^{-k}}{\s_k}\a_{k}\right),\\
&\braB{\vec v,\vec\l}=\bra{\vac}\prod_{l=1}^{n+m}\tup_\vac^+(v_l)\prod_{x\in\vec\l}\u^+(\g^{-3/2}\chi_x^{-1})^{-1},\quad\tup_\vac^+(v)=\exp\left(\sum_{k>0}\dfrac{(\g^{3/2}v)^{k}}{\s_k}\a_{k}\right),
\end{split}
\end{align}
they obey the relations\footnote{Note that
\begin{equation}
\u_\vac^+(v)\u^-(z)=\left[1-\g^{1/2}z/v\right]_-\ :\u_\vac^+(v)\u^-(z):,\quad \tup_\vac^+(v)\u^-(z^{-1})=\left[(1-\g^{5/2}v/z)^{-1}\right]_+ :\tup_\vac^+(v)\u^-(z^{-1}):
\end{equation}}
\begin{align}
\begin{split}
&\braA{\vec v,\vec\l}\vphi^-(\g^{-1/2}z)=\g^{-2(n+m)}\left[\PsiY(z)\right]_- \braA{\vec v,\vec\l},\\
&\braA{\vec v,\vec\l}\eta^+(\g^{-1}z)=\prod_{l=1}^{n+m}\left(\dfrac{z}{-q_3v_l}\right)\left[\CYY(q_3^{-1}z)\right]_- \braA{\vec v,\vec\l}\u^+(\g^{-3/2}z)^{-1},\\
&\braA{\vec v,\vec\l}\eta^-(z)=\prod_{l=1}^{n+m}\left(\dfrac{-v_l}{z}\right)\left[\CYY(z)^{-1}\right]_- \braA{\vec v,\vec\l}\u^+(\g^{-3/2}z)\\
&\braB{\vec v,\vec\l}\vphi^-(\g^{-1/2}z^{-1})=\left[\PsiY(z)\right]_+ \braB{\vec v,\vec\l}\\
&\braB{\vec v,\vec\l}\eta^+(\g^{-1}z^{-1})=\left[\CYY(z)^{-1}\right]_+\braB{\vec v,\vec\l}\u^+(\g^{-3/2}z^{-1})^{-1},\\
&\braB{\vec v,\vec\l}\eta^-(z^{-1})=\left[\CYY(q_3^{-1}z)\right]_+\braB{\vec v,\vec\l}\u^+(\g^{-3/2}z^{-1}).
\end{split}
\end{align}
Note also the remarkable inner product:
\begin{equation}
\braA{\vec v_1,\vec\l_1}\vec v_2,\vec\l_2\rangle_A=\braB{\vec v_2,\vec\l_2}\vec v_1,\vec\l_1\rangle_B=\dfrac{\Zbf(\vec v_2,\vec \l_2,\vec v_1,\vec \l_1|\g^{-1})}{\prod_{l=1}\prod_{l'=1}\CG(v_{l'}^{(2)}/\g v_l^{(1)})}.
\end{equation}

\subsection{Construction of $\Phi^{\ast}$}
The construction of $\Phi^{\ast}$ turns out to be a little easier than $\Phi$. Explicitly, the equation \ref{AFS_VHH} for each generating current of the DIM algebra read:
\begin{align}
\begin{split}\label{cond_phis}
&\Phi^{\ast}\rho_{\vec v}^{(0,n+m)}(x^+(z))=\left[u'z^{-m}\ 1\otimes\eta^+(z)-u^{-1}\g^{n+m}z^{n}\ \eta^-(\g^{-1}z^{-1})\otimes\vphi^-(\g^{1/2}z)\right]\Phi^{\ast},\\
&\Phi^{\ast}\rho_{\vec v}^{(0,n+m)}(x^-(z))=\left[(u')^{-1}\g^{-n-m}z^m\ \vphi^-(\g^{1/2}z^{-1})\otimes\eta^-(\g^{-1}z)-uz^{-n}\ \eta^+(z^{-1})\otimes1\right]\Phi^{\ast},\\
&\Phi^{\ast}\rho_{\vec v}^{(0,n+m)}(\psi^\pm(z))=\g^{\mp(n+m)}\left[\vphi^\mp(\g^{\pm1/2}z^{-1})\otimes\vphi^\pm(\g^{\mp1/2}z)\right]\Phi^{\ast}.
\end{split}
\end{align}
In order to solve these equations, we need to define the reflection operators acting on the tensor product of Fock modules:
\begin{equation}\label{def_CU}
\CU_\pm(z)=\exp\left(\sum_{k>0}\dfrac1{\s_k}\ \a_{\pm k}\otimes\a_{\pm k}\right).
\end{equation}
Here, we will employ $\CU_-$, while $\CU_+$ will be used in the next subsection in the construction of $\Phi_\CS$. This reflection operator satisfies the important properties
\begin{align}\label{prop_CU_I}
\begin{split}
&\left(1\otimes\u^+(z)\right)\CU_-=\CU_-\left(\u^-(q_3^{-1}z^{-1})\otimes\u^+(z)\right)\implies\left(1\otimes\vphi^+(z)\right)\CU_-=\CU_-\left(\vphi^-(z^{-1})^{-1}\otimes\vphi^+(z)\right),\\
&\left(\u^+(z)\otimes1\right)\CU_-=\CU_-\left(\u^+(z)\otimes\u^-(q_3^{-1}z^{-1})\right)\implies\left(\vphi^+(z)\otimes1\right)\CU_-=\CU_-\left(\vphi^+(z)\otimes\vphi^-(z^{-1})^{-1}\right).
\end{split}
\end{align}
Applied to the vacuum $\ket{\vac}\otimes\ket{\vac}$, it produces the reflector state defined in \ref{def_reflector}.

The intertwiner $\Phi^\ast$ is defined as the action of $\CU_-$ on the tensor product of the two coherent states constructed previously,
\begin{equation}
\ketS{\vec v,\vec\l}=n_{\vec\l}^\ast\ \CU_-\left(\ket{\vec v,\vec\l}_B\otimes\ket{\vec v,\vec\l}_A\right).
\end{equation}
We have introduced here the normalization factor $n_{\vec\l}^\ast=(\bra{\vac}\otimes\bra{\vac})\ketS{\vec v,\vec\l}$ that will be determined later. We observe the property
\begin{equation}\label{prop_I}
\g^{\mp(n+m)}\left(\vphi^\mp(\g^{\pm1/2}z^{-1})\otimes\vphi^\pm(\g^{\mp1/2}z)\right)\ketS{\vec v,\vec\l}=\g^{-(n+m)}\left[\PsiY(z)\right]_\pm\ketS{\vec v,\vec\l},
\end{equation}
obtained as follows. In the $+$ case, the reflection property \ref{prop_CU_I} applied to $\vphi^+(\g^{-1/2}z)$ eliminates $\vphi^-(\g^{-1/2}z^{-1})$ in the first space. This operator $\vphi^+(\g^{-1/2}z)$ acting on $\ket{\vec v,\vec\l}_A$ further produces the function $\PsiY(z)$ as a result of the property \ref{prop_coh_st}. This function has to be expanded in inverse powers of $z$, i.e. it corresponds to $\left[\PsiY(z)\right]_+$. Similarly, in the $-$ case, one considers $\vphi^+(\g^{-1/2}z^{-1})$, and its action on $\ket{\vec v,\vec\l}_B$ produces the function $\PsiY(z)$ expanded in powers of $z$. This shows the intertwining property for the Cartan currents $\psi^\pm(z)$.

Let us now turn to the action of $\rho^{(0,n+m)}(x^+(z))$, and, employing a similar technique, compute
\begin{align}\small
\begin{split}
&\left(1\otimes\eta^+(z)\right)\ketS{\vec v,\vec\l}=\left[\CYY(z)^{-1}\right]_+n_{\vec\l}^\ast\CU_-\left(\u^-(\g^{-3/2}z^{-1})^{-1}\otimes\u^-(\g^{-3/2}z)\right)\left(\ket{\vec v,\vec\l}_B\otimes\ket{\vec v,\vec\l}_A\right)\\
&\dfrac{z^{n+m}}{\prod_l(-v_l)}\left(\eta^-(\g^{-1}z^{-1})\otimes\vphi^-(\g^{1/2}z)\right)\ketS{\vec v,\vec\l}=\left[\CYY(z)^{-1}\right]_-n_{\vec\l}^\ast\CU_-\left(\u^-(\g^{-3/2}z^{-1})^{-1}\otimes\u^-(\g^{-3/2}z)\right)\left(\ket{\vec v,\vec\l}_B\otimes\ket{\vec v,\vec\l}_A\right).
\end{split}
\end{align}
Taking the difference, and using that
\begin{equation}
\left[\CYY(z)^{-1}\right]_+-\left[\CYY(z)^{-1}\right]_-=\sum_{x\in A(\vec\l)}\d(z/\chi_x)\res_{z=\chi_x}z^{-1}\CYY(z)^{-1},
\end{equation} 
we find
\begin{equation}\label{prop_II}
\left(1\otimes\eta^+(z)-\dfrac{z^{n+m}}{\prod_l(-v_l)}\eta^-(\g^{-1}z^{-1})\otimes\vphi^-(\g^{1/2}z)\right)\ketS{\vec v,\vec\l}=\sum_{x\in A(\vec\l)}\d(z/\chi_x)\res_{z=\chi_x}z^{-1}\CYY(z)^{-1}\ \dfrac{n_{\vec\l}^\ast}{n_{\vec\l+x}^\ast}\ketS{\vec\l+x}.
\end{equation} 
Finally, from
\begin{align}\small
\begin{split}
&\left(\vphi^-(\g^{1/2}z^{-1})\otimes \eta^-(\g^{-1}z)\right)\ketS{\vec v,\vec\l}=\left[\CYY(q_3^{-1}z)\right]_+n_{\vec\l}^\ast\CU_-\left(\u^-(\g^{-3/2}z^{-1})\otimes\u^-(\g^{-3/2}z)^{-1}\right)\left(\ket{\vec v,\vec\l}_B\otimes\ket{\vec v,\vec\l}_A\right)\\
&z^{-n-m}\prod_l(-q_3v_l)\left(\eta^+(z^{-1})\otimes1\right)\ketS{\vec v,\vec\l}=\left[\CYY(q_3^{-1}z)\right]_-n_{\vec\l}^\ast\CU_-\left(\u^-(\g^{-3/2}z^{-1})\otimes\u^-(\g^{-3/2}z)^{-1}\right)\left(\ket{\vec v,\vec\l}_B\otimes\ket{\vec v,\vec\l}_A\right),
\end{split}
\end{align}
we get
\begin{equation}\label{prop_III}\small
\left(\vphi^-(\g^{1/2}z^{-1})\otimes \eta^-(\g^{-1}z)-z^{-n-m}\prod_l(-q_3v_l)\eta^+(z^{-1})\otimes1\right)\ketS{\vec v,\vec\l}=\sum_{x\in R(\vec\l)}\d(z/\chi_x)\res_{z=\chi_x}z^{-1}\CYY(q_3^{-1}z)\ \dfrac{n_{\vec\l}^\ast}{n_{\vec\l-x}^\ast}\ketS{\vec\l-x}.
\end{equation}

The three properties \ref{prop_I}, \ref{prop_II} and \ref{prop_III} imply that $\ketS{\vec v,\vec\l}$ transforms as the states $\dket{\vec v,\vec\l}$ in the vertical representation. More precisely, the quantity
\begin{equation}
\Phi^{\ast}=\sum_{\vec\l}a_{\vec\l}\ketS{\vec v,\vec\l}\langle\bra{\vec v,\vec\l}\implies \Phi_{\vec\l}^{\ast}=\Phi^{\ast}\ket{\vec v,\vec\l}\rangle=\ketS{\vec v,\vec\l},
\end{equation} 
satisfies the equations \ref{cond_phis}, provided that we set
\begin{equation}\label{cond_weights}
uu'=\prod_{l=1}^{n+m}(-\g v_l),\quad n_{\vec\l}^\ast=(u')^{|\vec\l|}\prod_{x\in\vec\l}\chi_x^{-m}.
\end{equation} 
This condition upon the weights is in agreement with the weights conservation relation.

\subsection{Construction of $\Phi$}
In this case, the constraints \ref{AFS_VHH} for $\Phi$ take the explicit form
\begin{align}\label{cond_phi}
\begin{split}
&\rho_{\vec v}^{(0,n+m)}(x^+(z))\Phi=\Phi\left[-u^{-1}z^n\ \eta^-(z^{-1})\otimes1+u'\g^{n+m}z^{-m}\ \vphi^+(\g^{1/2}z^{-1})\otimes\eta^+(\g^{-1}z)\right],\\
&\rho_{\vec v}^{(0,n+m)}(x^-(z))\Phi=\Phi\left[-u\g^{-n-m}z^{-n}\ \eta^+(\g^{-1}z^{-1})\otimes\vphi^+(\g^{1/2}z)+(u')^{-1}z^m\ 1\otimes\eta^-(z)\right],\\
&\rho_{\vec v}^{(0,n+m)}(\psi^\pm(z))\Phi=\g^{\mp(n+m)}\Phi\left[\vphi^\mp(\g^{\mp1/2}z^{-1})\otimes\vphi^\pm(\g^{\pm1/2}z)\right].
\end{split}
\end{align}
The intertwiner is obtained from the dual states $\braS{\vec v,\vec\l}$ defined as the action of $\CU_+$ on the tensor product of dual coherent states,
\begin{equation}
\braS{\vec v,\vec\l}=\left(\braB{\vec v,\vec\l}\otimes\braA{\vec v,\vec\l}\right)\CU_+\ n_{\vec\l}.
\end{equation} 
Using the properties of the reflection operator $\CU_+$ introduced in \ref{def_CU},
\begin{align}\label{prop_CU_II}
\begin{split}
&\CU_+\left(1\otimes\u^-(z)\right)=\left(\u^+(q_3^{-1}z^{-1})\otimes\u^-(z)\right)\CU_+\implies\CU_+\left(1\otimes\vphi^-(z)\right)=\left(\vphi^+(z^{-1})^{-1}\otimes\vphi^-(z)\right)\CU_+,\\
&\CU_+\left(\u^-(z)\otimes1\right)=\left(\u^-(z)\otimes\u^+(q_3^{-1}z^{-1})\right)\CU_+\implies\CU_+\left(\vphi^-(z)\otimes1\right)=\left(\vphi^-(z)\otimes\vphi^+(z^{-1})^{-1}\right)\CU_+,
\end{split}
\end{align}
and following the same steps as before, it is possible to show that the intertwiner
\begin{equation}
\Phi=\sum_{\vec\l} a_{\vec\l}\ket{\vec v,\vec\l}\rangle\braS{\vec v,\vec\l}
\end{equation} 
obeys the relations \ref{cond_phi} provided that the weights observe the conservation relation \ref{cond_weights}, and that the normalization factor takes the form
\begin{equation}
n_{\vec\l}=\braS{\vec v,\vec\l}\left(\ket{\vac}\otimes\ket{\vac}\right)=u^{|\vec\l|}\g^{-(n+m)|\vec\l|}\prod_{x\in\vec\l}\chi_x^{-n}.
\end{equation} 

\section{Vacuum S-transformation}
In this appendix, we determine the action on the vacuum of some of the S-dual modes $y_k^\pm$, $\xi_{\pm k}^\pm$ using Miki's relations \ref{Miki_y} and \ref{Miki_xi}.

\subsection{Example I}\label{AppD1}
First, we introduce a shortcut notation for the following states in the tensor product $(-1,0)\otimes (0,1)$:
\begin{equation}
\ket{0}=\ket{\vac}\otimes\dket{v,\vac},\quad \ket{1}=\a_{-1}\ket{\vac}\otimes\dket{v,\vac}.
\end{equation}
Expanding the action of the coproduct of DIM generators, we obtain the following characterizations:
\begin{align}\label{ax_I}
\begin{split}
&\left(\rho_u^{(-1,0)}\otimes\rho_v^{(0,1)}\ \D(a_{-k})\right)\ket{0}=-\dfrac{\g^k-\g^{-k}}{k}(\g^{3/2}v)^{-k}\ket{0},\quad (k>0),\\
&\left(\rho_u^{(-1,0)}\otimes\rho_v^{(0,1)}\ \D(x^-_{k})\right)\ket{0}=\left\{
\begin{array}{cc}
0 & (k<0)\\
-u\g^{-1}\ket{0} & (k=0)\\
-u\g^{-1}(1-q_3)v\ket{0}-u\g^{-3/2}\ket{1} & (k=1)\\
\cdots & (k>1)
\end{array}\right.\\
&\left(\rho_u^{(-1,0)}\otimes\rho_v^{(0,1)}\ \D(x^-_{k})\right)\ket{1}=\left\{
\begin{array}{cc}
0 & (k<-1)\\
u\g^{-3/2}\s_1\ket{0} & (k=-1)\\
u\g^{-3/2}\s_1(1-q_3)v\ket{0}-u\g^{-1}(1-\s_1\g^{-1})\ket{1} & (k=0)\\
\cdots & (k>0)
\end{array}\right.
\end{split}
\end{align}
These properties leads to identify the state $\ket{0}$ with the state $\ketB{v,\vac}$ in the Fock module with representation $(-1,1)_{u'=-\g uv}$ (up to a norm). 

In order to obtain the action of the S-dual modes, we use Miki's relations \ref{Miki_y} and \ref{Miki_xi} to show that:
\begin{align}
\begin{split}\label{dual_I}
&\left(\rho_u^{(-1,0)}\otimes\rho_v^{(0,1)}\ \D(y^+_{-k})\right)\ket{0}=\d_{k,0}(\g^{3/2}v)^{-1}\ket{0},\quad (k\geq0)\\
&\left(\rho_u^{(-1,0)}\otimes\rho_v^{(0,1)}\ \D(\xi^-_{-k})\right)\ket{0}=\left(\g\d_{k,0}-(\g-\g^{-1})(-u\g^{-2})^k\right)\ket{0}.
\end{split}
\end{align}
The derivation of the first identity is straightforward, but the second one requires some extra explanation in the case of modes with index $k>1$. Starting from Miki's formula \ref{Miki_xi} and noticing that $\D(x_{-1}^-)$ annihilates $\ket{0}$, we find
\begin{align}
\begin{split}\label{D4}
\left(\rho_u^{(-1,0)}\otimes\rho_v^{(0,1)}\ \D(\xi^-_{-k})\right)\ket{0}&=\g^{-1}(\g-\g^{-1})\s_1^{-(k-1)}\left(\rho_u^{(-1,0)}\otimes\rho_v^{(0,1)}\ \D\left(\left(\left(\text{ad}_{x_0^-}\right)^{k-2} x_{-1}^-\right)x_1^-\right)\right)\ket{0}\\
&=-u\g^{-5/2}(\g-\g^{-1})\s_1^{-(k-1)}\left(\rho_u^{(-1,0)}\otimes\rho_v^{(0,1)}\ \D\left(\left(\text{ad}_{x_0^-}\right)^{k-2} x_{-1}^-\right)\right)\ket{1}.
\end{split}
\end{align}
Then, from \ref{ax_I}, we get
\begin{align}
\begin{split}
&\left(\rho_u^{(-1,0)}\otimes\rho_v^{(0,1)}\ \D(e^{z x_0^-})\right)\ket{0}=e^{-u\g^{-1}z}\ket{0},\\
&\left(\rho_u^{(-1,0)}\otimes\rho_v^{(0,1)}\ \D(e^{z x_0^-})\right)\ket{1}=\g^{1/2}(1-q_3)v\left(e^{-u\g^{-1}(1-\s_1\g^{-1})z}-e^{-u\g^{-1}z}\right)\ket{0}+e^{-u\g^{-1}(1-\s_1\g^{-1})z}\ket{1},
\end{split}
\end{align}
which gives, using the Baker-Campbell-Hausdorff formula \ref{BCH},
\begin{align}
\begin{split}
&\implies \left(\rho_u^{(-1,0)}\otimes\rho_v^{(0,1)}\ \D(e^{z x_0^-}x_{-1}^-e^{-z x_0^-})\right)\ket{1}=u\g^{-3/2}\s_1e^{-\s_1 u\g^{-2}z}\ket{0}\\
&\implies \left(\rho_u^{(-1,0)}\otimes\rho_v^{(0,1)}\ \D\left(\left(\text{ad}_{x_0^-}\right)^{k} x_{-1}^-\right)\right)\ket{1}=-\g^{1/2}(-\s_1 u\g^{-2})^{k+1}\ket{1}.
\end{split}
\end{align}
This shows \ref{D4}, and thus the second identity in \ref{dual_I}. Note that in the application to the first example, $u$ and $v$ become $\tu_2$ and $\tv_2$.

\subsection{Example II}\label{AppD2}
We start again by introducing the shortcut notations:
\begin{equation}
\ket{0}=\ket{\vac}\otimes\ket{\vac},\quad \ket{1}=\a_{-1}\ket{\vac}\otimes\ket{\vac},\quad \ket{1'}=\ket{\vac}\otimes\a_{-1}\ket{\vac}.
\end{equation} 
From
\begin{equation}
\left(\rho_{u}^{(-1,0)}\otimes\rho_{u'}^{(1,1)}\ \D(x^-(z))\right)\ket{0}=(u')^{-1}\ \ket{\vac}\otimes z\u^-(\g^{-1/2}z)^{-1}\ket{\vac}-u\g^{-1}\ \u^-(\g^{-5/2}z^{-1})\ket{\vac}\otimes\ket{\vac},
\end{equation} 
we deduce by expansion that:
\begin{equation}
\left(\rho_{u}^{(-1,0)}\otimes\rho_{u'}^{(1,1)}\ \D(x^-_k)\right)\ket{0}=\left\{
\begin{array}{cc}
\cdots & (k< -2)\\
-(u')^{-1}\g^{1/2}\ket{1'} & (k=-2)\\
(u')^{-1}\ket{0} & (k=-1)\\
-u\g^{-1}\ket{0} & (k=0)\\
-u\g^{-5/2}\ket{1} & (k=1)\\
\cdots & (k>1)
\end{array}\right.
\end{equation}
Using the relation \ref{Miki_y}, it implies
\begin{equation}
\left(\rho_{u}^{(-1,0)}\otimes\rho_{u'}^{(1,1)}\ \D(y^+_{-k<0})\right)\ket{0}=-\d_{k,1}(u')^{-1}\g^{-1/2}\ket{0}.
\end{equation} 

The action of $\xi_{-k}^-$ is more complicated to evaluate. First, we need the action of $x_0^-$ and $x_{-1}^-$ on the state $\ket{1}$ that can be obtained using a similar method as before:
\begin{align}
\begin{split}
&\left(\rho_{u}^{(-1,0)}\otimes\rho_{u'}^{(1,1)}\ \D(x^-_0)\right)\ket{1}=-u\g^{-1}(1-\g^{-1}\s_1)\ket{1},\\
&\left(\rho_{u}^{(-1,0)}\otimes\rho_{u'}^{(1,1)}\ \D(x^-_{-1})\right)\ket{1}=(u')^{-1}\ket{1}+u\g^{-1/2}\s_1\ket{0}.
\end{split}
\end{align}
From these identities, we deduce successively
\begin{align}
\begin{split}
&\left(\rho_{u}^{(-1,0)}\otimes\rho_{u'}^{(1,1)}\ \D(e^{z\ \text{ad}_{x_0^-}}x_{-1}^-)\right)\ket{0}=(u')^{-1}\ket{0},\\
&\left(\rho_{u}^{(-1,0)}\otimes\rho_{u'}^{(1,1)}\ \D(e^{z\ \text{ad}_{x_0^-}}x_{-1}^-)\right)\ket{1}=(u')^{-1}\ket{1}+u\g^{-1/2}\s_1e^{-zu\g^{-2}\s_1}\ket{0},\\
&\left(\rho_{u}^{(-1,0)}\otimes\rho_{u'}^{(1,1)}\ \D(\text{ad}_{x_1^-}\ e^{z\ \text{ad}_{x_0^-}}x_{-1}^-)\right)\ket{0}=u^2\g^{-3}\s_1e^{-zu\g^{-2}\s_1}\ket{0},\\
\end{split}
\end{align}
and finally, for $k>0$,
\begin{equation}
\left(\rho_{u}^{(-1,0)}\otimes\rho_{u'}^{(1,1)}\ \D(\xi_{-k}^-)\right)\ket{0}=(1-q_3)(-u\g^{-2})^k\ket{0}.
\end{equation} 
Note that the calculation for the mode $k=1$ must be performed separately. In the application to the transformation of the Lax matrix in the second example, $u$ and $u'$ become $\tu_2$ and $\tu'_2$.

\subsection{Case $U(2)$}\label{AppD3}
Starting with the state $\ket{0}=\ket{\vac}\otimes\ket{\vac}\in (1,1)_u'\times(-1,1)_u$, we observe that
\begin{equation}
\left(\rho_{u}^{(1,1)}\otimes\rho_{u'}^{(-1,1)}\ \D'(x_k^-)\right)\ket{0}=\left\{
\begin{array}{cc}
\cdots & (k<-1)\\
u^{-1}\ket{0} & (k=-1)\\
0 & (k=0)\\
-\g^{-2}u'\ket{0} & (k=1)\\
\cdots & (k>1)\\
\end{array}\right.
\end{equation} 
Using Miki's formulas \ref{Miki_y} and \ref{Miki_xi}, we deduce for $k>0$:
\begin{align}
\begin{split}
&\left(\rho_{u}^{(1,1)}\otimes\rho_{u'}^{(-1,1)}\ \D'(y_{-k}^+)\right)\ket{0}=-\d_{k,1}\g^{-1}u^{-1}\ket{0},\\
&\left(\rho_{u}^{(1,1)}\otimes\rho_{u'}^{(-1,1)}\ \D'(y_{-k}^-)\right)\ket{0}=-\d_{k,1}\g^{-1}u'\ket{0},\\
&\left(\rho_{u}^{(1,1)}\otimes\rho_{u'}^{(-1,1)}\ \D'(\xi^-(z))\right)\ket{0}=\ket{0}.
\end{split}
\end{align}
Here $u$ corresponds to $\tu'_2$ in the main text, and $u'$ to $\tu'_3$. Similarly, we observe for the dual state $\bra{0}=\bra{\vac}\otimes\bra{\vac}$, 
\begin{equation}
\bra{0}\left(\rho_{u}^{(-1,1)}\otimes\rho_{u'}^{(1,1)}\ \D(x_k^+)\right)=\left\{
\begin{array}{cc}
\cdots & (k<-1)\\
-\g^{2}u^{-1}\bra{0} & (k=-1)\\
0 & (k=0)\\
u'\bra{0} & (k=1)\\
\cdots & (k>1)\\
\end{array}\right.
\end{equation} 
and deduce from the same formulas \ref{Miki_y} and \ref{Miki_xi} that for $k>0$:
\begin{align}
\begin{split}
&\bra{0}\left(\rho_{u}^{(-1,1)}\otimes\rho_{u'}^{(1,1)}\ \D(y_{k}^+)\right)=-\d_{k,1}\g u^{-1}\bra{0},\\
&\bra{0}\left(\rho_{u}^{(-1,1)}\otimes\rho_{u'}^{(1,1)}\ \D(y_{k}^-)\right)=-\d_{k,1}\g u'\bra{0},\\
&\bra{0}\left(\rho_{u}^{(-1,1)}\otimes\rho_{u'}^{(1,1)}\ \D(\xi^+(z))\right)=\bra{0}.
\end{split}
\end{align}
This time $u$ and $u'$ denote $\tu'_1$ and $\tu'_4$ respectively in the application to the $U(2)$ theory.


\end{document}